\documentclass[table,twocolumn]{aastex62}
\usepackage{amstext}
\usepackage{amsmath}
\usepackage{apjfonts}
\usepackage[capbesideposition={top}]{floatrow}
\usepackage{float}
\usepackage{cancel}
\usepackage{floatrow}

\newcommand{\beqar}{\begin{eqnarray}}
\newcommand{\eeqar}{\end{eqnarray}}

\newcommand{\beq}{\begin{equation}}
\newcommand{\eeq}{\end{equation}}

\definecolor{nick}{HTML}{006400}

\newcommand{\hamr}{\texttt{H-AMR}}
\newcommand{\gizmo}{\texttt{GIZMO}}
\newcommand{\tetgen}{\texttt{TetGen}}
\newcommand{\divb}{\vec{\nabla}\cdot\vec{B}}
\begin{document}
\title{H-AMR FORGE'd in FIRE I: Magnetic state transitions, jet launching and radiative emission in super-Eddington, highly magnetized quasar disks formed from cosmological initial conditions}

\correspondingauthor{Nick Kaaz}
\email{nkaaz@u.northwestern.edu}

\author[0000-0002-5375-8232]{Nicholas Kaaz}
\affiliation{Department of Physics \& Astronomy, Northwestern University, Evanston, IL 60202, USA}
\affiliation{Center for Interdisciplinary Exploration \& Research in Astrophysics (CIERA), Evanston, IL 60202, USA}

\author[0000-0003-4475-9345]{Matthew Liska}
\affiliation{Center for Relativistic Astrophysics, Georgia Institute of Technology, Howey Physics Bldg, 837 State St NW, Atlanta, GA 30332, USA}

\author[0000-0002-9182-2047]{Alexander Tchekhovskoy}
\affiliation{Department of Physics \& Astronomy, Northwestern University, Evanston, IL 60202, USA}
\affiliation{Center for Interdisciplinary Exploration \& Research in Astrophysics (CIERA), Evanston, IL 60202, USA}

\author[0000-0003-3729-1684]{Philip F. Hopkins}
\affiliation{TAPIR, Mailcode 350-17, California Institute of Technology, Pasadena, CA 91125, USA}

\author[0000-0003-2982-0005]{Jonatan Jacquemin-Ide}
\affiliation{Center for Interdisciplinary Exploration \& Research in Astrophysics (CIERA), Evanston, IL 60202, USA}

\begin{abstract} 
Quasars are powered by supermassive black hole (SMBH) accretion disks, yet standard thin disk models are inconsistent with many observations. Recently, \citet{FORGE1} simulated the formation of a quasar disk feeding a SMBH of mass $M=1.3\times10^7\,M_\odot$ in a galaxy. The disk had surprisingly strong toroidal magnetic fields that supported it vertically from gravity and powered rapid accretion. What feedback can such a system produce? To answer this, we must follow the gas to the event horizon. For this, we interpolated the quasar into the general-relativistic radiation magnetohydrodynamics code $\verb|H-AMR|$ and performed 3D simulations with BH spins $a=0$ and $a=0.9375$. This remapping generates magnetic monopoles, which we erase using a novel divergence cleaning approach. Despite the toroidal magnetic field's dominance at large radii, vertical magnetic flux builds up near the event horizon, leading to a magnetic state transition within the inner $200$ gravitational radii of the disk. This powers strong winds and, for spinning BHs, relativistic jets that can spin-down the BH within $5-10\,{\rm Myrs}$. Sometimes, vertical magnetic fields of opposite polarity reach the BH, causing a polarity inversion event that briefly destroys the jets and, possibly, the X-ray corona. These strong fields power accretion at rates $5\times$ the Eddington limit, which can double the BH mass in $5-10\,{\rm Myrs}$. When $a=0.9375$ ($a=0$), the energy in mechanical outflows and radiation equals about $60\%$ ($10\%$) and $100\%$ ($3\%$) of the accreted rest mass energy, respectively. Much of the light escapes in cool, $\gtrsim1300\,{\rm au}$ photospheres, consistent with quasar microlensing and spectral energy distributions.
\end{abstract}


\section{Introduction}
\label{sec:intro}

It has been expected for decades that luminous quasars are powered by supermassive black hole (SMBH) accretion disks \citep{schmidt_1963,salpeter_1964,lyndenbell_1969}, which grow the mass \citep{soltan_1982,chokshi_1992} and alter the spin \citep{bardeen_1970,thorne_1974,gammie_2004,sasha_2012,sasha_2015,narayan_2022,bev_2024} of the SMBH significantly. A slew of observed correlations between SMBH masses and host galaxy properties \citep{magorrian_1998,ferrarese_2000,gebhardt_2000,hopkins_2007_observed,aller_richstone_2007,kormendy_2011} also suggests that quasars and other types of active galactic nuclei (AGN) strongly influence the evolution of their host galaxies, likely by injecting energy into their surroundings \citep{silk_rees_1998,king_2003,dimatteo_2005,murray_2005,phil_2005b,torrey_2020}, which is consistent with the widespread detection of outflows in quasars and Seyfert galaxies \citep{crenshaw_2000,dunn_2010,sturm_2011,zakamska_2016,williams_2017}. These outflows may be powered by radiation \citep{proga_2000,fabian_2009,phil_2010a,sadowski_narayan_2016,ricci_2017}, magnetized winds \citep{bp_1982,ferreira_1995,hawley_2015,jon_2019,jon_2021} or relativistic jets \citep{bz_1977,sasha_2011,wagner_2012,gaspari_2012,mukherjee_2016,kwan_2023}, all of which ultimately require an accretion disk \citep[though, see][]{aris_2024}. Although theoretical models have predicted the structure and emission of such disks for several decades \citep{ss73,nt73}, significant gaps between theory and observation remain.

``Classical'' accretion disks -- those that are geometrically thin, optically thick and mechanized by turbulent viscosity \citep[for a review see][]{pringle_1981} -- are thermally and viscously unstable in their inner regions \citep{lightman_eardley_1974,shakura_sunyaev_1976,piran_1978}. Yet, there is no observational evidence of the flickering that would indicate such an instability exists in either quasars or X-ray binary (XRB) disks \citep{gierlinski_done_2004,done_2004}. While there are many proposed mechanisms that could stabilize these disks \citep{krolik_1998,begelman_pringle_2007,neilsen_2011,jiang_2016,sadowski_2016,liska_kaaz_2023}, no consensus has been reached. Such disks are also unstable to gravitational fragmentation in their outer regions \citep{toomre_1964,goldreich_lyndenbell_1965,shlosman_1989,shlosman_1990,gammie_2001,goodman_2003}, which suggests that gas destined to power quasars instead forms stars. Proposed solutions include stabilization by strong magnetic fields \citep{begelman_pringle_2007} or the radiation pressure of the formed stars \citep{thompson_2005}.

\begin{figure*}[bht]
    \centering
    \includegraphics[width=\textwidth]{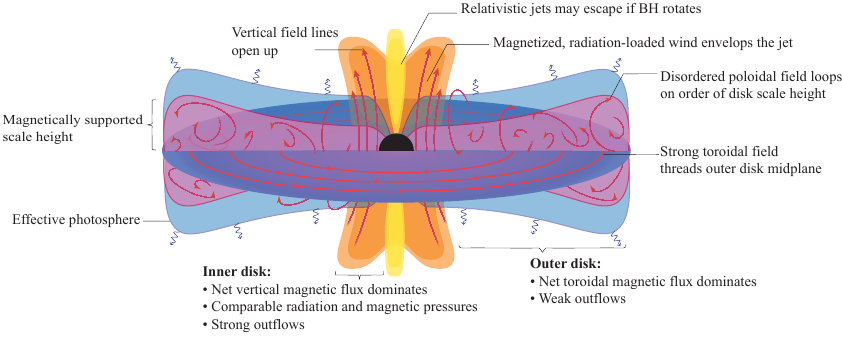}
    \caption{Cartoon illustrating the primary features of the quasar disk. Magnetic field lines are depicted in reddish orange. The velocity field is not depicted, but the radial velocity is $\gtrsim1-10\%$ of Keplerian throughout. \textbf{Outer disk.} The outer disk exhibits strong net toroidal flux which supports a density scale height of aspect ratio $H/r\gtrsim0.2-0.5$. The thermal and radiation pressure is subdominant. The disk contains ``intermediate scale'', $\mathcal{O}(H)$ poloidal field loops of alternating polarity. The effective photosphere is even more vertically extended than the disk. \textbf{Inner disk.} Intermediate scale   poloidal field loops in the outer disk are advected to the inner disk, where they are relatively large-scale. This results in a magnetic state transition where the disk switches from having dominant net toroidal magnetic flux to dominant net vertical magnetic flux. The thermal pressure is still weak, but the radiation and magnetic pressures are approximately equal. The net vertical flux threading the inner disk enables radiation-dominated magnetized winds. If the BH rotates, it launches a relativistic \citet{bz_1977} jet that is enveloped by these winds.}
    \label{fig:cartoon}
\end{figure*}

Even if the stability issues of classical thin disks were avoidable, there are multiple inconsistencies between theory and observations \citep[for a review of many of these inconsistencies, see][]{davis_tchekhovskoy_2020}. For instance, quasar microlensing observations \citep{pooley_2007,morgan_2010,blackburne_2011} indicate that the half-light radii of quasars can be an order of magnitude or more larger than what classical disk models predict and depends qualitatively differently on wavelength. Quasar spectra also commonly feature an excess of soft X-ray emission \citep{turner_pounds_1989,masnou_1992,walter_fink_1993}, which usually requires an unexplained soft contribution to the X-ray reflection spectrum \citep{ross_fabian_1993} or a warm, Comptonized corona \citep{magdziarz_1998}. There are also many commonly observed multi-phase structures in quasars \citep{netzer_2015}, including broad-line regions \citep{kaspi_2005,peterson_2006}, dusty torii \citep{antonucci_1993}, warm absorbers \citep{halpern_1984}, and molecular masers \citep{greenhill_1995}, which are unlikely to arise from a classical thin disk and mostly require a mechanism to levitate gas in a variety of thermal states above the disk.

Underlying both the challenges to quasar accretion disk theory and the mysteries of AGN-galaxy symbiosis is the question of how the disks form in the first place. This is a difficult question to answer because the disparate scales involved are hard to model. SMBHs gravitationally attract the surrounding gas on Bondi scales \citep[$\gtrsim10-10^4\,{\rm pc}$, ][]{bondi_1952}, the SMBH's gravity shapes the kinematics of stars/gas within its radius of influence ($\sim GM/\sigma_{\rm galaxy}^2\sim10\,{\rm pc}$, where $\sigma_{\rm galaxy}$ is the galactic stellar velocity dispersion and $M$ is the SMBH mass), yet most of the energy leaves the disk near the event horizon ($\gtrsim10^{-8}-10^{-6}\,{\rm pc}$) and can impact gas up to the scales of the intracluster medium \citep[$\gtrsim10^3-10^6\,{\rm pc}$, most notably in the Perseus cluster, e.g. ][]{boehringer_1993,churazov_2000,zhuravleva_2016}. Numerical studies thus have to make compromises to model such vast scales. Some authors have artificially reduced the Bondi radius to make the full-scale problem tractable \citep{sean_2021,aris_2022,me_2023,aris_2024} while  others have introduced ``multi-zone'' methods, which model the gas at each scale pseudo-independently and stitch together a global, steady-state solution \citep{hyerin_2023,hyerin_2024}. There are also many works that have followed the gas from the nuclear regions of the galaxy down to BH feeding scales via repeated ``zoom-ins'' \citep{phil_2010b, sean_2018,sean_2020,minghao_2023,minghao_2024} or via Lagrangian hyper-refinement \citep{angles-alcazar_2021}. 

In a recent series of papers \citep[][henceforth ``H24'', collectively]{FORGE3,FORGE1,FORGE2}, the authors
simulated the formation of a quasar accretion disk from first principles and incorporated a variety of multiphysics (e.g., radiation magnetohydrodynamics, thermochemistry, star formation). H24 accomplished this by beginning with cosmological initial conditions, zooming in on an individual galaxy, and then using the Lagrangian hyper-refinement method to resolve the nascent quasar disk down to $\gtrsim300\,r_{\rm g}$, where $r_{\rm g}\equiv GM/c^2$ is the gravitational radius of the SMBH. The resulting disk was distinct from tradiational accretion disks: it cooled rapidly on sub-orbital timescales and was dominated by toroidal magnetic fields with plasma $\beta\equiv p_{\rm t}/p_{\rm b}\ll 1$, where $p_{\rm t}$ and $p_{\rm B}$ are the gas thermal and magnetic pressures, respectively. These properties result from the mechanism of disk assembly: the SMBH accretes dynamically cold, already-magnetized gas from the interstellar medium (ISM), resulting in streams of gas that circularize with preferentially strong, $p_{\rm B} \gg p_{\rm t}$, toroidal magnetic fields supplied by the advected toroidal magnetic flux, which, in turn, drives rapid, super-Eddington accretion. This is also quite similar to \citet{gaburov_2012}, who performed the first of such simulations of SMBH disks formed from the disruption of cold, magnetized interstellar gas, and found similarly dominant toroidal magnetic flux. In both H24 and \citet{gaburov_2012}, these strong magnetic fields stabilized the gas from star formation \citep[as specifically focused on in][]{FORGE3}, resolving the long-standing issue of gravitational fragmentation in classical accretion disk models.

Recently, \citet{MDD} developed models of magnetically dominated disks (MDDs) to explain this new type of accretion flow analytically \citep[see also][]{johansen_levein_2008}. These MDDs are distinct from magnetically arrested disks \citep[MADs,][]{bisnovatyikogan_1974,igumenschchev_2003,narayan_2003,sasha_2011}, which are defined by having strong net vertical magnetic flux, but still have strong toroidal fields (although low \textit{net} toroidal flux) and maintain $\beta\sim1$ in their midplanes. Despite the nomenclature ``arrested'', MADs actually accrete extremely efficiently through a combination of magnetized turbulence and winds \citep[e.g., ][]{vikram_2024}, leading to inflow within a tenth of an orbital timescale \citep{igumenschchev_2003,narayan_2003,jon_2021,nico_2024}. Traditional MADs are radiatively inefficient and have usually been studied without explicit cooling or radiation. However, the MAD state may also extend to more luminous regimes where the net vertical magnetic flux is saturated \citep[e.g.,][ Lowell et al. in prep 2024]{avara_2016,curd_2023,nico_2024,liska_kaaz_2024}. 

In this work, we take the H24 simulation a step further, and follow the quasar accretion flow all the way to the SMBH event horizon. We do this by remapping the quasar accretion flow, which was simulated using \gizmo{} \citep{GIZMO}, into the general-relativistic radiation magnetohydrodynamics (GRRMHD) code, \hamr{} \citep{HAMR}. Here, we focus on the overall magnetic and outflow properties of the disk, and leave other important details such as angular momentum transport, the effects of misalignment between the disk and BH spin, and detailed observational predictions to later papers. To set the stage for our results, we show a cartoon in Figure \ref{fig:cartoon}, which highlights the main aspects of the accretion system. Magnetic field lines are depicted in reddish orange. The disk undergoes a ``magnetic state transition'', which distinguishes the inner disk from the outer disk \citep[similarly, see][]{liska_2020,jon_2024}. The outer disk has strong net toroidal magnetic flux that supports the disk vertically. The poloidal (i.e., radial and vertical) field loops in this region are unstructured and have extents that are on the order of the disk scale height. These loops are advected by the accretion flow to the event horizon, where they open up and become large scale. These inner regions have strong net vertical magnetic flux and launch radiation-loaded magnetized winds. If the BH is rotating, it launches a \citet{bz_1977} jet. The effective photosphere vertically extends to distances much larger than the disk scale height, except at the polar axis where the jet clears out gas and allows the photosphere to reach down to the innermost part of the disk. 

We summarize our simulation remapping approach, which includes a novel technique to clean magnetic field divergences, in Section \ref{sec:numerics}. In Section \ref{sec:results}, we describe our results, including the magnetic state transition (Sec.~\ref{sec:results:multiscale}-\ref{sec:results:topology}), the resulting outflows (Sec.~\ref{sec:results:outflows}), and the time evolution of horizon-scale properties and their implications for the cosmological evolution of SMBH (Sec.~\ref{sec:results:evolution}).  In Section \ref{sec:discussion}, we discuss some of the observational implications of our results (Sec.~\ref{sec:discussion:fluxflip}-\ref{sec:discussion:microlensing}), emphasize certain caveats of our work (Sec.~\ref{sec:discussion:caveats}) and summarize our findings (Sec.~\ref{sec:discussion:summary}). 

\begin{figure*}[bht]
    \centering
    \includegraphics[width=\textwidth]{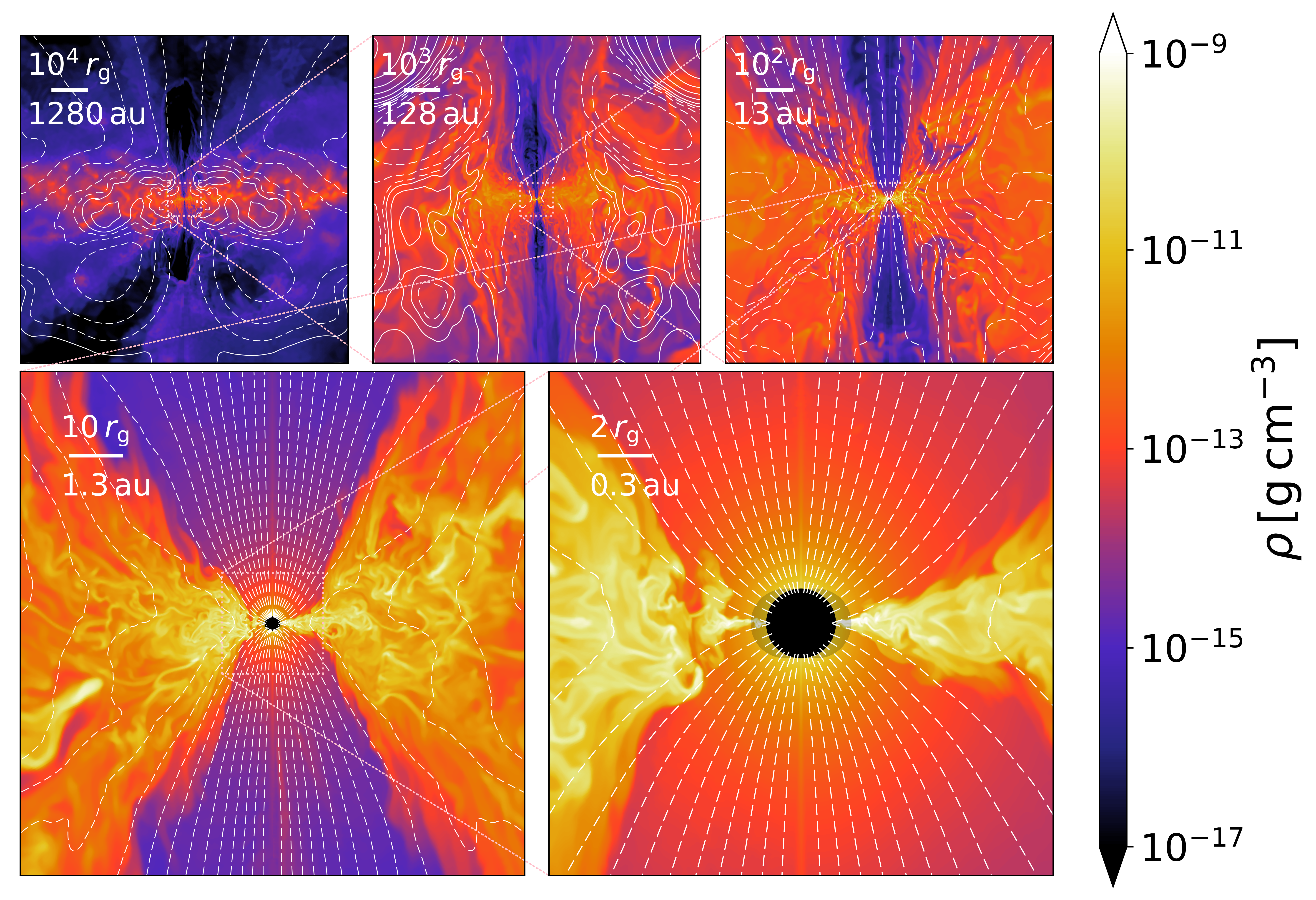}
    \caption{Density contours of the accretion flow depicted over five orders of magnitude in scale separation for run HS at time $t=111650\,r_{\rm g}/c$. The top left panel shows the full extent of the disk (outer radius $\approx5\times10^4\,r_{\rm g}=6400\,{\rm au}$). Subsequent zoom-ins follow the gas down to the event horizon (black). We also show the ergosphere (gray shaded pumpkin-shape) for the nearly maximally rotating ($a=0.9375$) BH. In each panel we also draw contours of the square root of the axisymmetrized poloidal magnetic flux (Eq. \ref{eq:magnetic_flux}), where solid/dashed lines indicate +/- values of $\Phi_{\rm P}$. The large scale ($\gtrsim200\,r_{\rm g}=26\,{\rm au}$) disk contains poloidal flux structures of alternating polarity, whereas closer to the event horizon the field is dipolar.}
    \label{fig:multiscale}
\end{figure*}

\section{Remapping From GIZMO to H-AMR}
\label{sec:numerics}

\subsection{Interpolation} 
\label{sec:numerics:interpolation}
H24 carried out their simulation using the \gizmo{} code \citep{GIZMO}, which is Newtonian and evolves gas quantities on an unstructured mesh of finite elements. \hamr{} \citep{HAMR} is a general-relativistic code and uses a non-uniform grid with both static and adaptively refined meshes, so our interpolation scheme must negotiate these differences. We describe this in detail in Appendix \ref{app:interpolation} and provide a brief overview here. 

We remap the following \gizmo{} quantities: the gas density ($\rho$), the thermal+radiation pressure ($p_{\rm t+r}\equiv p_{\rm t}+p_{\rm r}$), the velocity ($v_i$), and the magnetic field ($\mathcal{B}_i$). We neglect self-gravity since our region of interest ($<10^5\,r_{\rm g}$) is dominated by the SMBH gravity. \hamr{} evolves the gas internal energy density, $u_{\rm g}$, which we relate to the \gizmo{} $p_{\rm t+r}$ via the equation of state for a perfect, ideal gas,
\begin{equation}
    u_{\rm g}^{(\hamr{})} = p_{\rm t+r}^{(\gizmo{})}/(\gamma-1)
    \label{eq:eos_remap},
\end{equation}
where we are converting the sum of radiation and thermal pressure in \gizmo{} to the thermal pressure only in \hamr{}. This is justified since the emission timescales in the disk are short, so the radiation module in \hamr{} rapidly converts thermal energy into radiation. Since $p_{\rm r}\gg p_{\rm t}$ through much of the inner disk (the region that we primarily focus on here), we adopt an adiabatic index $\gamma=4/3$ in Equation \ref{eq:eos_remap}.  

We interpolate to \hamr{} by first organizing the \gizmo{} resolution elements into a tetrahedral mesh using the software \tetgen{} \citep{TETGEN}. We load this mesh into \hamr{}, and then find the tetrahedra that host each \hamr{} grid point. Then, we interpolate the data associated with each tetrahedron onto the enclosed grid point. Afterwards, we transform the interpolated vector quantities ($\mathcal{B}_i$, $v_i$) into spherical coordinates. We then transform the vectors from the Newtonian spherical coordinates to modified Kerr-Schild coordinates via the relation, 
\begin{equation}
\begin{aligned}
    \mathcal{B}_i &= \sqrt{g_{ii}}B^i, \\
    v_i &= \sqrt{g_{ii}}u^i,
\end{aligned}
\label{eq:mag_vel_phys}
\end{equation}
where $B^i$ is the contravariant magnetic field vector, $u^i$ is the contravariant four-velocity, $g_{\mu\nu}$ is the covariant metric, Latin indices span from $1$ to $3$, and Greek indices span from $0$ to $3$. This coordinate conversion is exact asymptotically far from the BH because the spatial components of our metric are asymptotically diagonal and the \gizmo{} data is non-relativistic. Because of this, $u^t\approx1$ throughout the domain of the \gizmo{} solution, and so we can regard each interpolated scalar quantity as a ``fluid-frame'' quantity and no further modification is necessary.

\subsection{Divergence Cleaning}
Once we have interpolated all quantities onto the \hamr{} grid, they have been faithfully translated from \gizmo{} up to the truncation error of the interpolation scheme. This truncation error is acceptably small for all quantities except for the magnetic fields. This is because the divergence of the magnetic fields, $\divb$, must be kept as close to zero as possible to reliably perform magnetohydrodynamic calculations. \hamr{} uses a staggered mesh to evolve magnetic fields and conserves the initial value of $\divb$ to machine precision. After the interpolation, the maximum value of $|\divb|$ is $\approx6000$ in code units (see Section \ref{sec:numerics:simulation}) and $\approx0.1$ in normalized units ($\Delta x |\divb/B|$, where $\Delta x$ is the average cell width) . This pollutes our solution with unphysical magnetic monopoles, which we must ``clean'' to obtain trustworthy results. Divergence cleaning usually involves an elliptic equation, which must be solved over the entire grid simultaneously. This is prohibitively expensive and especially complicated on our spherical, non-uniform grids with static and adaptive mesh refinement. To make progress, we have introduced a novel version of ``scalar'' divergence cleaning \citep[e.g.,][]{balsara_2004}, which splits the global problem up into many small local problems \citep[similar in spirit, but different in practice, is a cell-by-cell method introducd by][]{silberman_2019}. We describe this approach in detail in Appendix \ref{app:divb_cleaning}, where we have verified that the maximum value of $\Delta x |\divb/B|$ after cleaning is $\mathcal{O}(10^{-7})$. Importantly, our cleaning approach largely leaves the interpolated magnetic field the same. We provide a map of $|\divb|$ before and after cleaning in Figure \ref{fig:app:divb} in the Appendix. The combined process of interpolation and divergence cleaning takes a few hours for our problem.

\subsection{H-AMR Simulation Setup}
\label{sec:numerics:simulation}
\textit{Code, grid and boundary conditions.} \hamr{} \citep{HAMR} is a GPU-accelerated, three-dimensional, general-relativistic, radiation magnetohydrodynamics (GRRMHD) code.  We use a stationary metric and a spherical polar grid that is centered on the BH, where the base grid is uniform in the modified Kerr-Schild coordinates, ${\rm log}r$, $\theta$ and $\varphi$. The inner grid radius is sufficiently deep inside the event horizon ($r_{\rm in} = 0.9\,r_{\rm H}$) such that the region beyond the event horizon is causally disconnected from the inner radial grid boundary. The outer radius of the simulation is at $r_{\rm out}\approx10^5\,r_{\rm g}$ where
\begin{equation}
    r_{\rm g}=GM/c^2
\end{equation}
is the BH gravitational radius, $r_{\rm H}=r_{\rm g}(1+\sqrt{1-a^2})$ is the BH event horizon radius, and $M=1.3\times10^7M_\odot$ is the BH mass. We use 1D static mesh refinement, such that there are fewer cells in the $\varphi$ direction closer to the pole \citep{liska_2018,HAMR}. This keeps the aspect ratio of each cell roughly constant and avoids the so-called cell-squeezing problem at the poles. We employ local adaptive timestepping, which allows different blocks to evolve at different timesteps and increases the accuracy and speed of our simulations \citep{koushik_2019}. Our radial boundary conditions are outflowing, our polar boundary conditions are transmissive \citep[see supplementary information in][]{liska_2018}, and our azimuthal boundary conditions are periodic. We assume a perfect ideal gas with internal energy density $u_{\rm g}=p_{\rm t}/(\gamma-1)$. 

\textit{Radiation treatment.} \hamr{} uses the two-moment ``M1'' closure scheme for radiation \citep{levermore_1984}, which posits that there is always a frame in which the radiation field is isotropic. We separately evolve the radiation energy density, $E_{\rm r}$ (in the radiation frame), and the radiation velocity, $u_{\rm r}^\mu$ (which is the first moment of the frequency-integrated intensity). We operator split the Riemann problem to separately calculate the radiative and non-radiative fluxes. We evolve radiation using a second-order implicit-explicit (IMEX) time integrator, wherein the radiative flux is calculated during the explicit part of the timestep. We use absorption-averaged opacities that are valid from $T\gtrsim10^4\,{\rm K}$ to $T\lesssim10^{10}\,{\rm K}$ and the high densities (relative to, e.g., the interstellar medium) modeled in the disk. The opacities include free-free, bound-free and bound-bound processes fit over a range of density and temperatures, a Gaussian profile in temperature at $\approx1.5\times10^5\,{\rm K}$ to treat the iron line opacity, synchrotron opacities fitted for SMBHs, and electron scattering opacities which account for thermal Comptonization, as described in \citet{mckinney_2014,mckinney_2017}. Our opacities assume local thermodynamic equilibrium and depend on $\rho$, gas temperature $T_{\rm g}$, and radiation temperature $T_{\rm r}$. We assume thermal radiation such that $T_{\rm r}=(\hat{E}_{\rm r}/a)^{1/4}$, where $\hat{E}_{\rm r}$ is the fluid-frame radiation energy density and $a$ is the radiation constant. The gas temperature is given by $T_{\rm g} = (\gamma-1)u_{\rm g}\mu_{\rm g}m_{\rm H}c^2/k_{\rm B}\rho$, where we use the mean molecular weight $\mu_{\rm g}=\frac{4}{6X+Y+2}\approx0.61$ (where $X=0.70$ and $Y=0.28$ are the adopted hydrogen and helium mass fractions, respectively) and we set $\gamma=5/3$ whenever radiation is activated. While our opacities are reliable within the optically thick body of the disk, they are not well-motivated beyond the photosphere (which, we will see, is very large) or at the cold ($\lesssim10^4\,{\rm K}$) outer edges of our disk. However, the outer disk does not dynamically evolve in the runtime of our simulations, so they remain consistent with the \gizmo{} simulation that treated the thermodynamics of these regions reliably. Our photosphere does evolve, but the thermodynamics at the photosphere does not affect the dynamics of the accretion flow itself. 

We begin most of our simulations without evolving radiation and then restart them later with radiation. We do this because evolving radiation makes the code about $15$ times slower\footnote{Specifically, while our non-radiative runs operate at about $\gtrsim1.5\times10^8$ zone cycles per second per individual GPU on the AMD MI250X graphics card, the radiative runs operate at about $\gtrsim10^7$ zone cycles per second per GPU, on 16 GPUs total.}, which is exacerbated by the large optical depths in the inner disk wherein our IMEX integrator for radiation requires many iterations to converge. This approach is justified because our $\gtrsim10^5\,r_{\rm g}/c$ simulation runtime corresponds to an orbital period at $r\approx2300\,r_{\rm g}$. This means that only gas at smaller radii is evolved for more than a single orbital timescale. At these small radii, the disk is highly optically thick to electon scattering, such that we can employ a $\gamma=4/3$ equation of state to approximate the dynamics of the gas+radiation fluid. After restarting with radiation activated, we set $\gamma=5/3$. We also ran a simulation that evolved radiation starting from the interpolated conditions (not reported on in this work) for comparison at early times ($\lesssim2\times10^4\,r_{\rm g}/c$), and we did not find significant differences in our results.

\begin{table}[bht]
\begin{tabular}{|c|c|c|c|}
\hline
Sim. Name & BH Spin & Non-Rad. Runtime & Rad. Runtime  \\ \hline
HS       & 0.9375 & $100339\,r_{\rm g}/c$ & $11756\,r_{\rm g}/c$               \\ \hline
NS       & 0 & $100010\,r_{\rm g}/c$ & $11687\,r_{\rm g}/c$       \\ \hline
\end{tabular}
\caption{Simulations reported in this work. At $t=0$, we initialize non-radiative \hamr{} simulations with the initial data that was remapped from \gizmo{} and carry them out for the duration listed in the third column. Then, we activate our radiation module, and continue to evolve the simulations for the duration listed in the fourth column.}
\label{table:sims}
\end{table}

\textit{Simulations.} In this work, we report two simulations, which differ in the value of the dimensionless BH spin, $a$. We label the simulations ``non-spinning'' (NS), where $a=0$, and ``highly-spinning'' (HS), where $a=0.9375$; see Table \ref{table:sims}. We adopt a base grid of $1536$ logarithmically-spaced radial cells, $384$ polar cells, and $128$ azimuthal cells. Using 1D SMR, we gradually increase the azimuthal resolution to $512$ azimuthal cells within 60 degrees of the equator. We also use ``internal derefinement'', which gradually decreases the azimuthal resolution within the grid blocks adjacent to the poles \citep[for details see][]{HAMR}. This results in roughly cubical cells that resolve the characteristic density scale height of the disk, $(H/r)_\rho\sim0.2-0.5$, by $\sim24-60$ cells. Usually, magnetized disk simulations are considered resolved if their cell size is sufficiently small compared to the wavelength of the magnetorotational instability \citep{hawley_2011}. Since our disk is extremely magnetized, this wavelength is on the order of the disk scale height itself, so we consider our resolution adequate.

\begin{figure}[bht]
    \centering
    \includegraphics[width=\textwidth]{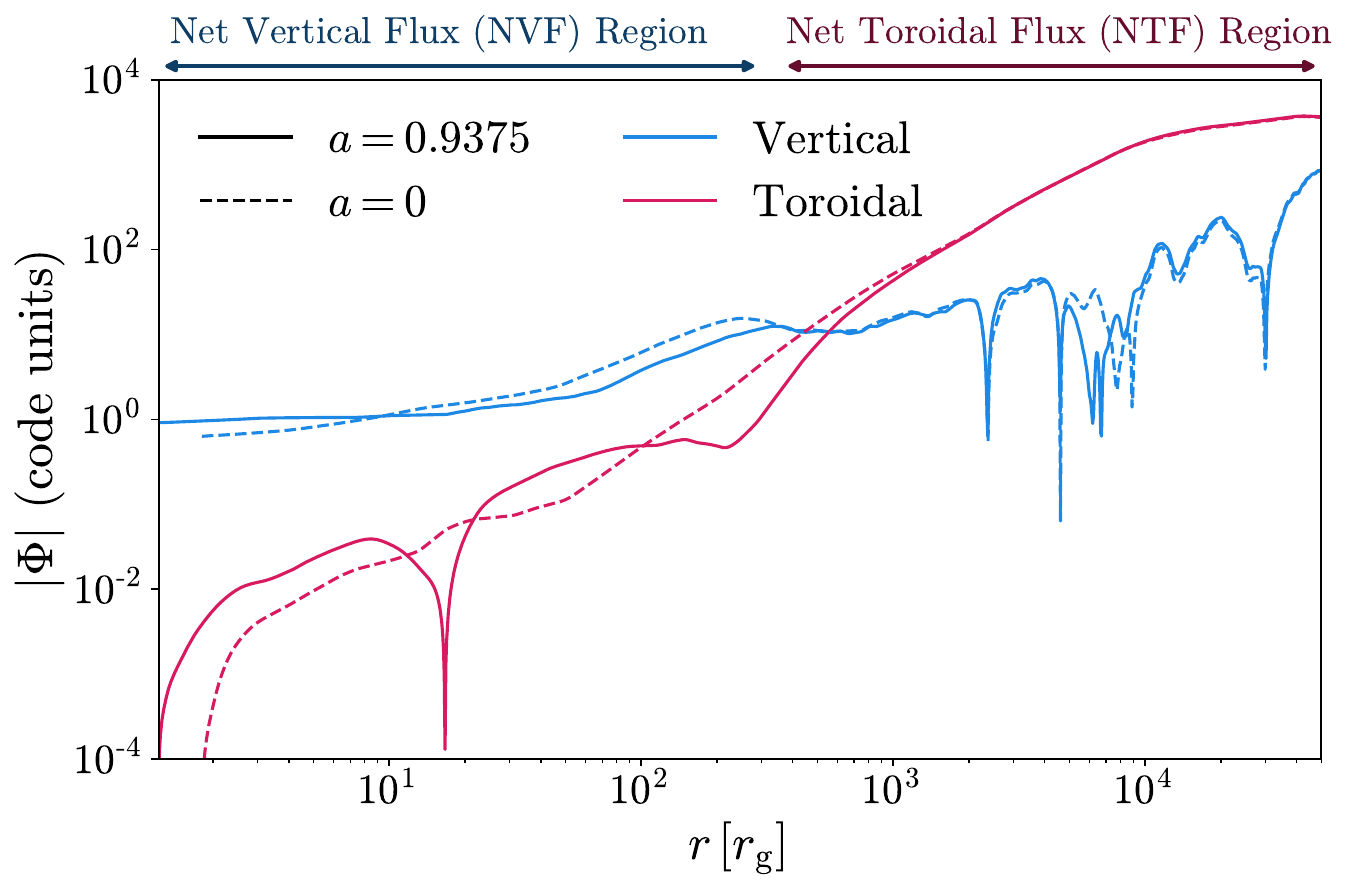}
    \caption{Net toroidal ($\Phi_{\rm T}$, Eq.~\ref{eq:phiT}) and vertical ($\Phi_{\rm V}$, Eq.~\ref{eq:phiV}) magnetic flux as a function of radius at $t=110700\,r_{\rm g}/c$. At $r\gtrsim200\,r_{\rm g}$, the disk is dominated by the net toroidal flux (NTF), and within it is dominated by the net vertical flux (NVF). So, we refer to these as the NTF and NVF regions, respectively.}
    \label{fig:netflux1d}
\end{figure}

\section{Results}
\label{sec:results}

\subsection{Multiscale accretion flow}
\label{sec:results:multiscale}
\begin{figure*}[bht]
    \centering
    \includegraphics[width=\textwidth]{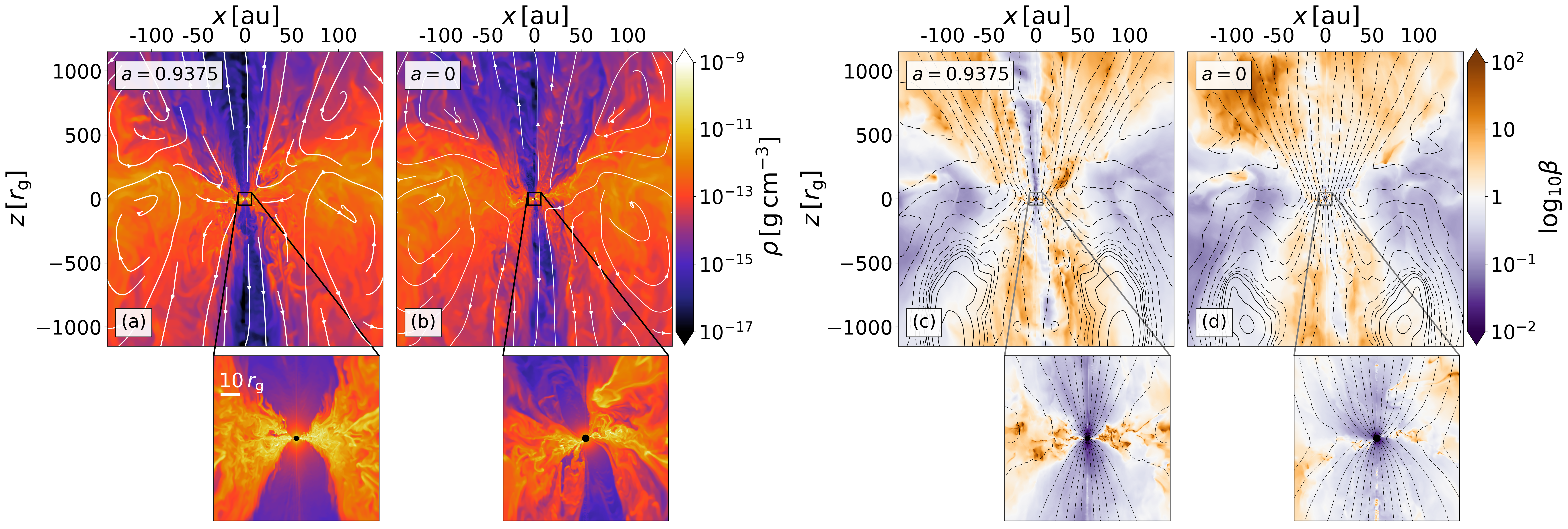}
    \caption{Snapshots of gas density $\rho$ and plasma $\beta=(p_{\rm t}+p_{\rm r})/p_{\rm B}$ in runs HS and NS. The outer disk is magnetically dominated with disordered poloidal fields, whereas the inner disk has comparable magnetic and radiation energy densities, negligible thermal energy and a highly structured poloidal magnetic field. Inset panels have width $100\,r_{\rm g}$. \textbf{Panels a-b.} Contours of $\rho$. We also show axisymmetrized mass flux streamlines ($\rho u^i$, white), which show outflows at high latitudes in both runs. \textbf{Panels c-d.} Contours of plasma $\beta$. While $\beta\approx10^{-1}$ at $\approx500-1000\,r_{\rm g}$, it reaches unity $\lesssim200\,r_{\rm g}$ (where radiation pressure is much larger than thermal).
    The BH in HS powers magnetized \citet{bz_1977} jets where $\beta<1$. We also show contours of $\sqrt{|\Phi_{\rm P}|}$ (Eq. \ref{eq:magnetic_flux}, black). In both runs, the magnetic field develops a single, large scale polarity (solid lines) closer to the BH but features structures of opposite polarity (dashed lines) in the southern hemisphere at larger radii. }
    \label{fig:bstate2x2}
\end{figure*}

Figure \ref{fig:multiscale} shows vertical slices of gas density, $\rho$, in run HS spanning a range of length scales at $t=111650\,r_{\rm g}/c$. The top left panels show the entire disk, which extends to $r\lesssim10^5\,r_{\rm g}$ ($\approx6200\,{\rm au}$). In successive panels, we zoom in towards event horizon scales at $r\sim0.1-10\,\,{\rm au}$. The disk is thick on all scales, with the typical $(H/r)_\rho\sim0.2-0.5$. The polar regions feature a pair of \citet{bz_1977} (BZ) jets that extract rotational energy from the BH. The jets wobble, and in the top right panel of Fig.~\ref{fig:multiscale} the northern jet bends in the $-x$ direction. We also show the square root\footnote{We plot contours of the square root of $\Phi_{\rm P}$ instead of $\Phi_{\rm P}$ itself in Figs. \ref{fig:multiscale} and \ref{fig:bstate2x2} to reduce contour crowding near the BH and improve plot clarity.} of the axisymmetrized poloidal magnetic ($\Phi_{\rm P}$) flux contours (white),
\begin{equation}
\begin{aligned}
\Phi_{\rm P,NH}(r,\theta,t) &= \int_0^{\theta}\int_0^{2\pi} \sqrt{-g}B^r d\theta'd\varphi'\\
\Phi_{\rm P,SH}(r,\theta,t) &=\int_{\theta}^{\pi}\int_0^{2\pi} \sqrt{-g}B^r d\theta'd\varphi',
\end{aligned}
\label{eq:magnetic_flux}
\end{equation}
where $g$ is the metric determinant. Here, NH and SH indicate the northern ($\theta<\pi/2$) and southern ($\theta>\pi/2$) hemispheres, respectively. The no-monopoles constraint (i.e., $\divb=0$) requires that $\Phi_{\rm NH}(\theta=\pi/2)=\Phi_{\rm SH}(\theta=\pi/2)$. Solid (dashed) lines indicate positive (negative) enclosed magnetic flux. At horizon scales, the poloidal magnetic field is ordered, while at large scales it is more disordered.

\begin{figure*}[bht]
    \centering
    \includegraphics[width=\textwidth]{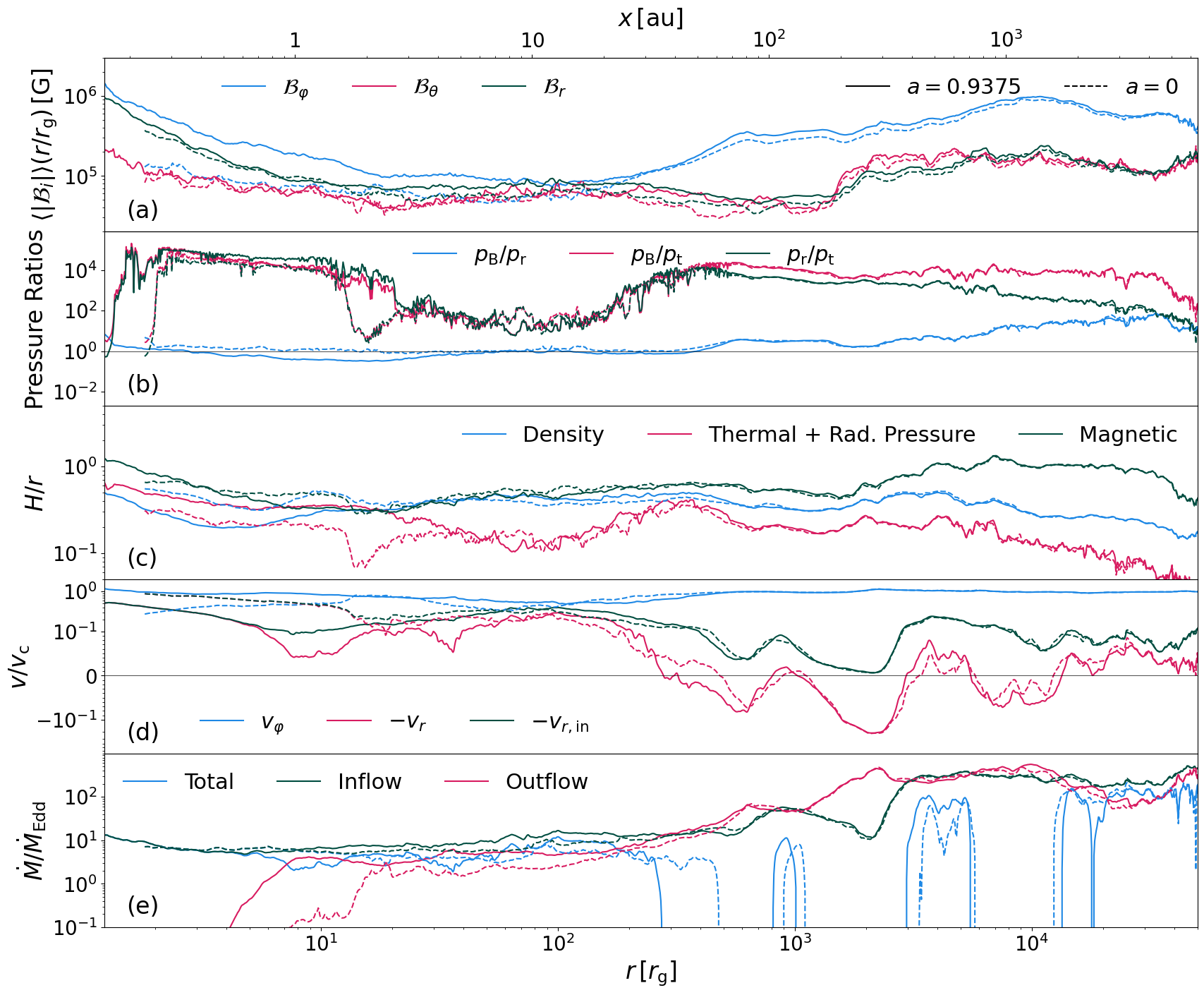}
    \caption{Radial profiles of various density-weighted quantities at $t=110700\,r_{\rm g}/c$. Many of the profiles change qualitatively at $\approx200\,r_{\rm g}$, where the disk transitions from NTF dominated to NVF dominated (Fig.~\ref{fig:netflux1d}).
    \textbf{Panel a.} Absolute magnetic field (``physical'' components, Eq. \ref{eq:mag_vel_phys}) strength in the disk, weighted by $r/r_{\rm g}$ for clarity. 
    \textbf{Panel b.} Ratios of magnetic ($p_{\rm B}$), radiation ($p_{\rm r}$) and gas thermal ($p_{\rm t}$) pressures in the disk. 
    \textbf{Panel c.}  Three different measures of the dimensionless scale height $H/r$: the density scale height (Eq. \ref{eq:hr_density}), which indicates how vertically distributed the gas is; the gas thermal plus radiation pressure scale height (Eq. \ref{eq:hr_thermal}), which indicates how much gas thermal and radiation pressure (where $p_{\rm r}\gg p_{\rm t}$) support the disk; and the magnetic scale height (Eq. \ref{eq:hr_magnetic}), which indicates how much magnetic pressure supports the disk. 
    \textbf{Panel d.} Disk velocities (``physical'' components, Eq. \ref{eq:mag_vel_phys}) normalized to the velocity of circular orbits ($v_{\rm c}$, Eq. \ref{eq:v_kep}). We depict the azimuthal velocity $v_\varphi$, the net radial velocity $-v_{\rm r}$, and the net radial velocity of all inflowing gas $-v_{\rm r,in}$. 
    \textbf{Panel e.} Mass accretion rate ($\dot{M}$, Eq. \ref{eq:mdot}) normalized to the Eddington accretion rate ($\dot{M}_{\rm Edd}$, Eq. \ref{eq:mdot_Edd}, assuming radiative efficiency $\eta=0.1$). We also plot separate profiles of the inflowing and outflowing mass transport rates.}
    \label{fig:radialProfs}
\end{figure*}


\subsection{A magnetic state transition in the inner disk}
\label{sec:results:transition}
The H24 disk had strong net toroidal magnetic flux (NTF),
\begin{equation}
    \Phi_{\rm T}(r,t) = \frac{1}{2\pi}\int_0^{2\pi}\int_0^\pi\int_{r_{\rm H}}^{r}\sqrt{-g}B^\varphi d\theta'd\varphi'dr'.
    \label{eq:phiT}
\end{equation}
We also define the net vertical magnetic flux (NVF),
\begin{equation}
    \Phi_{\rm V}(r,t) = \Phi_{\rm NH}(r,\theta=\pi/2,t).
    \label{eq:phiV}
\end{equation}
Figure \ref{fig:bstate2x2} shows $\Phi_{\rm T}$ and $\Phi_{\rm V}$ as a function of radius for runs NS and HS at $t=110700\times10^5\,r_{\rm g}/c$. At $r\approx200\,r_{\rm g}$, the disk transitions from having dominant NTF to having dominant NVF. We refer to this as the magnetic state transition and will refer to the inner and outer regions the NVF and NTF regions, respectively. \citet{minghao_2024} reported a similar transition, but in their case magnetic heating also altered the thermodynamic state of the disk, wherein their NTF region was cold and their NVF region was hot. These differences are expected, since they simulated accretion at strongly sub-Eddington rates, whereas we consider super-Eddington rates. Furthermore, \citet{gaburov_2012} also reported organized poloidal fields in the inner regions of their NTF-dominated disks, suggesting that such a transition may be common in accretion flows such as these.

Figures \ref{fig:bstate2x2}(a)-(b) show snapshots of $\rho$ for runs HS ($a=0.9375$) and NS ($a=0$) at $t\approx1.1\times10^5\,r_{\rm g}/c$. We also show axisymmetrized mass flux streamlines ($\rho u^i$, white). The snapshots have width $\approx2000\,r_{\rm g}$ and focus mainly on the NTF region. The inset panels have width $100\,r_{\rm g}$ and highlight only the NVF region. The disk ($\rho\gtrsim10^{-12}\,{\rm g\,cm^{-3}}$) remains geometrically thick throughout. The polar regions contain mass-loaded outflows ($\rho\lesssim10^{-14}\,{\rm g\,cm^{-3}}$) in both simulations. In HS, the BH launches BZ jets that carve out twin lower density cavities ($\rho\lesssim10^{-16}\,{\rm g\,cm^{-3}}$) within the surrounding wind. The inset panel shows the base of the jet. In NS, the inset panel also depicts a jet-like structure, but it is fundamentally different -- unlike BZ jets, which extract energy from the BH rotation, these outflows extract energy from the disk and are more strongly mass-loaded. 

Figure \ref{fig:bstate2x2}(c)-(d) shows snapshots of plasma $\beta=(p_{\rm t}+p_{\rm r})/p_{\rm B}$, where $p_{\rm B}=b^2/2$ is the fluid frame magnetic pressure. The disk is magnetically dominated at $r\approx500-1000\,r_{\rm g}$, with $\beta\approx10^{-1}$. The outflow region has $\beta\gtrsim1-10$. The inset panels show that closer to the BH ($\lesssim100\,r_{\rm g}$), the flow is multi-phase, with both $\beta\approx1$ and $\beta\gtrsim10$ patches. As we will see later, here and elsewhere the thermal pressure is significantly subdominant to the radiation pressure. 

Figs.~\ref{fig:bstate2x2}(c) and (d) also show $\sqrt{|\Phi_{\rm P}|}$ contours (black, Eq. \ref{eq:magnetic_flux}). At $r\gtrsim500\,r_{\rm g}$, $\Phi_{\rm P}$ displays patches of opposite sign above and below the disk, indicating that the poloidal field has less large-scale structure at these radii, which is consistent with the lack of NVF there. Conversely, the inset panels show that near the BH the poloidal field is highly structured, which is a necessary ingredient for launching strong outflows and (when the BH rotates) BZ jets. 

\subsection{Radial profiles}
\label{sec:results:radial}

Whereas some quantities change dramatically between the NTF and NVF regions, others do not. In Figure \ref{fig:radialProfs}, we plot various quantities as a function of radius at $t=110700\times10^5\,r_{\rm g}/c$. We do this for runs HS and NS, however the results meaningfully differ only near the event horizon. We use density-weighted shell averages to measure most quantities; e.g., the 1D profile of quantity $x$ in the disk is,
\begin{equation}
    \langle X \rangle_{\rm disk} = \int_0^\pi\int_0^{2\pi} \rho X \sqrt{-g}d\theta d\varphi \bigg/ \int_0^\pi\int_0^{2\pi} \rho \sqrt{-g}d\theta d\varphi.
    \label{eq:disk_avg}
\end{equation}

\textit{Magnetic field.} In Fig.~\ref{fig:radialProfs}(a), we show the average strength of each component of the magnetic field in the disk, $\langle|\mathcal{B}_i|\rangle_{\rm disk}$. At $r\gtrsim 200\,r_{\rm g}$, $\mathcal{B}_{\varphi}$ dominates by an order of magnitude over $\mathcal{B}_\theta$ and $\mathcal{B}_r$, which is consistent with Fig.~\ref{fig:netflux1d}. However, at small $r$ all field components are comparable, with $\mathcal{B}_\varphi$ being generally the largest component. At first, this may seem inconsistent with Fig.~\ref{fig:netflux1d}, which shows that NVF dominates in this region. Although the NTF is weak in this region, the absolute strength of the toroidal field remains high. This is because much of the toroidal flux has varying sign and does not contribute to the NTF, as we discuss in Section \ref{sec:results:topology}.


\textit{Thermal, radiation and magnetic pressures.} In Fig.~\ref{fig:radialProfs}(b), we show the ratios of the different pressures in the disk. Each component of the pressure is density-weighted to focus the average on the disk region (i.e., we show ratios of $\langle p_i \rangle_{\rm disk}$, where $i$ indicates the pressure component).  The pressure ratios in HS and NS are nearly identical. The thermal pressure, $p_{\rm t}$, is highly subdominant to both the magnetic pressure, $p_{\rm B}$, and the radiation pressure, $p_{\rm r}$, everywhere. However, $p_{\rm B}$ and $p_{\rm r}$ are more comparable, and in the inner disk ($\lesssim200\,r_{\rm g}$), $p_{\rm B}\approx p_{\rm r}$. In the outer disk, $p_{\rm B}$ exceeds $p_{\rm r}$ by a factor $\approx10-10^2$. This is roughly consistent with H24, who generally find $p_{\rm B}\approx 10^2 p_{\rm r}$ at large radii. This is also consistent with the analytic predictions of \citet{MDD} and \citet{hopkins_BLR}.


\textit{Scale heights.} In Fig.~\ref{fig:radialProfs}(c), we show three measures of the disk scale height. We show the (dimensionless) density scale height,
\begin{equation}
    (H/r)_\rho = \sqrt{\langle (\theta-\pi/2)^2 \rangle_{\rm disk}},
    \label{eq:hr_density}
\end{equation}
which is $\approx0.2-0.5$ throughout. The minimum value, $(H/r)_\rho\approx0.2$, occurs at $10\,r_{\rm g}\lesssim r\lesssim100\,r_{\rm g}$ in both simulations (ignoring features at $r>10^4\,r_{\rm g}$). This is somewhat slim. Before activating with radiation, $(H/r)_\rho$ was up to $50\%$ larger within the inner few hundred $r_{\rm g}$, indicating that this region of the disk can cool somewhat despite accreting at super-Eddington rates. While $(H/r)_\rho$ tells us how vertically extended the gas is, it does not tell us what is supporting the gas against vertical gravity. For this show the pressure scale height,
\begin{equation}
    (H/r)_{\rm pressure} = \langle c_s \rangle_{\rm disk}/v_{\rm c},
    \label{eq:hr_thermal}
\end{equation}
where we use the relativistic sound speed presuming local thermodynamic equilibrium and negligibly small radiation mean free paths \citep{mihalas_mihalas_1984},
\begin{equation}
    c_{\rm s} = \sqrt{\frac{(4/3)(p_{\rm t}+p_{\rm r})}{\rho c^2 + u_{\rm g} + \hat{E}_{\rm r}}},
    \label{eq:sound_speed}
\end{equation}
where we have assumed a $4/3$ adiabatic index for simplicity since most of the flow is radiation dominated. We have used the fluid-frame radiation energy density, $\hat{E}_{\rm r}$. We have also introduced the velocity of circular orbits,
\begin{equation}
    v_{\rm c} = c(r/r_{\rm g})/((r/r_{\rm g})^{3/2}+a)
    \label{eq:v_kep}
\end{equation}
which is approximately the Keplerian velocity except near the BH. The pressure scale height (which is dominated by the radiation pressure) is subdominant to the density scale height, except at very small $r\lesssim10-20\,r_{\rm g}$. This is expected since the disk was supported magnetically in H24. The magnetic scale height is,
\begin{equation}
    (H/r)_{\rm magnetic} = \langle v_{\rm A} \rangle_{\rm disk}/v_{\rm c},
    \label{eq:hr_magnetic}
\end{equation}
where $v_{\rm A}=\sqrt{b^2/(b^2+\rho + u_{\rm g})}$ is the relativistic Alfv\'en velocity\footnote{Note, that this is an inaccurate estimate where a large scale poloidal field is present, because in that case the large scale magnetic field compresses the disk instead of supporting it \citep[e.g.,][]{ferreira_1995}.}. The magnetic scale height clearly dominates at large radii. At $r\lesssim10-20\,r_{\rm g}$, the pressure and magnetic scale heights are comparable, so both radiation and magnetic pressure support the disk. 

\textit{Velocity fields.} In Fig.~\ref{fig:radialProfs}(d), we compare the orbital and radial velocities in the disk. We show disk averages of $v_\varphi$, $-v_r$ and $v_{r,{\rm in}}$, where $v_{r,{\rm in}}$ is averaged only over infalling gas with $v_r<0$. We normalize each velocity component to $v_{\rm c}$. At $r\gtrsim200\,r_{\rm g}$, $v_\varphi$ is almost exactly Keplerian, but dips slightly at $r\approx20\,r_{\rm g}-200\,r_{\rm g}$ where $v_\varphi$ is somewhat sub-Keplerian. This dip may be related to the runtime of the radiative portion of the simulation; before the radiation module was activated, the disk was sub-Keplerian at $r\lesssim200\,r_{\rm g}$, but once radiative cooling slimmed the disk somewhat we found that $v_\varphi/v_{\rm c}$ increased to unity near the BH in both simulations. This suggests that the innermost parts of the disk are almost entirely supported radially against gravity by rotation, not pressure. The radial velocities undergo a milder transition: $v_{r,\,{\rm in}}\gtrsim10^{-1}\,v_{\rm c}$ at $r\lesssim200\,r_{\rm g}$ and $v_{r,\,{\rm in}}\approx10^{-2}-10^{-1}\,v_{\rm c}$ at larger radii. The higher radial velocity in the inner region may be a sign of more vigorous turbulence and thus angular momentum transport.

Wide swaths of the disk beyond $\approx300-500\,r_{\rm g}$ have a bulk radial outflow. This is also seen in a spacetime diagram in Figure \ref{fig:app:vr_spacetime} in the Appendix. At the beginning of our simulation, gas near the event horizon is launched outwards on an orbital timescale. This phenomenon is likely both numerical and physical. It is numerical because it presumably results from different inner boundary conditions between the two codes: \gizmo{} uses a standard ``sink'' for the BH, representing an unresolved region on larger-than-horizon scales, which artificially removes NVF. \hamr{}'s inner boundary is inside the BH event horizon, which does not destroy the NVF and is much smaller than the \gizmo{} sink.  However, we also expect the dynamical state of the disk to be altered on physical grounds. Even if \gizmo{} had the exact same inner boundary conditions as \hamr{}, only when the nascent disk reached the horizon would the magnetic flux build up, thus altering the disk state. This description is consistent with \citet{minghao_2024}, who reported a bulk outflow where their cold, NTF-dominated disk with transitioned to a hot, NVF-dominated disk. 

\textit{Mass accretion.} We measure the mass accretion rate, 
\begin{equation}
    \dot{M} = \int_0^\pi\int_0^{2\pi} -\rho u^r \sqrt{-g}d\theta d\varphi,
    \label{eq:mdot}
\end{equation}
and use the Eddington luminosity,
\begin{equation}
    L_{\rm Edd} \approx 1.6\times10^{45}\,{\rm ergs\,s^{-1}}\,\left(\frac{M}{1.3\times10^7\,M_\odot}\right),
    \label{eq:Ledd}
\end{equation}
to define the Eddington accretion rate from the relation $L_{\rm Edd}=\eta_{\rm rad}\dot{M}_{\rm Edd}c^2$,
\begin{equation}
    \dot{M}_{\rm Edd} \approx 0.3\,M_\odot/{\rm yr}\left(\frac{\eta_{\rm rad}}{0.1}\right)^{-1}\left(\frac{M}{1.3\times10^7\,M_\odot}\right),
    \label{eq:mdot_Edd}
\end{equation}
where we have taken $\eta_{\rm rad}=0.1$ as a fiducial radiative efficiency. We show $\dot{M}(r)/\dot{M}_{\rm Edd}$ in Fig.~\ref{fig:radialProfs}(e), where we separately plot the total (Eq.~\ref{eq:mdot}), inflowing ($v_r<0$ only) and outflowing ($v_r>0$ only) rates. At $r\lesssim100-200\,r_{\rm g}$, the accretion rate is steady, and smoothly connects to the inflow at larger radii. The accretion rate here is $\gtrsim5-10\,\dot{M}_{\rm Edd}$, demonstrating that quasars can indeed accrete above the Eddington limit. This is not a surprise, as it has been established in other works \citep[e.g., ][]{sadowski_narayan_2016,jiang_2019} and is expected because the inner accretion flow traps photons and advects them into the BH \citep{begelman_1979,abramowicz_1988,king_begelman_1999}, lowering the radiative efficiency. The outflow rate in the steady state region ($r\lesssim100-200\,r_{\rm g}$) suggests that the inner disk launches nearly the same amount of mass in winds that it accretes. 

\begin{figure}[bht]
\centering\includegraphics[width=\textwidth]{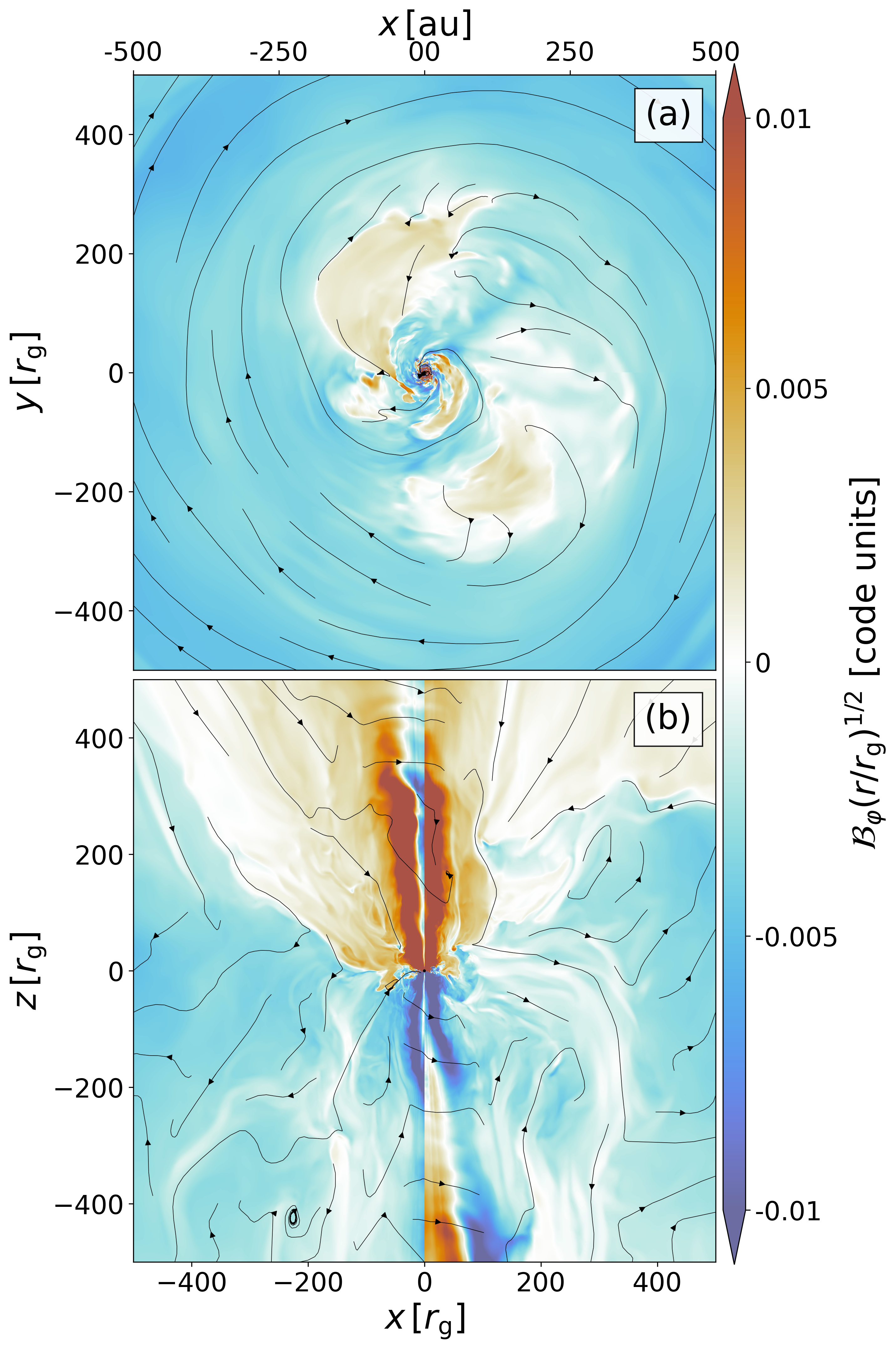}
\caption{Toroidal magnetic field weighted by $(r/r_{\rm g})^{1/2}$ in run HS at time $t=110700\,r_{\rm g}/c$. The toroidal field is roughly symmetric (anti-symmetric) about the midplane beyond (within) $\approx200\,r_{\rm g}$.  \textbf{Panel a.} Equatorial snapshot of the toroidal field within $\approx500\,r_{\rm g}$. Magnetic field lines are shown in black. \textbf{Panel b.} Same as panel a, except we show the $x-z$ plane.}
\label{fig:toroidal}
\end{figure}

\begin{figure}[bht]
\centering\includegraphics[width=\textwidth]{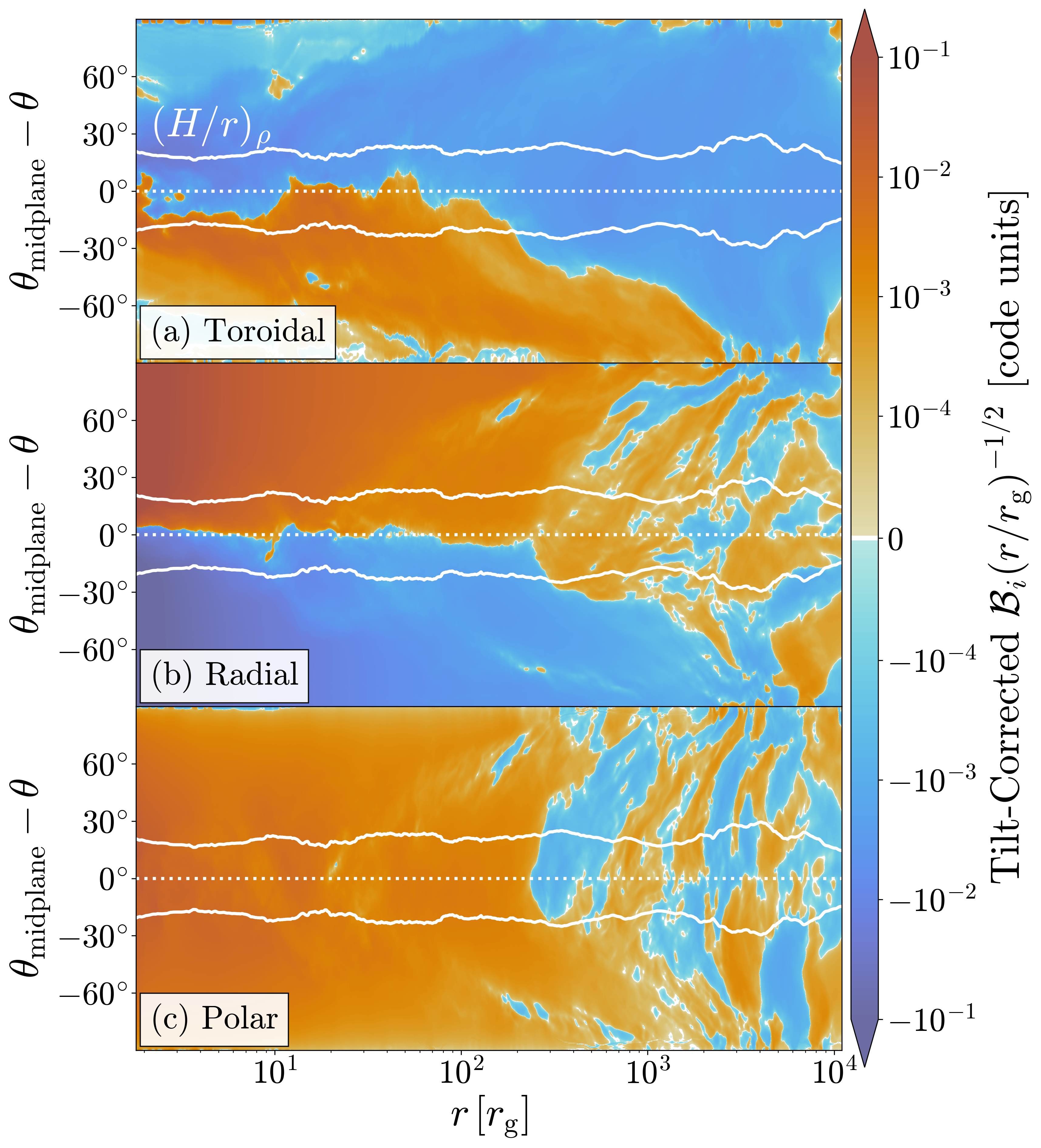}
\caption{Axisymmetrized components of ``tilt-corrected'' magnetic field in run NS ($a=0$) in the ${\rm log}r-\theta$ plane at $t=110700\,r_{\rm g}/c$, where we have drawn the density scale height ($(H/r)_\rho$, Eq.~\ref{eq:hr_density}) in white. We identify a transition in magnetic topology near $r\approx200\,r_{\rm g}$. \textbf{Panel a.} The toroidal field is symmetric about the midplane at large radii and becomes roughly antisymmetric about the midplane at small radii. \textbf{Panel b.} The radial field is unstructured at large radii and becomes roughly antisymmetric about the midplane at small radi. \textbf{Panel c.} The polar field is unstructured at large radii and becomes roughly symmetric about the midplane at small radii.}
\label{fig:baxisym}
\end{figure}

\subsection{Topology of the magnetic field}
\label{sec:results:topology}
\subsubsection{Symmetries}
\label{sec:results:topology:symmetry}
We can better understand the magnetic state transition by examining the magnetic field topology. In Figure \ref{fig:toroidal}, we show a pair of late time snapshots of $\mathcal{B}_\varphi(r/r_{\rm g})^{1/2}$ for run HS. Fig.~\ref{fig:toroidal}(a) shows an equatorial snapshot, where the field has the same (negative) sign at $r\gtrsim200-300\,r_{\rm g}$, where the NTF dominates. At smaller radii, the toroidal field is non-axisymmetric and switches sign several times an orbit in a spiral pattern. Magnetized spiral waves are a common signature of poloidal flux eruptions, which are usually associated with MADs \citep{sasha_2011}. We show magnetic field streamlines in black, which are mainly toroidal in the outer parts of the snapshot, but bend radially inwards within $200\,r_{\rm g}$. In Fig.~\ref{fig:toroidal}(b), we show an $x-z$ snapshot wherein the toroidal field undergoes a clear topological transition: at $r\gtrsim200\,r_{\rm g}$, in the NTF region, it is symmetric about the midplane, and at smaller radii in the NVF region it is antisymmetric. The antisymmetric component of the toroidal field does not contribute to $\Phi_{\rm T}$. We also notice that the midplane is tilted, which is common in strongly magnetized flows, likely because they produce large eddies that  may displace the disk midplane. 

Figure \ref{fig:baxisym} depicts the large-scale (anti-)symmetries of each magnetic field component. Here, we show run NS, where outflows are weaker because the BH is not rotating. We have also corrected for the $\mathcal{O}(10^\circ)$ radially dependent disk tilt angle by reorienting the disk such that the midplane is always at $\theta=\pi/2$ (see \citealt{kaaz_2023} for details). In the ``tilt-corrected'' frame, we show axisymmetrized profiles of each component of $\mathcal{B}_i(r/r_{\rm g})^{1/2}$. We also depict the density scale height $(H/r)_\rho$ (Eq.~\ref{eq:hr_density}) above and below the midplane in white. Fig.~\ref{fig:baxisym}(a) shows that the toroidal field is entirely symmetric about the midplane at large radii in the NTF region. However, within the $r\lesssim200\,r_{\rm g}$ NVF region, this symmetry is broken, and the toroidal field becomes antisymmetric about the midplane both at large scales and within the density scale height. This explains why the toroidal field is stronger than the poloidal field in Fig.~\ref{fig:radialProfs}(a) within the NVF region. Although NVF dominates over NTF, the toroidal field is still overall stronger, but its contributions to the NTF in the north/south hemispheres cancel. In Figs.~\ref{fig:baxisym}(b) and (c), we show the radial and polar fields, respectively. At large scales, both are unstructured, with ``intermediate-scale'' unipolar poloidal field regions on the order of the disk scale height. At smaller scales, the radial and polar magnetic fields develop symmetry properties. The radial field becomes antisymmetric about the midplane, whereas the polar field becomes symmetric. The symmetric polar field is consistent with this region having dominant NVF. 

We can assess the symmetry properties of the magnetic field quantitatively by measuring its parity. We define parity as the degree of reflection symmetry across the midplane (i.e., $z\rightarrow -z$). Magnetic fields with ``even'' parity are symmetric about the midplane, while fields with ``odd'' parity are antisymmetric about the midplane. Dipolar fields have odd parity, while quadrupolar fields have even parity. If, for example, the toroidal field is generated from the shearing of the poloidal field -- as in local studies of the magnetorotational instability with NVF \citep[e.g.,][]{hawley_1995} as well as global studies of the magnetorotational dynamo \citep{jon_2024} -- parity is preserved because the shear has an even dependence on $z$. However, in many other configurations, the toroidal field is not generated from the shearing of the vertical field \citep[e.g., in local zero net flux models of disks with strong toroidal fields,][]{johansen_levein_2008,squire_2024}. Toroidal fields with even parity have the same sign above/below the midplane, while toroidal fields with odd parity switch sign above/below the midplane and are zero at the midplane.


We measure parity quantiatively \citep[e.g.,][]{pariev_2007,flock_2012} by first defining the (anti)symmetric components of the magnetic field,
\begin{equation}
\begin{aligned}
    \mathcal{B}^{\rm S}_i &= \frac{1}{2}\left(\mathcal{B}^{\rm NH}_i + \mathcal{B}^{\rm SH}_i\right)\\
    \mathcal{B}^{\rm AS}_i &= \frac{1}{2}\left(\mathcal{B}^{\rm NH}_i - \mathcal{B}^{\rm SH}_i\right),
\end{aligned}
\end{equation}
where we have defined the volume-weighted physical components of the magnetic field in the northern and southern hemispheres\footnote{We do not density-weight the magnetic field averages used in the parity calculations because we want to measure the large scale field, not just that which threads the disk midplane.},
\begin{equation}
\begin{aligned}
    \mathcal{B}^{\rm NH}_i &= \langle \mathcal{B}_i\Theta(\pi/2-\theta)
    \rangle_{\theta,\varphi}\\
    \mathcal{B}^{\rm SH}_i &= \langle \mathcal{B}_i\Theta(\theta-\pi/2)
    \rangle_{\theta,\varphi},
\end{aligned}
\end{equation}
where $\Theta(x)$ is the Heaviside step function and,
\begin{equation}
\langle X\rangle_{\theta,\varphi}\equiv \int_0^\pi\int_0^{2\pi}\sqrt{-g}Xd\theta d\varphi/\int_0^\pi\int_0^{2\pi}\sqrt{-g}d\theta d\varphi,
\label{eq:shell_average}
\end{equation}
is the volume-weighted average over a spherical shell.
Then, the energy densities of the odd (``D'' for dipole-like) and even (``Q'' for quadrupole-like) components of the field are,
\begin{equation}
\begin{aligned}
    U^{\rm D} &= (\mathcal{B}_r^{\rm AS})^2 + (\mathcal{B}_\theta^{\rm S})^2 + (\mathcal{B}_\varphi^{\rm AS})^2\\
    U^{\rm Q} &= (\mathcal{B}_r^{\rm S})^2 + (\mathcal{B}_\theta^{\rm AS})^2 + (\mathcal{B}_\varphi^{\rm S})^2\\
\end{aligned}
\label{eq:quad_dip_energies}
\end{equation}
We compare these quantities using the function,
\begin{equation}
    {\rm C}(X,Y) = \frac{X-Y}{X+Y},
\label{eq:compare_func}
\end{equation}
such that ${\rm C}(X,0)=1$ and ${\rm C}(0,Y)=-1$. So, ${\rm C}(U^{\rm D},U^{\rm Q})=1$ ($-1$) for an odd (even) parity magnetic field.

\begin{figure}[bht]
    \centering
    \includegraphics[width=\textwidth]{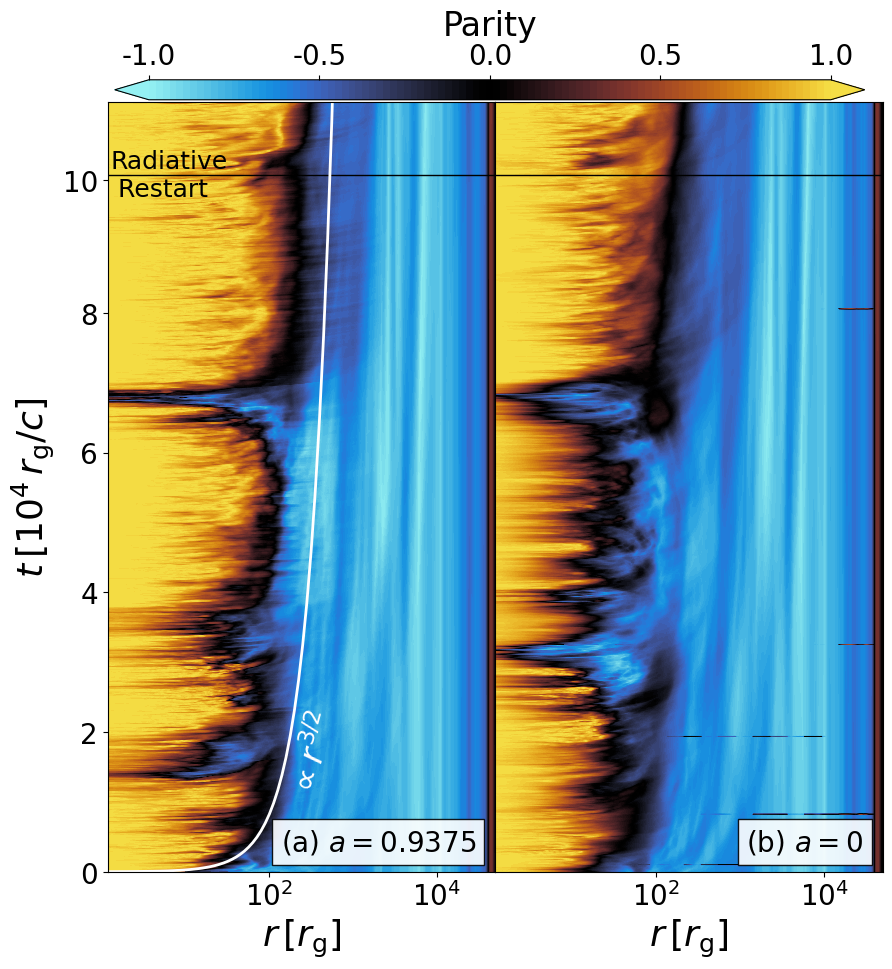}
    \caption{Spacetime (${\rm log}r-t$) diagram of the magnetic field parity ($C(U^{\rm D},U^{\rm Q})$ see Eqs. \ref{eq:quad_dip_energies}-\ref{eq:compare_func}) in runs HS (panel a) and NS (panel b), where $1$, $-1$ and $0$ indicate dipole-like/odd parity, quadrupole-like/even parity, and mixed parity respectively. By late times, the disk is odd at $r\lesssim200\,{\rm rg}$ and even at larger radii in both runs. We have also marked the time when our radiation module is activated with a black line.}
    \label{fig:spacetime}
\end{figure}

We show the volume-weighted parity, ${\rm C}(U^{\rm D},U^{\rm Q})$, for run HS in a ${\rm log}r-t$ spacetime diagram in Fig.~\ref{fig:spacetime}(a). Near the beginning of the simulation, the volume-filling parity is dipole-like at radii $\lesssim30\,r_{\rm g}$ and quadrupole-like at larger radii. The original H24 disk was quadrupole-like everywhere, as shown by Figure \ref{fig:app:bquantities} in the Appendix. During the initial evolution, the dipolar region expands radially outwards on the accretion timescale, which obeys a $\propto r^{3/2}$ scaling (shown in white). In Fig.~\ref{fig:spacetime}(b), we show the same but for run NS. At early times, the dipole-like region is smaller than in HS. This may be due to the stronger outflows for $a=0.9375$ reinforcing the parity in the volume-filling bicones above and below the disk. However, after $t\approx7\times10^4\,r_{\rm g}/c$ (when the NVF switches sign, which we discuss in the next section), the dipole-like region in NS expands to size $\gtrsim200\,r_{\rm g}$, as in HS. We conclude that the expansion of the dipole-like region does not follow a simple power law behavior, and instead grows in nonlinear bursts, which we will explain shortly.

At late time, our results closely follow the symmetry properties that are depicted in Fig.~\ref{fig:baxisym} and are consistent with the NVF and NTF regions described in Fig.~\ref{fig:netflux1d}. The magnetic field in the NVF region has odd parity, while in the outer NTF region it has even parity. We note that although here the dipolar, NVF region developed within the non-radiative part of our simulation, we verified that this also occurs in shorter test simulations that were radiative from the onset. After the radiation module is activated, the even parity region in HS increases slightly. This is likely because the addition of radiation broadens dipolar, biconical outflow region shown in Fig.~\ref{fig:toroidal}(b).

\subsubsection{What causes the magnetic state transition?}
\label{sec:results:topology:whatcausesit}
We have described a magnetic state transition where the disk switches from NTF-dominated to NVF-dominated (Fig.~\ref{fig:netflux1d}). In Figs.~\ref{fig:baxisym} and \ref{fig:spacetime}, we have also shown that the parity of all field components changes in the NVF region. What causes this transition?

\begin{figure}[bht]
\centering\includegraphics[width=\textwidth]{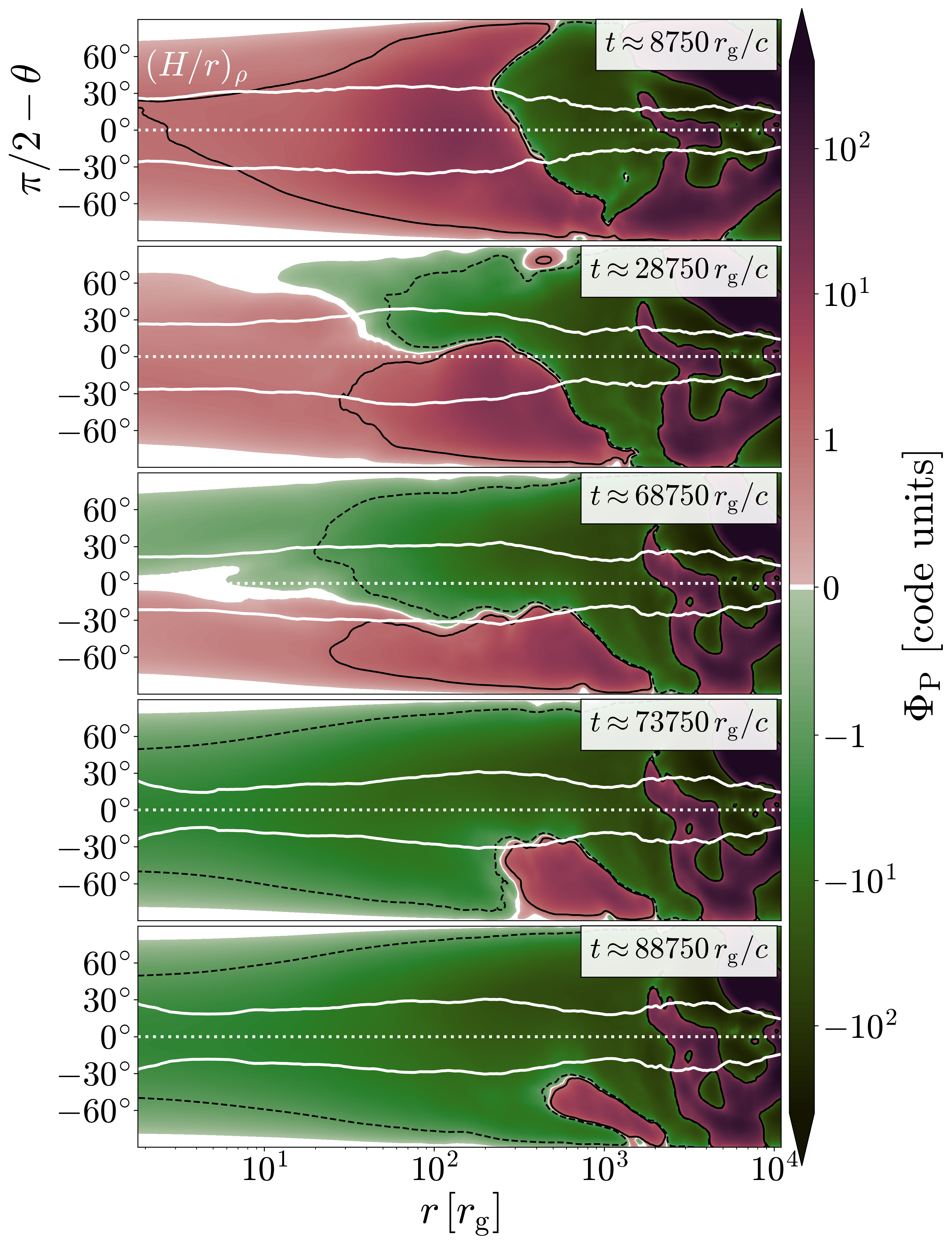}
\caption{Sequence of magnetic flux ($\Phi_{\rm P}$, Eq.~\ref{eq:magnetic_flux}) contours in the accretion flow, with green and red indicating opposite polarity. Black solid (dashed) lines are $\Phi_{\rm P}=1$ ($-1$) contours (in code units). Here, we highlight a polarity inversion event, where at $t=8750\,r_{\rm g}/c$ the BH is saturated with positive flux and at larger radii $\approx200-2000\,r_{\rm g}$ there is a larger region of negative flux. By $t=68750\,r_{\rm g}/c$, the negative flux has reached the event horizon, and is ejecting the positive flux. By $t=73750$ and $88750\,r_{\rm g}/c$, the negative flux has completely saturated the disk out to $\approx300$ and $400\,r_{\rm g}$, respectively. We also show the density scale height ($(H/r)_\rho$, Eq.~\ref{eq:hr_density}) above and below the midplane in white.}
\label{fig:fluxfliploops}
\end{figure}

While the toroidal field is organized in the NTF region, the poloidal field is not. It is composed of intermediate-scale structures of alternating polarity, as seen in Figs.~\ref{fig:baxisym}(b) and (c). Here, we define ``polarity'' as the sign of $\Phi_{\rm P}$. As the disk accretes, the poloidal field is advected inwards by accretion and diffused outwards by turbulent resistivity \citep{guilet_2012,guilet_2013}. We can gain insight into this process by examining Figure \ref{fig:fluxfliploops}, where we show contours of the poloidal flux $\Phi_{\rm P}(r,\theta)$ (Eq.~\ref{eq:magnetic_flux}) in the $r-\theta$ plane at five different times. We also show the density scale height ($(H/r)_\rho$, Eq.~\ref{eq:hr_density}) with white lines. At $t\approx8750\,r_{\rm g}/c$, $\Phi_{\rm P}$ has a single polarity (red) in the inner disk; this corresponds to the dipole-like region at the same time in Fig.~\ref{fig:spacetime}(a). Beyond this region at radii $\lesssim10^3\,r_{\rm g}$, there is a $\approx2H$ region of poloidal flux of opposite polarity (green). This feature is advected inwards with the accretion flow, and by $t\approx68750\,r_{\rm g}/c$ it has filled the northern hemisphere of the horizon and competes with the previously accreted flux, which dominates on the southern hemisphere. However, the new flux is larger and ultimately wins, ejecting the old flux. The result can be seen at $t\approx73750\,r_{\rm g}/c$, where the inner region has the polarity of the new flux. This suggests that the size of the NVF region is, in part, determined by the size of the poloidal flux structures advected from large radii. 

Once the poloidal field becomes structured and dipolar, it quickly transforms toroidal field in kind. This is because the orbital shear generates a toroidal field of odd parity from  the dipolar poloidal field on an orbital timescale. In the outer disk, the even parity, quadrupole-like toroidal field is supplied mainly by flux-freezing \citep[see arguments in ][]{FORGE2}, which occurs on an inflow timescale \citep[the Parker dynamo likely also plays a role, e.g.][]{johansen_levein_2008,gaburov_2012,squire_2024}. Therefore, the parity of the toroidal field will be determined by a competition between the advection of quadrupole-like $\mathcal{B}_\varphi^{\rm S}$ and the shear of dipole-like $\mathcal{B}_r^{\rm S}$ and $\mathcal{B}_\theta^{\rm S}$. The ratio of these two generation mechanisms (shear $v_\phi$ vs. advection $v_r$) is thus a condition for the dominant parity. Indeed, if the condition,
\begin{equation}
|\mathcal{B}_{\theta/r}^{\rm S}|\gtrsim |B_\varphi^{\rm S}v_r/v_\varphi|
\end{equation}
is met (where we write $\mathcal{B}_{\theta/r}^{\rm S}$ to refer to either $\mathcal{B}_{\theta}^{\rm S}$ or $\mathcal{B}_{r}^{\rm S}$), then it implies that the generation of odd parity toroidal field by the shearing of the dipolar poloidal field can win over radial advection of even parity toroidal field. Since $|v_r|\sim10^{-1}v_\varphi$ at most, this condition is realizable even if $\mathcal{B}_{\theta/r}^{\rm S}$ is subdominant to $\mathcal{B}_\varphi^{\rm S}$. We argue that this is why the toroidal field develops odd parity (e.g., Fig.~\ref{fig:baxisym}(a)) when the poloidal field does.

A similar NVF region also develops in \citet{jon_2024}, where in a GRMHD simulation of a torus with an initially toroidal field, the advection of intermediate-scale turbulent poloidal field loops enabled the development of a strong, dipolar, poloidal field in the inner $\gtrsim300\,r_{\rm g}$. In this work, the toroidal field evolution depended strongly on magnetic buoyancy.
Magnetic buoyancy may eject toroidal field lines and thus reduce the strength of the NTF in the outer disk. However, 
NTF may also sustained when $\beta<1$ if a Parker dynamo regenerates the field \citep{johansen_levein_2008,squire_2024}. In H24, the toroidal field was sustained by continued inwards advection of NTF from the tidally disrupted giant molecular cloud that formed the disk \citep{FORGE2}. Magnetic buoyancy presumably plays an important role in our disk as well, but requires a more detailed analysis that is beyond the scope of this paper. 

\subsection{Outflows \& Emission}
\label{sec:results:outflows}

\begin{figure*}[bht]
    \centering
    \includegraphics[width=\textwidth]{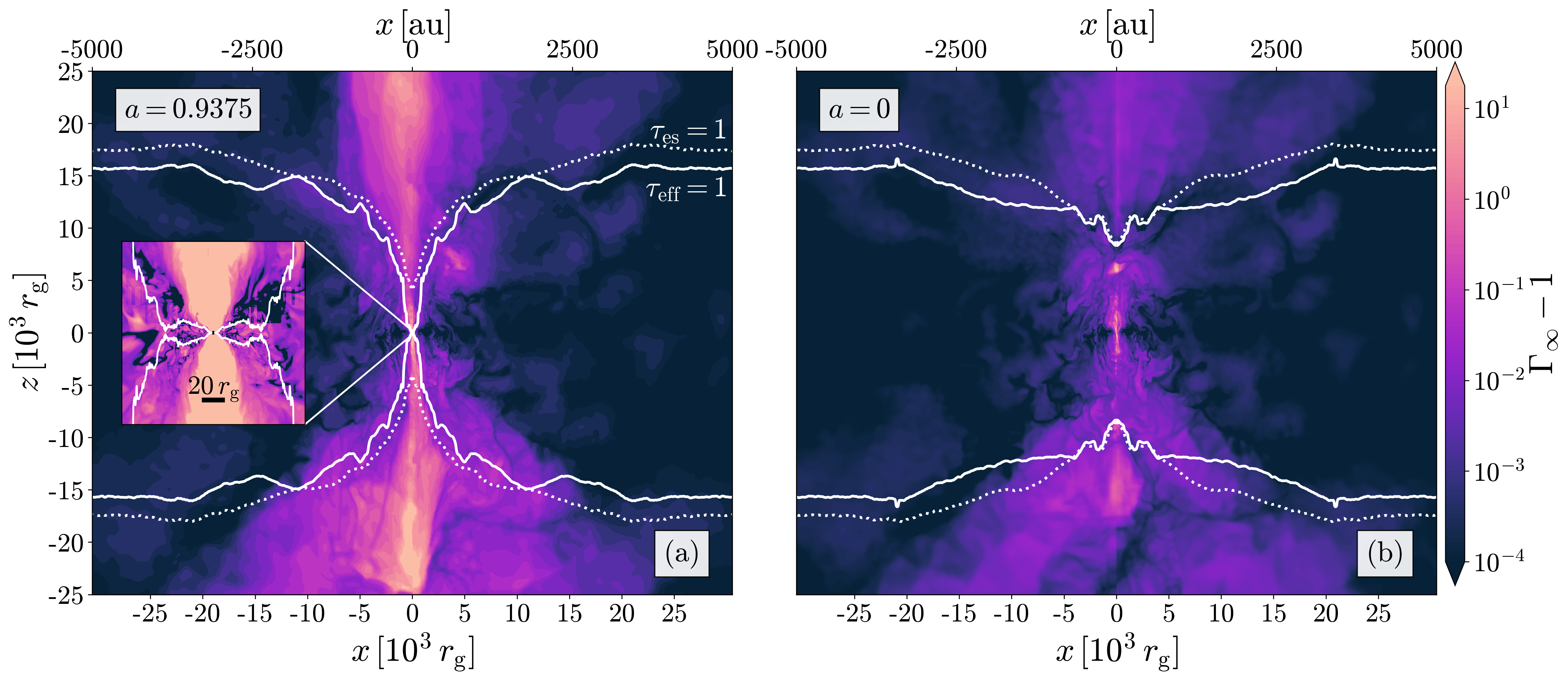}
    \caption{Energy per unit mass ($\Gamma_\infty$, Eq.~\ref{eq:lorentz_infty}) snapshot on large scales for runs NS and HS at time $t=110700\,r_{\rm g}/c$. When $a=0.9375$, the BZ jets clear out polar funnels and drive highly relativistic ($\Gamma_\infty\gtrsim2-10$) outflows. When $a=0$, the outflows are mildly relativistic ($\Gamma_\infty\approx10^{-2}-10^{-1}$) and do not clear out a funnel. \textbf{Panel a.} $\Gamma_\infty$ in $x-z$ plane for HS. We show the axisymmetrized effective ($\tau_{\rm eff}=1$, Eq.~\ref{eq:effective_optical_depth}, white solid lines) and scattering ($\tau_{\rm es}=1$, Eq.~\ref{eq:scattering_optical_depth}, white dotted lines) photospheres. We also show a $200\,r_{\rm g}\times200\,r_{\rm g}$ inset panel of the effective photosphere reaching down to the BH. \textbf{Panel b.} Same as panel a, but for NS.}
\label{fig:mu2panel}
\end{figure*}

\subsubsection{The structure of the outflow and emission regions}

Runs HS and NS both feature strong, magnetized outflows. We depict these outflows qualitatively by using the energy per unit mass of the flow, 
\begin{equation}
    \Gamma_\infty = -u_t(p_{\rm t}+u_{\rm g} + b^2 + 1)/\rho c^2 - u_{{\rm r},t}\frac{4}{3}E_r/\rho c^2\Theta(\tau_{\rm eff}-1)
    \label{eq:lorentz_infty}
\end{equation}
If this quantity was converted to kinetic energy, then the gas would be accelerated to Lorentz factor $\Gamma_\infty$. Indeed, in non-radiative GRMHD, the first term on the right hand side is constant along streamlines in steady state \citep[e.g., ][]{sasha_2009}. We added the radiative term to generalize $\Gamma_\infty$ to the M1 formalism; in the optically thick limit,  this term accounts for the (sometimes appreciable) trapped radiation energy per unit mass. So, we only include it in our calculation of $\Gamma_\infty$ where $\tau_{\rm eff}>1$, where $\tau_{\rm eff}$ is the ``effective'' optical depth, which we will define shortly.

We depict $\Gamma_\infty$ on large scales for run HS in Fig.~\ref{fig:mu2panel}(a). The BH launches strong, biconical outflows reaching relativistic energies with $\Gamma_\infty\gtrsim10$. These are BZ jets, powered by the rapid rotation of the BH. The BZ jet power scales as $\propto\Phi_{\rm V}^2$ and is thus enabled by the accumulation of NVF. The jet heads are just outside the depicted snapshot at $\approx4\times10^4\,r_{\rm g}\,(\approx\,5000\,{\rm au})$. These jets were relaunched after the polarity inversion event at $t\approx7\times10^4\,{r_{\rm g}/c}$ (e.g., Fig.~\ref{fig:fluxfliploops}), which destroyed the previous jet. So, in this snapshot, the jet heads have been propagating for $\approx4\times10^4\,r_{\rm g}/c$. Their current position is thus consistent with the jet propagating at the speed of light since it was launched. The highly relativistic jets are surrounded by mildly relativistic outflows. In Fig.~\ref{fig:mu2panel}(b), we show the same for run NS. Here, the outflows are still biconical, but weaker, with a maximum $\Gamma_\infty\approx10^{-1}$. This is consistent with heavier, mass-loaded winds launched from the disk. The winds are also at least partially a result of the NVF, since large-scale poloidal fields are required to launch winds magnetically \citep[][]{bp_1982}.


We define the ``effective'' optical depth,
\begin{equation}
    \tau_{\rm eff} = \int_{z}^{z_{\rm max}} \kappa_{\rm eff}\rho dz,
\label{eq:effective_optical_depth}
\end{equation}
where the medium is only considered opaque if absorption can remove photons. We have set $z_{\rm max}=\pm4\times10^4\,r_{\rm g}$ as a proxy for ``infinity'' in the integral. We have neglected the length contraction factor which is a negligible correction because most of the photospheric gas is non-relativistic. We have used the effective opacity,
\begin{equation}
    \kappa_{\rm eff} = \sqrt{\kappa_{\rm abs}(\kappa_{\rm abs} + \kappa_{\rm es})},
    \label{eq:kappa_eff}
\end{equation}
where $\kappa_{\rm abs}$ and $\kappa_{\rm es}$ are the absorption and electron scattering opacities, respectively. 

The white contours in Figure \ref{fig:mu2panel} show the axisymmetrized effective photosphere ($\tau_{\rm eff}=1$) averaged over both hemispheres, which extends to $r\gtrsim10^4\,r_{\rm g}$. In HS, the BZ jet carves out a cavity, allowing the effective photosphere to reach down to the BH. This is important, as it suggests that emission from the ISCO -- including, for instance, the iron K-$\alpha$ line which is often used to infer the BH spin \citep{reynolds_2021} -- can escape at very high viewing angles. If viewed face-on, this system would likely appear blazar-like. However, at more intermediate viewing angles, the observed emission is significantly reprocessed by the time it escapes into the optically thin regions. In NS, there is no BZ jet, and the effective photosphere is at $r\gtrsim10^4\,r_{\rm g}$ even directly above the BH. This indicates that all emission escaping the disk will be reprocessed. We also note the flared structure of the photosphere at $x\approx5-15\times10^3\,r_{\rm g}$, which likely plays an important role in shaping the broad-line region \citep[e.g.,][]{hopkins_BLR}. 

We also show the scattering photosphere ($\tau_{\rm es}=1$) in white dotted lines using the same method, where
\begin{equation}
    \tau_{\rm es} = \int_{z}^{z_{\rm max}} \kappa_{\rm es}\rho dz.
\label{eq:scattering_optical_depth}
\end{equation}
Generally, the scattering photosphere extends slightly farther than the effective photosphere. Interestingly, in the jet funnel in HS, the scattering photosphere reaches down to $\approx5\times10^3\,r_{\rm g}$. This means that although light emitted near the ISCO can escape unabsorbed, it will still scatter. This likely Comptonizes the emission and distorts its spectrum.


\begin{figure}[bht]
    \centering
    \includegraphics[width=\textwidth]{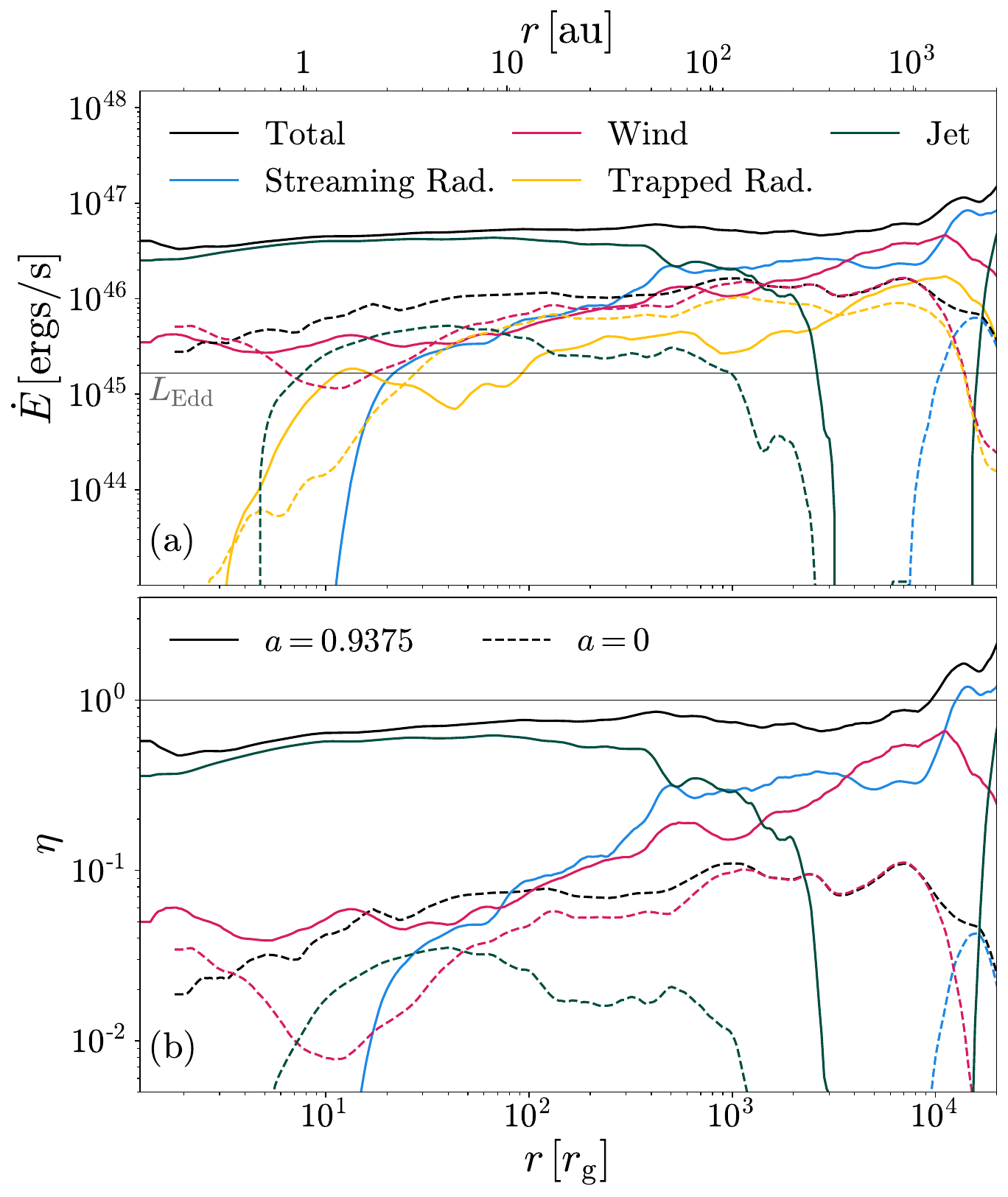}
\caption{Radial profiles of outflows for runs HS (solid lines) and NS (dashed lines) at $t=110700\,r_{\rm g}/c$. The disks exhibit super-Eddington luminosities, radiation-loaded winds, and if the BH rotates, a relativistic BZ jet. \textbf{Panel a.} Total luminosity (Eq.~\ref{eq:edot}), wind luminosity (Eq. \ref{eq:edot_wind}), jet luminosity (Eq.~\ref{eq:edot_jet}), luminosity of radiation trapped in the wind (Eq.~\ref{eq:edot_rad_trapped}), and luminosity of free-streaming radiation (Eq.~\ref{eq:edot_rad_streaming}). We have also labeled the Eddington luminosity ($L_{\rm Edd}$, Eq.~\ref{eq:Ledd}). \textbf{Panel b.} Efficiencies of each luminosity (except the trapped radiation) with respect to the time-averaged mass accretion rate (Eq.~\ref{eq:efficiencies_each}).}
    \label{fig:outflowprofs}
\end{figure}

\subsubsection{Energy inventory of the outflow \& emission components}
\label{sec:results:outflows:inventory}

We can study the outflows quantitatively by using the conservation of energy. The non-radiative stress-energy tensor is, 
\begin{equation}
    T^\mu_\nu = (\rho c^2 + u_{\rm g} + p_{\rm t} + b^2)u^\mu u_\nu + (p_{\rm t}+p_{\rm B})\delta^\mu_\nu - b^\mu b_\nu,    \label{eq:stress_energy_tensor}
\end{equation}
which accounts for (magneto-)hydrodynamic transport. The radiative stress-energy tensor is,
\begin{equation}
    R^\mu_\nu = \frac{4}{3c^2}E_{\rm r}u_{\rm r}^\mu u_{{\rm r},\nu} + p_{\rm r}\delta^\mu_\nu   \label{eq:rad_stress_tensor}
\end{equation}
The total luminosity of the system is then, 
\begin{equation}
    \dot{E} = \int_0^\pi\int_0^{2\pi} -\left(T^r_t+R^r_t+\rho c^2 u^r\right) \sqrt{-g}d\theta d\varphi 
    \label{eq:edot}
\end{equation}
where we have subtracted off the flux of rest mass energy, $\rho c^2 u^r$, which is separately conserved. On its own, $\dot{E}$ is not very illuminating, so we will decompose it into separate components. Broadly, we expect contributions from winds, relativistic jets, and radiation. In Figure \ref{fig:outflowprofs}(a), we show the various contributions to $\dot{E}$, and in Figure \ref{fig:outflowprofs}(b), we show the corresponding efficiencies, 
\begin{equation}
    \eta_i = \frac{\dot{E}_i}{\langle \dot{M}(r=5\,r_{\rm g})\rangle_{\Delta t=10^3\,r_{\rm g}/c}c^2},
    \label{eq:efficiencies_each}
\end{equation}
where $i$ is shorthand for a given component of $\dot{E}$ and $\langle\dot{M}(r=5\,r_{\rm g})\rangle_{\Delta t=10^3\,r_{\rm g}/c}$ is the mass accretion rate (Eq.~\ref{eq:mdot}) at\footnote{We sometimes measure $\dot{M}$ and other conserved quantities at $r=5\,r_{\rm g}$ instead of the event horizon because we use numerical floors that limit $\sigma$ to be no more than $25$, which is achieved by injecting gas artificially. In practice, these floors are only activated very near the horizon. This is, for instance, why the total $\dot{M}$ increases near the horizon in Fig.~\ref{fig:radialProfs}(e).} $r=5\,r_{\rm g}$ averaged over a time interval $\Delta t=10^3\,r_{\rm g}/c.$

\textit{Total.} In steady state, $\dot{E}$ is constant as a function of radius. $\dot{E}$ varies by a factor of a few between the event horizon and $\approx5\times10^3\,r_{\rm g}$ in both simulations. In this region, $\dot{E}$ is $\lesssim5-10\,L_{\rm Edd}$ in NS and $\lesssim30-50\,L_{\rm Edd}$ in HS. $\dot{E}$ is relatively steady to much larger radii than $\dot{M}$ (which is steady to $\gtrsim200\,r_{\rm g}$, see Fig.~\ref{fig:radialProfs}(e)) because most outflows escape at velocities near $c$. We have activated the radiation module only in the final $\approx10^4\,r_{\rm g}/c$ of the simulations (Table \ref{table:sims}), so our radiation field has not evolved too much beyond the photosphere (which is $\gtrsim10^4\,r_{\rm g}$, see Fig.~\ref{fig:mu2panel}).  The total efficiency is $\lesssim5-10\%$ in NS and $\approx50-100\%$ at most radii in HS. The high efficiency in HS is enabled by the extraction of rotational energy from the BH via the BZ mechanism. 

\textit{Jets.} 
Jets are defined by their relativistic velocities. So, we calculate the jet luminosity by only including gas where $\Gamma_\infty>2$,
\begin{equation}
    \dot{E}_{\rm jet} = \int_0^\pi\int_0^{2\pi} -\left(T^r_t+R^r_t+\rho c^2u^r\right)\Theta(\Gamma_\infty-2) \sqrt{-g}d\theta d\varphi,  
    \label{eq:edot_jet}
\end{equation}
where we classify outflows with smaller $\Gamma_\infty$ as ``winds''. In HS, we have $\Gamma_\infty>2$ near the poles (see Fig.~\ref{fig:mu2panel}(a)). $\dot{E}_{\rm jet}$ dominates the energetics of the outflows and is roughly constant out to $\lesssim2\times10^3\,r_{\rm g}$. Here, the jet decelerates, and the decrease in its luminosity is compensated by an increase in the wind luminosity. The jet luminosity increases again at $\approx2\times10^4\,r_{\rm g}$, which is where the effective photosphere is. This is not necessarily a persistent feature and is a result of the violent, time-variable nature of the funnel wherein the jet must push through slower, mass-loaded material. 

In HS, the high jet luminosity is enabled by the BZ mechanism. In NS, we also measure a super-Eddington jet luminosity, but it is much weaker than in HS and dies out 
 at $\lesssim1-2\times10^3\,r_{\rm g}$. As seen in Fig.~\ref{fig:mu2panel}(b), the $\Gamma_\infty<2$ region is small in volume and does not escape the photosphere. Since $a=0$, this is not a BZ jet, and is a high energy per unit mass component of the outflows launched by the disk. While for simplicity we have not split the jet into radiative and non-radiative components, we have verified that the NS jet luminosity is radiation dominated.

\textit{Winds.} We define winds as being mildly relativistic, with energy per unit mass $1.0002<\Gamma_\infty<2$. Here, the lower limit corresponds to a maximum possible velocity $v/c\approx0.02$. This cut is to avoid pollution from turbulent disk gas and from the transient launched from the initial conditions (see Fig.~\ref{fig:app:vr_spacetime} in the Appendix). The wind luminosity is,
\begin{equation}
\begin{aligned}
    \dot{E}_{\rm wind} &= \int_0^\pi\int_0^{2\pi} -\biggl[T^r_t+R^r_t\Theta(\tau_{\rm eff}-1)\Theta(u_{\rm r}^r)\\&+\rho c^2 u^r\biggr]\Theta(\Gamma_\infty-1.0002)\Theta(2-\Gamma_\infty)\sqrt{-g}d\theta d\varphi
    \label{eq:edot_wind},
    \end{aligned}
\end{equation}
where we have only included the radiative energy flux $R^r_t$ where the wind is optically thick. We have also only included outgoing ($u_{\rm r}^r>0$) radiation to avoid the inner disk region where radiation is advected into the BH. At small radii ($\lesssim20\,r_{\rm g}$), the wind is magnetohydrodynamic, as evidenced by the gap between the wind luminosity and the luminosity of trapped radiation (see below). In NS, the wind luminosity stabilizes above $\approx200\,r_{\rm g}$, which is roughly the size of the NVF region. The trapped radiation is within about a factor of two of the wind luminosity, suggesting that the wind is at least partially radiation-driven. At $\approx3\times10^3\,r_{\rm g}$, the jet decelerates, lowering $\dot{E}_{\rm jet}$ and increasing $\dot{E}_{\rm wind}$. 

\textit{Trapped radiation.}
Most of the emitted light is initially trapped within  the optically thick winds. We define the trapped radiative luminosity as,
\begin{equation}
\begin{aligned}
    \dot{E}_{\rm rad,trapped} &= \int_0^\pi\int_0^{2\pi} -R^r_t\Theta(\tau_{\rm eff}-1)\\&\Theta(u^r)\Theta(2-\Gamma_\infty)\sqrt{-g}d\theta d\varphi,
    \label{eq:edot_rad_trapped}
\end{aligned}
\end{equation}
where we have excluded the radiation in the jet. The trapped radiation is a subset of the winds in our definition. The trapped radiation rapidly increases within $r\lesssim10-20\,r_{\rm g}$ in both simulations, where most of the escaping light in the accretion flow is produced. At smaller radii, the radiation is advected directly into the BH. At larger radii, $\dot{E}_{\rm rad,trapped}$ closely traces $\dot{E}_{\rm wind}$. This indicates that the trapped radiation is loaded into the wind. We also notice that at $20\,r_{\rm g}\lesssim r\lesssim10^3\,r_{\rm g}$ $\dot{E}_{\rm rad,trapped}$ is larger in NS than in HS; this is because some of the trapped radiation in HS can escape into the optically-thin jet funnel. Just beyond the photosphere at $r\gtrsim10^4\,r_{\rm g}$, $\dot{E}_{\rm rad,trapped}$ decays, and there is a commensurate rise in the free-streaming radiation. 

\textit{Free-streaming radiation.}
Eventually, most of the light produced by the accretion flow will escape. We quantify this by measuring the free-streaming radiation,
\begin{equation}
\begin{aligned}
    \dot{E}_{\rm rad,streaming} &= \int_0^\pi\int_0^{2\pi} -R^r_t\Theta(1-\tau_{\rm eff})\\&\Theta(2-\Gamma_\infty)\sqrt{-g}d\theta d\varphi,
    \label{eq:edot_rad_streaming}
\end{aligned}
\end{equation}
In NS, $\dot{E}_{\rm rad,streaming}$ is zero out until $\gtrsim10^4\,r_{\rm g}$, because that is where the photosphere is located (Fig.~\ref{fig:mu2panel}(b)). At this radius, the escaping radiation is mildly super-Eddington with a peak radiative efficiency $\approx3\%$. This is a reasonable efficiency to expect, since it is somewhat smaller than the \citet{nt73} radiative efficiency which for $a=0$ is $\eta_{\rm NT}(a=0)\approx5.7\%$. We expect our efficiency to be lower, because $\eta_{\rm NT}$ assumes an efficiently cooling, sub-Eddington disk, but our disk is super-Eddington and traps radiation. However, because the photosphere is so large, the escaping radiation may require longer simulation runtimes for convergence. In HS, $\dot{E}_{\rm rad,streaming}$ is much larger, and is non-zero at $r\gtrsim10\,r_{\rm g}$. This is because the effective photosphere in the jet funnel reaches down to the near-ISCO region. However, we exclude the jet in our measurement of $\dot{E}_{\rm rad,streaming}$. This indicates that the cavity cleared out by the jet is larger than the jet itself, allowing space for light to escape down sub-relativistic channels (as seen in the inset panel of Fig.~\ref{fig:mu2panel}(a)). When the remaining light escapes the photosphere outside the jet funnel at $r\gtrsim10^4\,r_{\rm g}$, $\dot{E}_{\rm rad,streaming}$ increases, and the radiative efficiency rises to $\approx100\%$. Such a large efficiency is only possible via the extraction of BH rotational energy by the BZ mechanism. However, we caution that the conversion of the jets' Poynting flux to radiation is sensitive to our opacity treatment, where our assumptions are crude for optically thin regions such as the jet funnel.

\subsubsection{Summarized outflow \& emission properties}
Before continuing, we will briefly summarize the most important outflow and emission properties that we have just described,
\begin{itemize}
    \item When $a=0.9375$, the BH powers relativistic BZ jets that carve out a cavity and allow the effective photosphere to reach the event horizon (Fig.~\ref{fig:mu2panel}(a)) near the jet base. Outside the jet funnel, the photosphere is at $r\gtrsim10^4\,r_{\rm g}$ ($\approx1300\,{\rm au}$) above the disk. When $a=0$, the outflows are mass-loaded and mildly relativistic, and the effective photosphere is at $r\gtrsim10^4\,r_{\rm g}$ everywhere. 
    \item The mechanical (jet+wind) luminosities exceed the Eddington luminosity in both simulations. The mechancial luminosity is higher in HS, where the BZ jet contributes the most to the energetics, resulting in mechanical outflow efficiencies that reach $\sim60\%$. The outflow efficiency in NS is still high at $\sim10\%$. 
    \item The radiative luminosity is highly (mildly) super-Eddington when $a=0.9375$ ($a=0$). The radiative efficiency when $a=0.9375$ reaches $\sim100\%$, which we primarily attribute to the conversion of the Poynting flux in the jet to radiation. The radiative efficiency in NS reaches $\approx3\%$, which is a factor $\sim2$ smaller than the \citet{nt73} radiative efficiency for $a=0$.  
\end{itemize}

While we have focused on the outflow energetics, their momentum is also important for understanding the resulting feedback. We do not plot momentum, but note that when $a=0$, the momentum in the wind is $\approx2\times$ the momentum in the escaping radiation. When $a=0.9375$, the momentum in the wind is much less than the momentum in the escaping radiation, but this is because the jet boosts the radiation. Overall, the momentum in the winds is almost the same in both simulations. 

\subsection{Time evolution}
\label{sec:results:evolution}

\begin{figure}[bht]
    \centering
    \includegraphics[width=\textwidth]{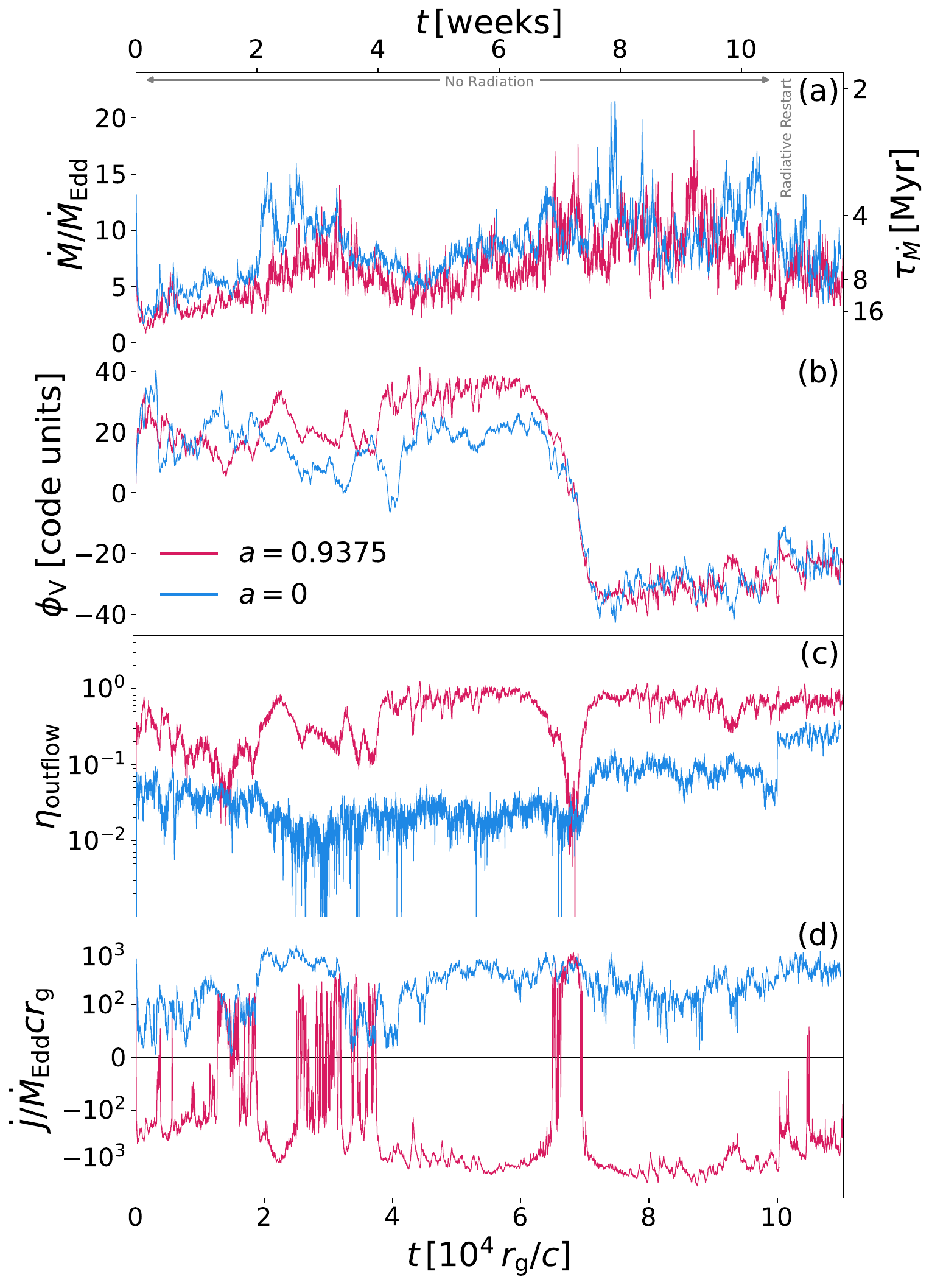}
    \caption{Time evolution of quantities measured near the event horizon for runs HS and NS. Super-Eddington accretion doubles the mass of the SMBH on $\approx5-10\,{\rm Myr}$ timescales. In HS, large horizon-scale values of $\phi_{\rm V}$ facilitate strong BZ jets that spin down the BH. After radiation is activated, we include radiative contributions in
    $\eta_{\rm outflow}$ and $\dot{J}$ only where $\sigma<1$. \textbf{Panel a.} Dimensionless magnetic flux ($\phi_{\rm V}$, Eq.~\ref{eq:dimensionless_magnetic_flux}) threading the event horizon. \textbf{Panel b.} Efficiency of horizon-scale mechanical outflows ($\eta_{\rm outflow}$, Eq.~\ref{eq:efficiencies_each} except neglecting $R^r_t$). \textbf{Panel c.} Mass accretion rate ($\dot{M}$, Eq. \ref{eq:mdot}), measured at $5\,r_{\rm g}$, normalized to the Eddington accretion rate ($\dot{M}_{\rm Edd}$, Eq. \ref{eq:mdot_Edd}). We also show the corresponding mass doubling timescale $\tau_{\rm M}=M/\dot{M}$. \textbf{Panel d.} Rate of change of BH angular momentum ($\dot{J}$, Eq.~\ref{eq:jdot})), normalized to $\dot{M}_{\rm Edd}cr_{\rm g}$.
    }
    \label{fig:timeseries}
\end{figure}

In Figure \ref{fig:timeseries}, we show quantities measured near the event horizon as a function of time for runs NS and HS. We mark the moment when radiation is activated ($t\approx10^5\,r_{\rm g}/c$, see Table \ref{table:sims}) with a vertical line.

\textit{Mass accretion.} In Fig.~\ref{fig:timeseries}(a), we plot $\dot{M}$ (Eq. \ref{eq:mdot}) normalized to the Eddington accretion rate ($\dot{M}_{\rm Edd}$, Eq. \ref{eq:mdot_Edd}). Once radiation is on, the BHs sustain super-Eddington accretion rates $\dot{M}\approx5\,\dot{M}_{\rm Edd}$. The disk is capable of super-Eddington accretion rates because the rapid inflow rate (e.g., Fig.~\ref{fig:radialProfs}(d)) traps photons and advects them into the BH, reducing the amount of work they can do on the accretion flow. We have also expressed $\dot{M}$ in terms of the mass doubling timescale, $\tau_{\dot{M}}=M/\dot{M}$, which is usually\footnote{Note here we neglect the energetic contribution to the mass evolution - namely, the extraction of rotational energy, which contributes to the rest-mass energy, by BZ jets.} $\approx5-10\,{\rm Myr}$. This is quite rapid, and is well within the typical range of quasar lifetimes $\approx1-100\,{\rm Myr}$ \citep{martini_2004,hopkins_2009}. This is consistent with the expectation that the mass evolution of SMBHs is dominated by accretion during quasar phases \citep{soltan_1982,chokshi_1992,hopkins_2006}. 

\textit{Flux saturation.} The ``dimensionless'' vertical magnetic flux at the event horizon,
\begin{equation}
    \phi_{\rm V}(r_{\rm H},t) = \frac{\Phi_{\rm V}(r=r_{\rm H},\theta=\pi/2,t))}{\langle \dot{M}(r=5\,r_{\rm g})\rangle_{\Delta t=10^3\,r_{\rm g}/c}c},
\label{eq:dimensionless_magnetic_flux}
\end{equation}
measures how saturated the accretion flow is with NVF. It is commonly used to diagnose the MAD state, in which $\phi_{\rm V}$ is saturated at the value $\phi_{\rm V,\,MAD}$, which is sustained by periodically ejecting excess field in poloidal flux eruptions. We note that $\phi_{\rm V,\,MAD}$ is defined empirically. Often, MADs are associated with radiatively inefficient accretion flows such as low-luminosity AGN, where $\phi_{\rm V,\,MAD}\gtrsim50$ \citep[e.g.,][]{sasha_2011}. However, the MAD state may also be extended to disks in any thermodynamic state that are fully saturated with NVF \citep[e.g.,][]{avara_2016,curd_2023,nico_2024}. These works find $\phi_{\rm V,MAD}$ to be somewhat lower, usually near $30$, but the exact value depends on the cooling or radiative prescription used. In Fig.~\ref{fig:timeseries}(b), we plot the dimensionless magnetic flux $\phi_{\rm V}$ at the event horizon. We can see that $\phi_{\rm V}\approx20-40$ until $t\approx7\times10^4\,r_{\rm g}/c$, after which vertical flux of opposite polarity accretes (see Fig.~\ref{fig:fluxfliploops}), and the BH recovers flux up to $|\phi_{\rm V}|\approx20-40$ once again. The value of $\phi_{\rm V}$ does not change significantly when radiation is turned on and settles at values $\gtrsim20$, although we may need to evolve the simulations to longer time to assess whether $\phi_{\rm V}$ is steady. Regardless, such values are within order unity of $\phi_{\rm V,MAD}$, indicating that the innermost region of the disk is nearly saturated with the maximum amount of NVF that it can hold. 


\textit{Mechanical outflow efficiency.} 
 In Fig.~\ref{fig:timeseries}(c), we show the mechanical outflow efficiency ($\eta_{\rm outflow}$, Eqs.~\ref{eq:edot} and \ref{eq:efficiencies_each}) at $5\,r_{\rm g}$. We have neglected the radiative term $R^r_t$ in Eq.~\ref{eq:edot} when our radiation module is active since at these radii radiation is mainly advected into the BH. In HS, $\eta_{\rm outflow}$ reaches values $\sim100\%$. As described in Section \ref{sec:results:outflows}, this is mostly in the form of BZ jets. In NS, the BH is not rotating, so there is no BZ mechanism and $\eta_{\rm outflow}\gtrsim5-10\%$ for most of the simulation runtime. When radiation is turned on, the outflow efficiency increases to $\gtrsim20\%$. We attribute this to a radiation-driven outflow.

\textit{Spin evolution.} Quasar disks may also either spin up SMBHs via prograde accretion or spin down SMBHs via the extraction of rotational energy by BZ jets \citep[e.g.,][]{bardeen_1970,gammie_2004}. We can use the conservation of spin-aligned angular momentum to define the angular momentum transport rate,
\begin{equation}
    \dot{J}(r,t) = \int_0^\pi\int_0^{2\pi} T^r_\varphi + R^r_\varphi\Theta(1-\sigma)\sqrt{-g}d\theta d\varphi.
    \label{eq:jdot}
\end{equation}
Here, when our radiation module is activated, we only include radiative contributions to the angular momentum flux where the magnetization $\sigma=b^2/\rho c^2)$ is less than $1$ to avoid the optically thin jet funnel since our numerical density floors are active in this region and may artificially scatter light back into the BH. We show this quantity at $5\,r_{\rm g}$ in Fig.~\ref{fig:timeseries}(d), normalized to $\dot{M}_{\rm Edd}cr_{\rm g}$. In NS, the BH spins up due to the accretion of angular momentum. In HS, the BH generally spins down, due to the extraction of rotational energy by the BZ mechanism. When radiation is activated, the disk slims somewhat and the gas orbits at velocities closer to Keplerian, so $\dot{J}$ increases.

As the mass and angular momentum of the BH evolves, it will tend towards an equilibrium spin $a_{\rm eq}$
at which spin extraction by BZ jets is balanced by the accretion of mass and angular momentum. \citealt{bev_2024} reported that in non-radiative MADs, $a_{\rm eq}\approx0.07$, whereas \citet{narayan_2022} reported $a_{\rm eq}\approx0.035$. However, our $\phi_{\rm V}$ is about a factor of two smaller than such systems, so our $a_{\rm eq}$ is likely somewhat larger. Our disk is also Keplerian down to the ISCO (e.g., Fig.~\ref{fig:radialProfs}(d)), whereas non-radiative MADs are sub-Keplerian. So, our disk will also spin-up more from accreted angular momentum. Both of these effects are accounted for in the spin up parameter \citep{gammie_2004},
\begin{equation}
    s = \frac{da}{dt}\frac{M}{\dot{M}},
    \label{eq:spindown_parameter}
\end{equation}
where negative (positive) values indicate spin-down (-up). In non-radiative MADs, $s\lesssim-7$ for $a=0.9$ and $s\lesssim1$ for $a=0$ \citep{bev_2024}. We measure the following median spin-down parameters during the radiative portion of our simulation: $s=-1.22^{0.52}_{-0.68}$ for HS and $s=1.23^{+0.18}_{-0.19}$ for NS, where the superscript (subscript) indicates the difference between the median and the third (first) quantiles. 
Since $|s|$ is order unity, it suggests that the spin-down timescale $(\frac{da}{dt})^{-1}\sim\tau_{\rm M}\sim5-10\,{\rm Myr}$. This, again, is well within the quasar duty cycle, suggesting that the BH will reach $a_{\rm eq}    $ relatively quickly. Since $s$ in HS is roughly equal and opposite to $s$ in NS, we expect the BH spin to equilibrate at an intermediate value of $a$. However, it is difficult to predict $a_{\rm eq}$ without conducting a larger parameter space survey.

\begin{figure*}[bht]
    \centering
    \includegraphics[width=\textwidth]{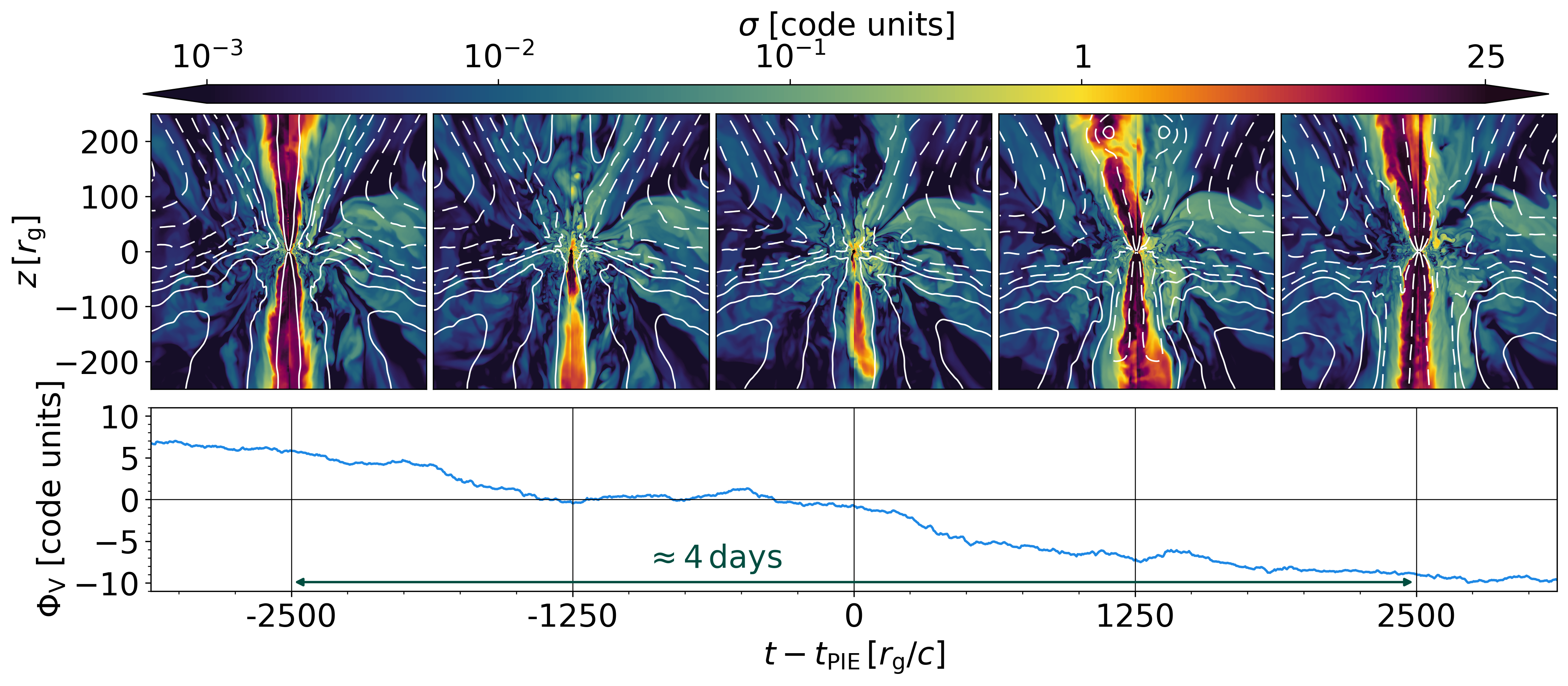}
    \caption{Depiction of a polarity inversion event in run HS at time $t_{\rm PIE}\approx7\times10^4\,r_{\rm g}/c$. The polarity inversion causes the relativistic BZ jet ($\sigma>1$ regions) to turn off and on with different polarities. In the top panel, we show a sequences of $500\,r_{\rm g}$ slices of magnetization $\sigma=b^2/\rho c^2$ separated by time $1250\,r_{\rm g}/c$ ($\approx 1$ day), with $\sqrt{|\Phi_{\rm P}|}$ contours (Eq. \ref{eq:magnetic_flux}) shown in white. Solid and dashed flux contours have opposite polarity. In the bottom panel, we show $\Phi_{\rm V}$ as a function of time, where each vertical line marks the time of each $\sigma$ slice.}
    \label{fig:fluxflip}
\end{figure*}

\section{Discussion}
\label{sec:discussion}

\subsection{Polarity inversion events}
\label{sec:discussion:fluxflip}

Poloidal field loops advected from large distances to the event horizon have no preferred direction. So, it is inevitable that as poloidal field loops are accreted, the polarity (i.e., sign) of the NVF on the event horizon will sometimes invert. This is exactly what we observed at $t\approx7\times10^4\,r_{\rm g}/c$ in Figs.~\ref{fig:spacetime}, \ref{fig:fluxfliploops} and \ref{fig:timeseries}(a). We note that this occurred during the non-radiative portion of the simulation, but we do not expect the occurrence of the inversion to be sensitive to radiation, because in our system radiation does not significantly change the disk dynamics. In recent years, several authors have proposed that such polarity inversion events may explain observed phenomena such as state transitions in X-ray binaries \citep{livio_2003,igumenshchev_2009,dexter_2014} or certain changing-look active galactic nuclei \citep{nico_2024}. 

In Figure \ref{fig:fluxflip}, we show a sequence of magnetization $\sigma=b^2/\rho c^2$ snapshots in the $x-z$ plane before and after the polarity inversion event at $t_{\rm PIE} = 7\times10^4\,r_{\rm g}/c$. We also plot, in white, contours of $\sqrt{|\Phi_{\rm P}|}$ (Eq. \ref{eq:magnetic_flux}), where solid (dashed) lines indicate positive (negative) $\Phi_{\rm P}$ values. There is a distinctly quadrupolar field geometry, where the polarity of $\Phi_{\rm P}$ is antisymmetric about the disk midplane as indicated by the lines switching from solid to dashed (also seen in Fig.~\ref{fig:fluxfliploops}). At the beginning of the sequence, a BZ jet is active ($\sigma>1$ regions), and the jet has negative polarity. Soon after, the positive polarity field loops above the disk overtake the negative polarity field. The jet in the north hemisphere turns off first because it reconnects with encroaching field lines of negative polarity. As the negative polarity field wins, the jet is relaunched. In the bottom of Fig.~\ref{fig:fluxflip}, we show the evolution of $\Phi_{\rm V}$ during this event, where solid vertical lines mark the times of the above panels. The entire transition spans $\approx4\,{\rm days}$.

Since the jet is briefly off during the polarity inversion, it is possible that any coronal gas associated with the jet base (e.g., in classic ``lamp-post'' style models) may temporarily collapse or weaken \citep[which may, for example, manifest as the destruction of the X-ray corona,][]{nico_2024}. Then, the newborn jet must propagate through the ejecta of the previous jet. Since the previous jet had the opposite polarity, its fossil field lines may reconnect with the field lines composing the new jet, forming a ``striped'' jet. 
The formation of striped jets by the advection of field loops of alternating polarity has been proposed by other authors \citep[e.g., ][]{parfrey_2015,mahlmann_2020,anna_2021,emma_2023} and magnetic reconnection at the stripes may accelerate the jets themselves \citep{zhang_giannios_2021,giannios_uzdensky_2019}, accelerate particles to high energies \citep{lyubarsky_liverts_2008,lorenzo_2011}, and produce variable emission that may explain some X-ray and $\gamma$-ray variability in blazars \citep{zhang_2020,zhang_2021,zhang_2022}. We also note that the emission during the polarity inversion event may significantly vary depending on the hemisphere it was observed from, since the quadrupolarity of the encroaching poloidal field can cause one jet to reconnect promptly and the other to reconnect more gradually. Our results demonstrate that polarity inversion events occur naturally in quasar disks where the disk and magnetic field were realistically assembled, which suggests that such events are common and deserve further study. 

\begin{figure}[bht]
    \centering
    \includegraphics[width=\textwidth]{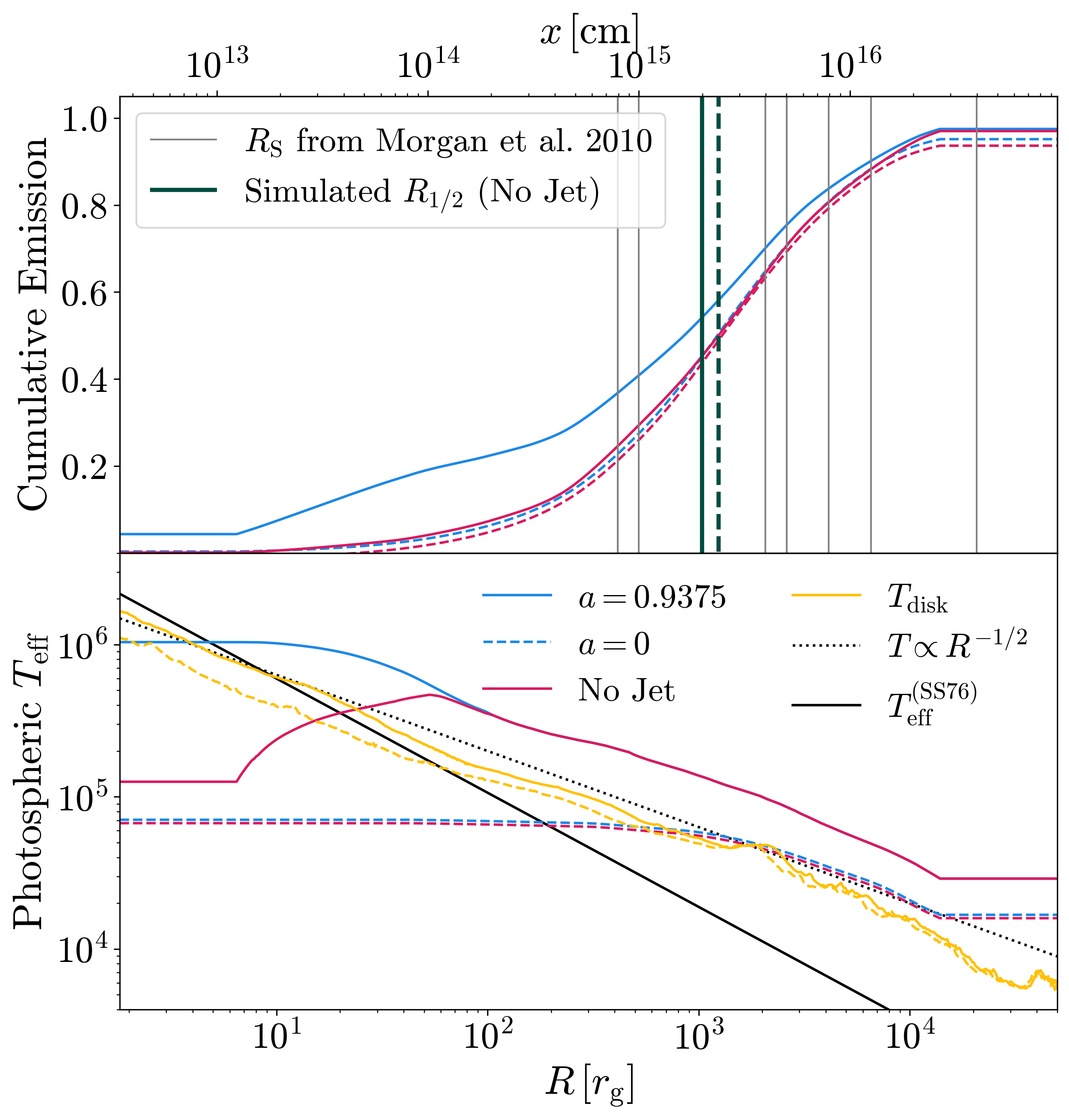}
    \caption{Profiles of the cumulative emission and the effective temperature ($T_{\rm eff}$) of the disks at $t=111250\,r_{\rm g}/c$. The half-light radii, $R_{1/2}$, are large, which is consistent with observations. The temperature profiles peak at lower values than classical thin disks \citep{shakura_sunyaev_1976,nt73} and have shallower radial profiles. \textbf{Panel a.} Cumulative emission leaving the effective photosphere as a function of cylindrical radius, normalized to the emission at $R=5\times10^4\,r_{\rm g}$. We measure $R_{1/2}\approx940\,r_{\rm g}$ and $1180\,r_{\rm g}$ for runs HS and NS, respectively (green vertical lines). We also include curves that exclude regions where $\Gamma_\infty>2$ (red) to avoid relativistic jet emission. We also mark the accretion disk sizes ($R_{\rm S}$, gray) reported by \citet{morgan_2010}. \textbf{Panel b.} Axisymmetrized $T_{\rm eff}(R)$ with same line and color styles as panel a. We also show the thin disk prediction, $T_{\rm eff}^{\rm (SS76)}$ (Eq.~\ref{eq:teff_t76}). At radii $\gtrsim 10^2\r_{\rm g}$ ($\gtrsim 10^3\,r_{\rm g}$) in HS (NS), the temperature profiles obey a power law that is $\propto R^{-1/2}$ (black solid line). We also show the density-weighted radiation temperature of the disk ($T_{\rm disk}$, Eq.~\ref{eq:tmid}).}
    \label{fig:photosphereprofs}
\end{figure}

\subsection{Implications for observed accretion disk sizes}
\label{sec:discussion:microlensing}

Quasar microlensing observations suggest that the half-light radii of quasars are about three to ten times larger \citep{pooley_2007,morgan_2010,blackburne_2011,ren_2024} than predicted by classical \citep[][]{ss73,nt73} models. Such large disk sizes are consistent with reverberation mapping studies \citep{kaspi_2000,peterson_2004,zu_2011} and optical interferometry of the broad-line region in 3C 273 \citep{GRAVITY_2018}. 
Additionally, classical models predict effective temperature profiles $T_{\rm eff}\propto R^{-3/4}$, where $R$ is the cylindrical radius. However, many quasar microlensing studies have inferred different profiles in integrated light. While some observations infer shallower profiles \citep{rojas_2014,bate_2018,cornachione_2020}, others predict steeper profiles \citep{blackburne_2011,munoz_2011,jimenez-vicente_2014,blackburne_2015,munoz_2016,motta_2017}, although the differences may depend on the method of analysis \citep[e.g., ][]{cornachione_2020}. While the ``real'' effective temperature profile of quasar disks remains unclear, it is well-established that the peak effective temperatures are cooler than expected. For instance, the effective temperature of a \citealt{shakura_sunyaev_1976} disk is,
\begin{equation}
\begin{aligned}
    T_{\rm eff}^{({\rm SS76})} &\approx 2\times10^6\,{\rm K}\,\left(\frac{M}{1.3\times10^7\,M_\odot}\right)^{1/4}\\&\times\left(\frac{\dot{M}}{10\dot{M}_{\rm Edd}}\right)^{1/4}\left(\frac{R_{\rm inner}}{2\,r_{\rm g}}\right)^{-3/4}
    \label{eq:teff_t76}
\end{aligned}
\end{equation}
where $R_{\rm inner}$ is the inner cutoff radius of the disk, usually taken between the event horizon and ISCO. Yet, the spectral energy distributions of quasars peak in the blue to ultraviolet, with observed $T_{\rm eff} \sim {\rm a\ few}\times 10^{4}$\,K \citep[the ``big blue bump'', e.g.][]{shields_1978,czerny_1987,czerny_2003,koratkar_1999,richards_2006,bonning_2013}.

In Fig.~\ref{fig:photosphereprofs}(a), we show the cumulative emission, normalized to the net emission at $5\times10^4\,r_{\rm g}$ leaving the photosphere. We do this by projecting $R^\mu_t$ at the photosphere (where $\tau_{\rm eff}=1$, Eq. \ref{eq:effective_optical_depth}) in the $\hat{e}_z$ direction and integrating over $\varphi$ and the cylindrical radius of the axisymmetrized effective photosphere. We do this for both hemispheres and average the result. We also show curves where we only include light where $\Gamma_\infty<2$ to avoid emission from the relativistic jet. We show the result for run HS (NS) in solid (dashed) lines and mark the resulting ($\Gamma_\infty<2$) half-light radii ($R_{1/2}$) where we find $R_{1/2}\approx940\,r_{\rm g}$ and $1180\,r_{\rm g}$ for HS and NS, respectively. We also show the accretion disk size, $R_{S}$, for the sample of microlensed quasars studied by \citet{morgan_2010}. Our simulated $R_{1/2}$ values are well within the range of observed disk sizes. This result demonstrates that the emitting region of our quasar is much larger than in classical disk models. This is because of the extended photospheres in our simulations. While thin disks emit directly from their inner regions, the light in our disk scatters many times before escaping, which spreads the emitted light out over a larger region. Such large scattering surfaces may explain why most of the  disk sizes inferred by quasar microlensing observations have a weak dependence on wavelength, as Thompson scattering (dominant here) is achromatic.

In Fig.~\ref{fig:photosphereprofs}(b), we show the axisymmetrized effective temperature, $T_{\rm eff}(R)$. We measure $T_{\rm eff}$ by assuming the radiation flux leaving the photosphere is thermal and using the Stefan-Boltzmann law, $F=\sigma_{\rm SB}T_{\rm eff}^4$. Here, we again measure the radiative flux, $F$ by taking the $\hat{e}_z$ component of $R^\mu_t$ at the photosphere (averaged over both hemispheres). We can see that $T_{\rm eff}$ does not obey a single power law. Instead, $T_{\rm eff}$ at small radii is relatively constant. The reason for this is evident in Fig.~\ref{fig:mu2panel}: the photosphere extends to large heights above the BH, especially in NS. So, although Fig.~\ref{fig:photosphereprofs}(b) exhibits a constant temperature over 2-3 orders of magnitude in cylindrical radius, this is actually a relatively narrow range of spherical radii. Additionally, the peak temperature for NS is $\approx8\times10^4\,{\rm K}$. This is well below the peak temperatures predicted by Equation \ref{eq:teff_t76}. Here, we regard the $a=0$ results as more representative of the disk, since much of the emission in HS (even where $\Gamma_\infty<2$) originates from the jet funnel (see Section \ref{sec:results:outflows}). The cool effective temperatures that we measure are consistent with the spectral energy distributions of quasars \citep{{shields_1978,czerny_1987,czerny_2003,koratkar_1999,richards_2006,bonning_2013}}. Beyond the polar emitting region, the temperature profiles begin falling off roughly as $\propto R^{-1/2}$, which is similar to a slim disk \citep{abramowicz_1988,beloborodov_1998}. These results lend credence to analyses which infer shallow effective temperature profiles in observed quasars \citep{cornachione_2020}.
We can also compare the effective temperature to the midplane temperature,
\begin{equation}
    T_{\rm disk} = \langle T_{\rm r} \rangle_{\rm disk},
    \label{eq:tmid},
\end{equation}
where we assume that the radiation and gas temperatures closely follow each other\footnote{In practice, our numerical scheme struggles to accurately predict $T_{\rm gas}$ in the inner regions where $p_{\rm t}\ll p_{\rm r}$ and the opacity is dominated by electron scattering. However, since radiation energy dominates over gas internal energy here, $T_{\rm rad}$ matters for the disk dynamics, not $T_{\rm gas}$.}. In Fig.~\ref{fig:photosphereprofs}, we see that $T_{\rm disk}$ falls off as $\propto R^{-1/2}$ in both simulations. This is consistent with slim disk models, which are more appropriate for super-Eddington accretion flows \citep{abramowicz_1988,beloborodov_1998}. $T_{\rm disk}$ reaches much higher peak temperatures than $T_{\rm eff}$ when $a=0$, which is expected for such extended photospheres.

\subsection{Caveats}
\label{sec:discussion:caveats}
Before summarizing, we emphasize a few caveats of our work,
\begin{itemize}
    \item The disk was initially evolved in the \gizmo{} code, which included different physics than the \hamr{} code and different numerical treatment of some of the same physics. This is especially true for radiation. While both \gizmo{} and \hamr{} use two-moment ``M1'' closure schemes \citep{levermore_1984} for radiation, \gizmo{} uses a multiband method whereas \hamr{} uses a gray (frequency-integrated) method. Additionally, the opacities in \gizmo{} are more sophisticated at low temperatures and have been well-tested in the interstellar medium in conditions more similar to the very outer regions of the disk and the photosphere. 
    \item As we have discussed, radiation was inactive for the majority of our \hamr{} simulation runtime. We activated radiation in the final $\approx10\%$ of the simulations. During the non-radiative portion, some of the physics that we have highlighted -- namely, the magnetic state transition -- occurred. However, in shorter test simulations with radiation activated from the start, we also saw the same transition, so we expect this result to be robust.
    \item Our opacities assume that the radiation is thermal and in local thermodynamic equilibrium with the gas. In optically thin or marginally optically thick regions, these assumptions are inappropriate. Additionally, since our radiation scheme is grey (frequency-integrated), we are forced to choose an averaging procedure for our opacities. This prevents our opacities from being accurate in all regimes; we have used an absorption average \citep{mckinney_2017}, intended to be accurate in the highly optically thin or optically thick regimes, but this likely fall short in transitional regimes of moderate optical depth. While we still expect our effective temperatures at the photosphere to be correct at the order of magnitude level, they remain crude. This may also affect the high rate of conversion of the BZ jet's Poynting flux into radiation, since this happens in the optically thin jet funnel. We also neglect dust opacities and partial ionization effects, which can both be important where the gas temperature is $<10^4\,{\rm K}$.
    \item We have neglected radiation viscosity. This can be important in AGN disks since the radiation pressure far exceeds the gas pressure \citep[e.g., ][]{jiang_2019b}. However, in the super-Eddington disks we have simulated, radiation viscosity is less important because the mean free path of the photons is small \citep{jiang_2019}. However, radiation viscosity may still play an important role in regions of moderate optical depth (i.e., where the mean free path is comparable to the local length scales of interest) such as the jet sheath or in lower-density, highly magnetized filaments, and deserves future study. 
    \item Finally, we emphasize that our flow is not statistically steady. Even though the disk was originally evolved on larger scales in \gizmo{}, the outer accretion flow responds to the inner accretion flow on timescales that remain computationally inaccessible. Our mass accretion rate is steady out to $r\gtrsim200\,r_{\rm g}$ (Fig.~\ref{fig:radialProfs}(e)) and our energy outflow rate is steady out to $r\approx5000\,r_{\rm g}$. Although the size of the NVF region appears to be set by the size of the poloidal flux structures advected from the outer disk (see Section \ref{sec:results:topology:whatcausesit}), it is not yet clear that the size of the NVF region is converged. Additionally, radiation produced by the inner accretion flow may diffuse out of the photosphere on longer timescales than modeled here. While our disk comfortably accretes above the Eddington limit during the radiative portion of the simulation, it is still possible that radiation feedback may drive duty cycle like behavior on longer timescales.
\end{itemize}
\subsection{Summary}
\label{sec:discussion:summary}
We report the results of two general-relativistic, radiation magnetohydrodynamics simulations of a quasar disk feeding a $1.3\times10^7\,M_\odot$ SMBH, with $a=0$ and $0.9375$. We carried out the simulation using the \hamr{} code. Our initial conditions were remapped from \gizmo{}, in which the disk was self-consistently formed within a galaxy \citep{FORGE1,FORGE2,FORGE3}. This is the first time a quasar has been assembled from cosmological initial conditions and evolved down to the event horizon of the SMBH. We have specifically focused on the magnetic evolution of the inner accretion disk and the resulting radiative and outflow properties. We have also gleaned insights into the cosmological evolution of the SMBH. Our main findings are,
\begin{itemize}
    \item The quasar undergoes a magnetic state transition at $\approx200\,r_{\rm g}$ ($\approx26\,{\rm au}$). We define the ``magnetic state'' by the relative strength of the net toroidal magnetic flux (NTF), $\Phi_{\rm T}$, and the net vertical magnetic flux (NVF), $\Phi_{\rm V}$. In H24, the disk was NTF-dominated throughout. Here, we found that the inner disk quickly transitioned to being NVF-dominated within $r\lesssim200\,r_{\rm g}$ (Fig.~\ref{fig:netflux1d}).  We have argued that the NVF region forms by advecting $\mathcal{O}(H)$ poloidal flux structures from large radii to the inner disk, wherein they become large-scale. Once the poloidal field becomes large-scale, the disk rotation shears it into a toroidal field that is anti-symmetric about the midplane (Fig.~\ref{fig:baxisym}). The NVF region is also associated with strong, biconical winds and, when the BH rotates, \citet{bz_1977} jets.
    \item The poloidal magnetic flux structures  advected from the NTF region have random polarity, which naturally leads to ``polarity inversion events'', wherein the NVF threading the inner accretion disk switches sign. This causes the jet to turn off and on within $\approx5000\,r_{\rm g}/c$ ($\approx4$ days). This likely results in rapid variability \citep{livio_2003,igumenshchev_2009,dexter_2014,parfrey_2015,nico_2024}, possibly by briefly turning off the X-ray corona or by magnetic reconnection when the new jets interact with the old jets of opposite polarity. 
    \item The SMBH sustains super-Eddington accretion, with $\dot{M}/\dot{M}_{\rm Edd}\approx5-10$. However, we caution that radiation feedback may still play a role on longer timescales than we have simulated. The resulting radiative efficiency is $\approx3\%$ when $a=0$, which is about a factor of two smaller than predicted by thin disk theory \citep{nt73}. When $a=0.9375$, the radiative efficiency reaches $\sim100\%$; this is enabled by the conversion of Poynting flux in the BZ jets into radiation, although we caution that this number may be sensitive to our opacity treatment. The mechanical (combined wind and jet) luminosities are also strong, reaching efficiencies of $\sim60\%$ when $a=0.9375$ and $\sim10\%$ when $a=0$.
    \item The photosphere is extremely extended, reaching radii $\gtrsim10^4\,r_{\rm g}$ ($\gtrsim1300\,{\rm au}$). The effective temperature of the photosphere is $\sim10^4-10^5\,{\rm K}$, which is cooler than predicted for thin accretion disks but in line with AGN SEDs and the ``big blue bump'' \citep[e.g.,][]{shields_1978,czerny_1987,koratkar_1999}. When $a=0.9375$, the jet clears out a funnel, lowering the effective photosphere near the polar axis to the event horizon. We have measured large half-light radii $R_{1/2}\approx10^3\,r_{\rm g}$ $(\approx2\times10^{15}\,{\rm cm})$, which is much larger than predicted for thin disks but consistent with the half-light radii inferred from observations.
    \item The super-Eddington accretion rate will double the mass of the SMBH in $\approx5-10\,{\rm Myrs}$, which is consistent with the expectation that quasar phases dominate the mass growth of SMBHs \citep{soltan_1982,chokshi_1992}. The spin evolution occurs on the same timescale, after which the BH will reach an equilibrium spin $a_{\rm eq}$ where the accretion of angular momentum, which spins up the SMBH, is balanced by the extraction of rotational energy by \citet{bz_1977} jets, which spins down the SMBH. Although we expect intermediate values of $a_{\rm eq}$, we cannot determine what $a_{\rm eq}$ is until we have evolved the disk at a wider range of BH spin values.

\end{itemize}

 Such strongly magnetized disks as we have studied here represent a frontier in accretion theory that is not well understood. The initial conditions we have used here represent a single data point, and a wider variety of studies -- on both disk assembly from galaxy scales, as done here, and more controlled numerical experiments -- are necessary to understand how generic this type of accretion flow is. There are several future directions that we can highlight. Firstly, the MHD turbulence mechanisms -- especially in the outer disk -- are not well understood. Yet, the resulting turbulence provides the reservoir of poloidal magnetic flux which dictates the magnetic state of the inner disk via advection physics that is also poorly understood. While we have not studied angular momentum transport and turbulence in this work, we plan to do so in the future. Secondly, it is also essential to understand the interplay of the MHD physics with thermodynamics, especially in the cold outer disk. The dominance of the toroidal field over the cold disk may be inhibiting the buoyant loss of toroidal flux \citep[see also][]{squire_2024}. Also, the turbulence in the outer disk may be enhanced and/or strongly influenced by multi-phase gas physics and optically-thin cooling effects \citep{FORGE2} not modeled herein. Thirdly, related to the question of thermodynamics, it is necessary to study similar accretion flows in a wider range of Eddington ratios to understand how well our results connect to other types of AGN. 

\begin{acknowledgments}
We thank Minghao Guo, Eliot Quataert and Ethan Vishniac for useful conversations. NK is supported by an NSF Graduate Research Fellowship.  ML was supported by the John Harvard, ITC and NASA Hubble Fellowship Program fellowships, and NASA ATP award 21-ATP21-0077. Support for PFH was provided by NSF Research Grants 20009234, 2108318, NASA grant 80NSSC18K0562, and a Simons Investigator Award. AT acknowledges support by NASA 
80NSSC22K0031, 
and 80NSSC18K0565 
grants, and by the NSF grants 
AST-2107839, 
AST-1815304, 
AST-1911080, 
AST-2206471, 
AST-2407475, 
OAC-2031997. 
JJ and AT acknowledge support by the NSF AST-2009884, NASA 80NSSC21K1746 and NASA XMM-Newton  80NSSC22K0799 grants.
This research was supported in part by grant NSF PHY-2309135 to the Kavli Institute for Theoretical Physics (KITP). An award of computer time was provided by the Innovative and Novel Computational Impact on Theory and Experiment (INCITE) and ASCR Leadership Computing Challenge (ALCC) programs under award AST178. This research used resources of the Oak Ridge Leadership Computing Facility at the Oak Ridge National Laboratory, which is supported by the Office of Science of the U.S. Department of Energy under Contract No. DE-AC05-00OR22725. The authors acknowledge the Texas Advanced Computing Center (TACC) at The University of Texas at Austin for providing computational resources that have contributed to the research results reported within this paper.
\end{acknowledgments}

\appendix

\twocolumngrid

\section{Interpolation}
\label{app:interpolation}
\subsection{Preparing the data}
We first modify the \gizmo{} data before remapping it to \hamr{}.  The \gizmo{} quantities that we remap are gas density ($\rho$), thermal+radiation pressure ($p_{\rm t+r}\equiv p_{\rm t}+p_{\rm r}$), velocity ($v^i$), and the magnetic field ($\mathcal{B}^i$).  We start by converting the data from c.g.s. to code units, where $G=M_{\rm SMBH}=c=1$. We also normalize the maximum density to $1$. Since the radiative physics is not scale-free, we must record the black hole mass $M_{\rm SMBH}=1.3\times10^7\,M_\odot$ and the maximum value of density $\rho_{\rm max}=1.36\times10^{-8}\,{\rm g\,cm^{-3}}$ to use in our scale-dependent calculations.

\hamr{} evolves the gas internal energy density, $u_{\rm g}$, which we relate to the \gizmo{} $p_{\rm t+r}$ via the equation of state for a perfect, ideal gas,
\begin{equation}
    u_{\rm g}^{(\hamr{})} = p_{\rm t+r}^{(\gizmo{})}/(\gamma-1)
    \label{eq:app:eos_remap},
\end{equation}
We specifically interpolate pressure, rather than internal energy, to ensure initial force balance. We initialize the radiation variables to zero (which, in the M1 closure scheme, are the radiation energy density and the three spatial components of radiation velocity). This is necessary in the simulations that are initially non-radiative (see Table \ref{table:sims}) and acceptable in initially radiative simulations since emission processes automatically inject any absent radiation within a handful of timesteps. 

The \gizmo{} variables are defined in Cartesian coordinates and the disk is not aligned with any particular axis. When we interpolate to \hamr{}, we want our disk to be aligned with the BH spin, which is oriented along the $\hat{e}_z$ axis. To do this, we calculate the total angular momentum of the \gizmo{} disk, which is oriented along the unit vector $\hat{l}_{\texttt{G}}$. Then, we use Rodrigues' rotation formula to rotate the position, velocity and magnetic field vectors in such a way that the rotated $\hat{l}_{\texttt{G}}$ is parallel to $\hat{e}_z$. Once this is done, the \gizmo{} data is ready to be interpolated. Vectors are transformed from Cartesian coordinates to general-relativistic coordinates after interpolation.

\subsection{Construction of Tetrahedral Mesh}
Resolution elements in \gizmo{} are unstructured, while the resolution elements in \hamr{} are structured irregularly due to the combination of polar 1D SMR (e.g., the azimuthal resolution increases closer to the equator) and 3D AMR. To interpolate accurately, we organize the \gizmo{} resolution elements into a tetrahedral mesh, wherein each point is a ``vertex'' which is connected to other vertices via faces and edges. We do this using the software \texttt{TetGen} \citep{TETGEN}, which generates Delaunay tetrahedral meshes out of input data. The ``Delaunay'' properties -- namely, that the circumsphere of each tetrahedron in the mesh does not overlap with the vertices of other tetrahedra -- helps ensure that the mesh is generated in a way that avoids undesirable tetrahedra, such as those with very little volume. Once we construct the tetrahedral mesh, we record all of the vertices, at which we know the \gizmo{} data, along with all of the tetrahedra faces which connect the vertices, and read this information into \hamr{}.

\subsection{Construction of Binary Space Partitioning Tree}
Once the tetrahedral mesh is read into \hamr{}, we can interpolate the vertex data to \hamr{} grid cells. To do this, we need to find the tetrahedron within which each cell lies. This is computationally expensive, so we start by building a ``binary space partitioning'' (BSP) tree. Specifically, we begin by defining a cubical domain with the minimum and maximum Cartesian bounds of the tetrahedral mesh in the $x$, $y$ and $z$ directions. This is our ``root node'' of the BSP tree and it contains the center of all tetrahedra. Then, we pick the longest length of the root node, and subdivide the root node in half along that directions to create two leaf nodes. We then check every tetrahedron in the parent node to see if its circumsphere overlaps with the bounding box of either leaf node. If it does overlap, we add the tetrahedron to that leaf nodes' list of tetrahedra. This method will over count the tetrahedra in each node, but not by too much. Next, we apply this algorithm recursively, continuing to subdivide leaf nodes unless one has $500$ tetrahedra or less. Once we finish subdividing leaf nodes, the BSP tree is finished. 

\subsection{Interpolation from Tetrahedral Mesh}

Now, we can use the BSP tree to quickly find which leaf node each \hamr{} grid cell lives in. Once we have found the node for a given cell, we search through the list of $<500$ tetrahedra that are associated with that node to find the tetrahedron that hosts the \hamr{} grid cell. This takes much less time than it would take to search the $\mathcal{O}(10^8)$ total number of tetrahedra. Then, given the coordinates and data located at the four vertices of the tetrahedron, we use barycentric coordinates to linearly interpolate to the \hamr{} cell. We can do this by calculating the four $C_i$ coefficients that satisfy,
\begin{equation}
\vec{P}=\displaystyle\sum_0^3\vec{P}_iC_i
\end{equation}
where $\vec{P}_i$ is a vector to vertex $i$ of the tetrahedron and $\vec{P}$ is the position vector to the \hamr{} cell. We use volume weights $C_i\equiv V_i/V$, where $V$ is the volume of the tetrahedron, and $V_i$ is the volume of a ``sub''-tetrahedron, wherein vertex $\vec{P}_i$ is replaced with $\vec{P}$. We can then interpolate any scalar quantity $S$ as
\begin{equation}
S=\displaystyle\sum_0^3S_iC_i,
\end{equation}
We iterate through the grid and perform this interpolation on each cell center. However, magnetic field components exist on cell faces in \hamr{}. So, we also iterate through the ``staggered'' grid to interpolate the magnetic fields. Additionally, we in some cases want to apply layers of AMR on our initial conditions. In this case, after interpolating to our initial grid, we add AMR levels one by one. Whenever we add an AMR level to our initial condition, we interpolate once again to both the unstaggered and staggered mesh on the added blocks. 

Since \hamr{} is a relativistic code, we have to turn the interpolated Newtonian quantities into relativistic quantities. Since the Newtonian data is non-relativistic, we can neglect the Lorentz factor of the gas and regard the interpolated scalars as fluid-frame quantities and no modification is necessary. For the velocity and magnetic field vectors, we begin by transforming them from (Newtonian) Cartesian coordinates into (Newtonian) spherical coordinates. Then, in the coordinate frame, we can transform them into relativistic modified Kerr-Schild coordinates using the relation,
\begin{equation}
\begin{aligned}
    &v_i = \sqrt{g_{ii}}u^i\\
    &\mathcal{B}_i = \sqrt{g_{ii}}B^i,
\end{aligned}
\end{equation}
where $g_{\mu\nu}$ is the covariant metric tensor and $i$ refers to the usual Latin indices $1,2$ and $3$. This step is justified because in the large $r$ limit, the spatial components of the modified Kerr-Schild metric are equal to diagonal spherical coordinates. We do not need to worry about the time component of the four-velocity as it is automatically given by the condition $u^\mu u_\mu=-1$. Our interpolation is now complete, but the remap is not done as interpolated magnetic fields will have a small, non-zero divergence which must be cleaned (Appendix \ref{app:divb_cleaning}).

\section{Divergence Cleaning}
\label{app:divb_cleaning}

\subsection{General 
considerations}
\label{app:divb_cleaning:general}

The interpolation of data results in truncation errors. For most quantities, these errors are acceptable. However, this results in a non-zero divergence of the magnetic field. The artificial magnetic monopoles that result from this have poorly constrained behavior, so we want to remove the divergence from the magnetic field. This is, in general, arduous. Here we will motivate our approach, and if the reader is only interested in the details of the method, they can continue to Section \ref{app:divb_cleaning:boundaries}.

Since the divergence of the magnetic field, $\vec{B}$, is (supposed to be) zero, we can describe it with a vector potential, $A$, via the relation $\vec{B}=\vec{\nabla}\times\vec{A}$. If instead the field is polluted, we instead decompose it as
\begin{equation}
    \vec{B} = -\vec{\nabla}P + \vec{\nabla}\times\vec{A},
\label{eq:app:newtonian_helmholtz_decomposition}
\end{equation}
where $P$ is the magnetic scalar potential, which is associated with the unwanted part of the magnetic field\footnote{Note here we are using non-relativistic definitions -- this is for simplicity and we will introduce relativistic analogoues of the relevant equations shortly.}. From Eq.~\ref{eq:app:newtonian_helmholtz_decomposition}, there are two standard methods to removing magnetic monopoles. The first method involves taking the curl of Equation \ref{eq:app:newtonian_helmholtz_decomposition} to eliminate $\vec{\nabla}P$,
\begin{equation}
    \vec{\nabla}\times\vec{B}=\vec{\nabla}^2\vec{A}
\label{eq:app:newtonian_vector_potential_solve}
\end{equation}
wherein we solve for $\vec{A}$ and ignore $P$. The second possibility is taking the divergence of Equation \ref{eq:app:newtonian_helmholtz_decomposition} to eliminate $\vec{\nabla}\times\vec{A}$, 
\begin{equation}
    \vec{\nabla}\cdot\vec{B}=\nabla^2P,
\label{eq:app:newtonian_scalar_potential_solve}
\end{equation}
in which we solve for $P$ directly and ignore $\vec{A}$. Both Equations \ref{eq:app:newtonian_vector_potential_solve} and \ref{eq:app:newtonian_scalar_potential_solve} are similar in that they require the inversion of a linear operator, $\mathcal{L}$. The general problem can be written,
\begin{equation}
    \mathcal{L}x = b
\label{eq:app:linear_form}
\end{equation}
Here, we know coulmn vector $b$ and must invert $\mathcal{L}$ to solve for column vector $x$. In the former case, $\mathcal{L}=\vec{\nabla}^2$, and in the latter case, $\mathcal{L}=\nabla^2$. Since these matrixes span the entire grid, solving Equation \ref{eq:app:linear_form} can be prohibitively expensive, both in terms of memory and computational time.

The memory constraints can be eased by using sparse matrices, wherein only the number of nonzero elements, $N_\mathcal{L}$, of $\mathcal{L}$ are stored. The time constraints can be eased by solving Equation \ref{eq:app:linear_form} iteratively, rather than directly. Iterative solves are cheaper and more scalable, but are approximate. Direct solves are guaranteed to exactly clean the field, but are much more expensive, and the computation time can easily scale as $N_\mathcal{L}^3$ depending on the number and distribution of nonzeros in $\mathcal{L}$. While iterative solvers can parallelize more effectively, it is difficult to make significant gains parallelizing direct solves. In the vector potential approach, any errors in the iterative solver appear in $\vec{A}$, so iterative solvers are preferred. However, in the scalar potential approach these errors appear in $\vec{B}$ (and $\divb$), so direct solvers are preferred.

The arguments about scalablity might make using iterative solvers on a vector potential solve seem like the better approach. However, vector potentials have three components to solve for. Equation \ref{eq:app:newtonian_vector_potential_solve} is separable for Cartesian grids in the Coulomb gauge. So, one can solve $\mathcal{L}x=b$ thrice for each component and in each case $N_\mathcal{L}=N^2$, where $N$ is the number of grid cells. However, we use a spherical grid\footnote{This is also, for instance, why we can't pursue a Fourier approach for solving Equation \ref{eq:app:newtonian_scalar_potential_solve}, which can be much faster. (CITE)}, so the terms in $\vec{\nabla}^2$ mix and we must solve for each component of $\vec{A}$ simultaneously. This means the total number of elements in $\mathcal{L}$ is $N_\mathcal{L}=(3N)^2$, and the vector potential solve scales poorly. The mixing in $\vec{\nabla}^2$ also complicates the distribution of nonzeros in $\mathcal{L}$, which likely impacts the cost scaling but is difficult to assess a priori. Given these concerns, we have opted to perform direct solves of the magnetic scalar potential. 

However, our grids easily have $10-100$ million elements, which is too large for a direct solve. Even if the grid were smaller, our initial conditions are structured irregularly due to static and adaptive mesh refinement (SMR/AMR). This complicates the construction of $\mathcal{L}$, which is dependent on our discretization scheme. It is difficult to write down a general discretization scheme for AMR grids and encompass it within $\mathcal{L}$. To make progress, we introduce a method to break up the global solve into many smaller solves. Our grid is made of blocks of cells of uniform resolution. SMR and AMR routines affect the distribution of blocks, but the total number of cells per block is fixed, which simplifies the construction of $\mathcal{L}$ on a block level. Our method solves (the relativistic analogue of) Equation \ref{eq:app:newtonian_scalar_potential_solve} on the block level and stitches together a solution on the entire domain, as we describe in the following section. 

\subsection{Cleaning magnetic fields on block boundaries}
\label{app:divb_cleaning:boundaries}

Solving for the magnetic scalar potential requires boundary conditions. During a solve on an individual block, we regard the magnetic flux crossing the boundaries to be fixed. This boundary condition is ``good'' if the magnetic flux integrated over the block surface is (nearly) zero. However, after interpolation, this condition is not met. So, we need to adjust the magnetic fluxes leaving the boundary of each block to satisfy this condition. 

\subsubsection{Boundary Cleaning Algorithm}
\label{app:divb_cleaning:boundaries:algorithm}

\begin{figure*}[bht]
    \centering
    \includegraphics[width=0.75\textwidth]{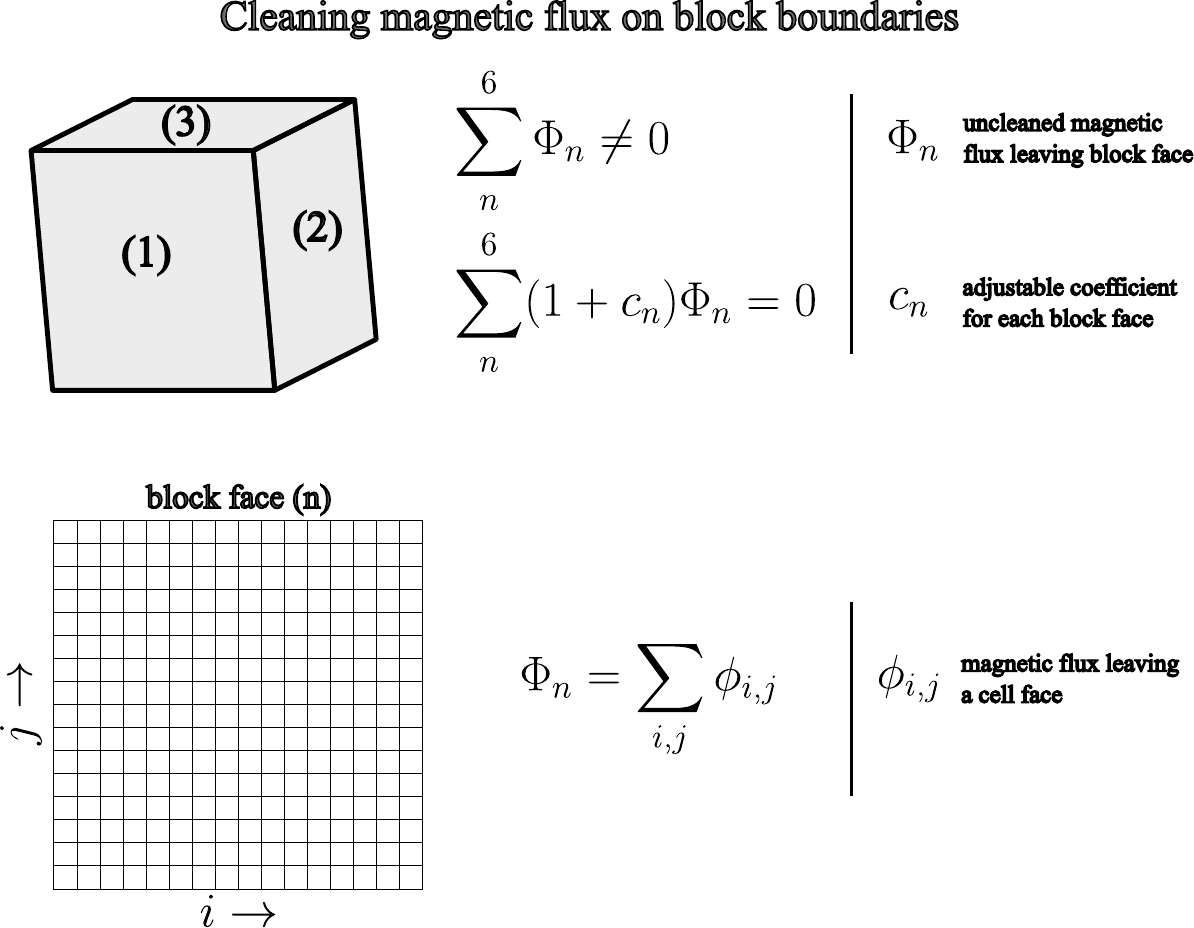}
    \caption{Schematic diagram for the boundary cleaning step in our divergence cleaning numerical routine.}
    \label{app:fig:boundary_cleaning_diagram}
\end{figure*}

Let us use Figure \ref{app:fig:boundary_cleaning_diagram} as a reference. First, recall that we are using a staggered mesh, so our magnetic field is defined on cell faces. Now, consider a block, as in the top left of Fig. \ref{app:fig:boundary_cleaning_diagram}. It has six faces, which we identify with $n$. The integrated magnetic flux leaving a given face is
\begin{equation}
    \Phi_n = \sum\limits_{i,j}\phi_{i,j}^{(n)},
\end{equation}
where we define $\phi^{(n)}_{i,j}$ as 
\begin{equation}
n = 
\begin{cases}
1, & -x_2,\,\,\phi^{(1)}_{i,j} = -B^{(i,-1/2,j)}_2\,dx_1dx_3 \\
2, & +x_1,\,\,\phi^{(2)}_{i,j} = B^{(N_1-1/2,i,j)}_1\,dx_2dx_3 \\
3, & +x_2,\,\,\phi^{(3)}_{i,j} = B^{(i,N_2-1/2,j)}_2\,dx_1dx_3 \\
4, & -x_1,\,\,\phi^{(4)}_{i,j} = -B^{(-1/2,i,j)}_1\,dx_2dx_3 \\
5, & +x_3,\,\,\phi^{(5)}_{i,j} = B^{(i,j,N_3-1/2)}_3\,dx_1dx_2  \\
6, & -x_3,\,\,\phi^{(6)}_{i,j} = -B^{(i,j,-1/2)}_3\,dx_1dx_2 
\end{cases}
\end{equation} 
This convention is specific to \hamr{} and may seem pathological, but follows from the right-hand rule with ones thumb pointing in the $+x_3$ direction and fingers pointing in the $-x_2$ direction. Our convention is that $x_1$, $x_2$ and $x_3$ are the radial, polar and azimuthal directions, but the distinction is unimportant. We proceed by assuming that the total magnetic flux leaving a block is non-zero, 
\begin{equation}
 \sum\limits_n^6\Phi_n\neq0,
 \label{eq:app:boundary_neq_0}
\end{equation}
We want to adjust $\Phi_n$ to obtain new magnetic fluxes $\Phi_n'$ such that $\sum\limits_n^6\Phi_n'=0,$. We do this by introducing adjustable coefficients $c_n$ for each face of the block,
\begin{equation}
\sum\limits_n^6(1+c_n)\Phi_n=0,
\label{eq:app:boundary_condition}
\end{equation}
where we must determine $c_n$. We treat this as a nonlinear optimization problem and use the ``Sequential Least Squares Programming'' (SLSQP) algorithm in the \verb|NLopt| software \citep{NLopt}. This is a local gradient-based algorithm that admits both a nonlinear equality constraint and an objective function to minimize. The objective function to minimize is,
\begin{equation}
    \sum\limits_n^6c_n^2 < \verb|OBJ_TOL|,
\label{eq:app:objective_function}
\end{equation}
where we set the tolerance \verb|OBJ_TOL| to $10^{-2}$. This objective function minimizes the coefficients $c_n$, because we want to minimally adjust the initial fluxes. Simultaneously, we use the following nonlinear equality constraint, 
\begin{equation}
\sum\limits_n^6(1+c_n)\Phi_n<\verb|CONSTR_TOL|,
\label{eq:app:boundary_condition}
\end{equation}
where \verb|CONSTR_TOL| is the tolerance to which we admit an integrated $\vec{\nabla}\cdot\vec{B}$ error which we set to $100\times$\verb|DBL_EPSILON|, where \verb|DBL_EPSILON| is the machine precision for double precision number. Additionally, we also bound $c_n$,
\begin{equation}
    |c_n|<\verb|BND_LIM|
\end{equation}
where \verb|BND_LIM| is set to be $0.5$ but in practice $c_n$ is usually minimized to values much smaller than this. Once the coefficients $c_n$ are found for a given block, then the magnetic flux leaving each cell on a block face is updated as $\phi_{i,j}^{(n)}\rightarrow(1+c_n)\phi_{i,j}^{(n)}$. 

\subsubsection{Iterating through each block on the grid}
\label{app:divb_cleaning:boundaries:gridwalk}

The block boundary cleaning algorithm introduced in Section \ref{app:divb_cleaning:boundaries:algorithm} must be repeated for every block in the grid, but blocks share boundaries. If we first clean a block somewhere in our grid and then move to a neighbor, their shared boundary is already cleaned. So, as we ``walk'' through the grid, cleaning each block, there are fewer and fewer uncleaned boundaries remaining. We run the risk of some blocks having zero remaining boundaries to clean. Additionally, our grid has a non-trivial structure due to the combination of SMR and AMR. So, we must proceed carefully. We use the following approach,
\begin{enumerate}
    \item First, we choose how to iterate through the grid. At any refinement level above the base level, only some blocks are active. We treat every refinement level as a uniform grid composed of inactive and active blocks to make iteration simple, and skip any inactive blocks. In \hamr{}, SMR complicates iteration because it only acts on the $\varphi$ direction as a function of $\theta$, while AMR acts on each dimension as a function of the refinement criterion. So, we first begin on the highest SMR level, which exists at the equator. Within this level, we begin at the highest AMR level, and iterate first in the radial, then polar, then azimuthal directions, beginning at the innermost block in each direction. We then jump to the next AMR refinement level, and repeat. Once we've exhausted the AMR refinement levels on a given SMR level, we move to the next lowest SMR level, and repeat. 
    \item Each MPI process holds different blocks. When we clean a blocks' boundaries, its neighboring blocks may be held by different MPI processes. So, after iterating on a block, we send the cleaned boundary data to the MPI processes holding the neighboring blocks on each refinement level. Every time we move to a new block, we must check if any of its neighbors have been cleaned, and receive the data from the MPI processes that hold the neighboring cleaned blocks. This includes the possibility of coarse-fine boundaries.
    \item Finally, we have some edge cases to consider at our grid boundaries. The azimuthal boundary conditions on periodic and require no special handling. We use transmissive polar boundary conditions and enforce the magnetic flux exiting the poles to be zero. We use outflow radial boundaries, so magnetic flux can freely leave the domain. Wherever we have a block at a radial boundary, we use a different approach than outlined in Section \ref{app:divb_cleaning:boundaries:algorithm}. Instead of optimizing for the coefficients $c_n$, we calculate the excess magnetic flux $\delta \Phi = \sum\limits_i^6\Phi_n$, and update the magnetic flux leaving the radial boundary to be $\Phi\rightarrow\Phi-\delta \Phi$; e.g., we dispose of the magnetic monopoles by depositing them outside the domain.
\end{enumerate}
After iterating through the entire grid and updating each of the block boundaries, we can perform scalar divergence cleaning on~ each block independently.

\subsection{Sub-Block Boundary Cleaning}
\label{app:divb_cleaning:boundaries:subblocks}

We have also introduced an additional step in our cleaning procedure, which is technically optional but we have included it in this work as an intermediary step between the  block boundary cleaning and the magnetic scalar potential cleaning. If we use blocks of size say $64^3$, a serial direct solve of the magnetic scalar potential is still extremely expensive. So, we take individual blocks and break them into a $2x2x2$ grid of sub-blocks. The external faces of these sub-blocks are already cleaned, but now we want to clean their internal faces. We do this via the exact same procedure done in Section \ref{app:divb_cleaning:boundaries:algorithm}, so there is no significant complication. We do note that the last sub-block to be cleaned has no remaining uncleaned faces and thus no remaining degrees of freedom, but if all its neighboring faces have been cleaned it will automatically be clean as well. We can then perform direct solves of the magnetic scalar potential on individual sub-blocks, which is much faster. 

\subsection{Scalar Divergence Cleaning}
\label{app:divb_cleaning:boundaries:scalar_divergence_cleaning}
We now introduce our scalar divergence cleaning routine. The perform a general-relativistic Helmholtz decomposition of the (uncleaned) magnetic field into scalar and vector potentials is,
\begin{equation}
    B^i = -\nabla^iP + \epsilon_{ijk}\nabla^jA^k,
\end{equation}
where here Latin indices span $i=1$ to $i=3$ and indicate spatial directions. As before, $P$ is the magnetic scalar potential, and $A^k$ is the contravariant magnetic vector potential. Then, the divergence of the magnetic field is,
\begin{equation}
    \nabla_iB^i=-\nabla^2P,
\label{eq:app:scalar_potential_poisson_gr}
\end{equation}
where $\nabla_i$ is the covariant derivative over spatial indices.
We can write the divergence operator as
\begin{equation}
    \nabla_iB^i = \frac{1}{\sqrt{-g}}\partial_i(\sqrt{-g}B^i),
\end{equation}
where $g$ is the metric determinant. We can expand the gradient of the scalar potential,
\begin{equation}
    \nabla^iP = \gamma^{ij}\partial_jP,
\end{equation}
where $\gamma^{ij}$ is the spatial component of the contravariant metric. In this last step we are able to take $\nabla_j\rightarrow\partial_j$ because $P$ is a scalar and thus does not depend on the basis vectors. We can then expand the Laplacian,
\begin{equation}
    \nabla^2P = \nabla_i\gamma^{ij}\partial_jP = \frac{1}{\sqrt{-g}}\partial_i(\sqrt{-g}\gamma^{ij}\partial_jP)
\end{equation}
We can combine this with Equation \ref{eq:app:scalar_potential_poisson_gr} to find,
\begin{equation}
\frac{1}{\sqrt{-g}}\partial_i(\sqrt{-g}B^i) = -\frac{1}{\sqrt{-g}}\partial_i(\sqrt{-g}\gamma^{ij}\partial_j P)
\end{equation}
Then, the ``cleaned'' component of the magnetic field is
\begin{equation}
    B^i_{(c)} = B^i + \nabla^iP
    \label{eq:gr_clean_field}
\end{equation}
These equations now need to be discretized properly. When writing down discretized equations, we will change our notation. The components of the vector or tensor quantities will explicitly  be labeled $1$, $2$ or $3$, and $i$, $j$ and $k$ will indicate cell centers. We will also use half-steps, e.g. $i+1/2$, to indicate cell faces. We have defined $P$ at cell centers. When discretized, the radial component of Equation \ref{eq:gr_clean_field} becomes
\begin{equation}
\begin{aligned}
&B_{(c),i+1/2,j,k}^{1} = B_{i+1/2,j,k}^{1} + \gamma^{11}_{i+1/2,j,k}\left(\frac{P_{i+1,j,k} - P_{i,j,k}}{\Delta x_1}\right)\\& + 
\gamma^{12}_{i+1/2,j,k}\left(\frac{P_{i+1,j+1,k}+P_{i,j+1,k} - P_{i+1,j-1,k} - \phi_{i,j-1,k}}{4\Delta x_2}\right)\\& + 
\gamma^{13}_{i+1/2,j,k}\left(\frac{P_{i+1,j,k+1} + P_{i,j,k+1} - P_{i+1,j,k-1} - P_{i,j,k-1}}{4\Delta x_3}\right),
\end{aligned}
\label{eq:app:bclean1}
\end{equation}
the polar component becomes
\begin{equation}
\begin{aligned}
&B^2_{{\rm (c)},i,j+1/2,k} = B^2_{i,j+1/2,k}\\& +
\gamma^{21}_{i,j+1/2,k}\left(\frac{P_{i+1,j+1,k} + P_{i+1,j,k} - P_{i-1,j+1,k} - P_{i-1,j,k}}{4\Delta x_1}\right)\\& + 
\gamma^{22}_{i,j+1/2,k}\left(\frac{P_{i,j+1,k}-P_{i,j,k}}{\Delta x_2}\right)\\& + 
\gamma^{23}_{i,j+1/2,k}\left(\frac{P_{i,j+1,k+1} + P_{i,j,k+1} - P_{i,j+1,k-1} - P_{i,j,k-1}}{4\Delta x_3}\right),
\end{aligned}
\label{eq:app:bclean2}
\end{equation}
and the azimuthal component becomes
\begin{equation}
\begin{aligned}
&B^3_{{\rm (c)},i,j,k+1/2} = B^3_{i,j,k+1/2}\\& + 
\gamma^{31}_{i,j,k+1/2}\left(\frac{P_{i+1,j,k+1} + P_{i+1,j,k} - P_{i-1,j,k+1} - P_{i-1,j,k}}{4\Delta x_1}\right)\\& + 
\gamma^{32}_{i,j,k+1/2}\left(\frac{P_{i,j+1,k+1} + P_{i,j+1,k} - P_{i,j-1,k+1} - P_{i,j-1,k}}{4\Delta x_2}\right)\\& + 
\gamma^{33}_{i,j,k+1/2}\left(\frac{P_{i,j,k+1}-P_{i,j,k}}{\Delta x_3}\right)
\end{aligned}
\label{eq:app:bclean3}
\end{equation}
We also need to discretize the divergence of the magnetic field. Here and in the following equations, we write the discretized $\sqrt{-g}$ as $g_{i,j,k}$ for brevity,
\begin{equation}
\begin{aligned}
&{\rm divB}_{i,j,k} = \frac{1}{g_{i,j,k}}\biggl(\frac{g_{i+1/2,j,k}B^1_{i+1/2,j,k}-g_{i-1/2,j,k}B^1_{i-1/2,j,k}}{\Delta x_1} +  \\&\frac{g_{i,j+1/2,k}B^2_{i,j+1/2,k}-g_{i,j-1/2,k}B^2_{i,j-1/2,k}}{\Delta x_2} + \\&\frac{g_{i,j,k+1/2}B^3_{i,j,k+1/2}-g_{i,j,k-1/2}B^3_{i,j,k-1/2}}{\Delta x_3}\biggr)
\end{aligned}
\end{equation}
We can use this expression and take $B\rightarrow B_{\rm (c)}$, for which we assume ${\rm divB}_{{\rm (c)}}=0$, and then input Equations \ref{eq:app:bclean1}-\ref{eq:app:bclean3} to acquire
\clearpage
\begin{widetext}
\begin{equation}
\begin{aligned}
&-g_{i,j,k}{\rm div}B_{i,j,k}=\\
+&\frac{g_{i+1/2,j,k}}{\Delta x_1}\left( \gamma^{11}_{i+1/2,j,k}\left(\frac{P_{i+1,j,k} - P_{i,j,k}}{\Delta x_1}\right) + 
\gamma^{12}_{i+1/2,j,k}\left(\frac{P_{i+1,j+1,k}+P_{i,j+1,k} - P_{i+1,j-1,k} - P_{i,j-1,k}}{4\Delta x_2}\right) + 
\gamma^{13}_{i+1/2,j,k}\left(\frac{P_{i+1,j,k+1} + P_{i,j,k+1} - P_{i+1,j,k-1} - P_{i,j,k-1}}{4\Delta x_3}\right)\right) \\
-& \frac{g_{i-1/2,j,k}}{\Delta x_1}\left( \gamma^{11}_{i-1/2,j,k}\left(\frac{P_{i,j,k} - P_{i-1,j,k}}{\Delta x_1}\right) + 
\gamma^{12}_{i-1/2,j,k}\left(\frac{P_{i,j+1,k}+P_{i-1,j+1,k} - P_{i,j-1,k} - P_{i-1,j-1,k}}{4\Delta x_2}\right) + 
\gamma^{13}_{i-1/2,j,k}\left(\frac{P_{i,j,k+1} + P_{i-1,j,k+1} - P_{i,j,k-1} - P_{i-1,j,k-1}}{4\Delta x_3}\right)\right) \\
+&\frac{g_{i,j+1/2,k}}{\Delta x_2}\left(\gamma^{21}_{i,j+1/2,k}\left(\frac{P_{i+1,j+1,k} + P_{i+1,j,k} - P_{i-1,j+1,k} - P_{i-1,j,k}}{4\Delta x_1}\right) + 
\gamma^{22}_{i,j+1/2,k}\left(\frac{P_{i,j+1,k}-P_{i,j,k}}{\Delta x_2}\right) + 
\gamma^{23}_{i,j+1/2,k}\left(\frac{P_{i,j+1,k+1} + P_{i,j,k+1} - P_{i,j+1,k-1} - P_{i,j,k-1}}{4\Delta x_3}\right)\right) \\
-&\frac{g_{i,j-1/2,k}}{\Delta x_2}\left(\gamma^{21}_{i,j-1/2,k}\left(\frac{P_{i+1,j,k} + P_{i+1,j-1,k} - P_{i-1,j,k} - P_{i-1,j-1,k}}{4\Delta x_1}\right) + 
\gamma^{22}_{i,j-1/2,k}\left(\frac{P_{i,j,k}-P_{i,j-1,k}}{\Delta x_2}\right) + 
\gamma^{23}_{i,j-1/2,k}\left(\frac{P_{i,j,k+1} + P_{i,j-1,k+1} - P_{i,j,k-1} - P_{i,j-1,k-1}}{4\Delta x_3}\right)\right)\\
+&\frac{g_{i,j,k+1/2}}{\Delta x_3}\left(\gamma^{31}_{i,j,k+1/2}\left(\frac{P_{i+1,j,k+1} + P_{i+1,j,k} - P_{i-1,j,k+1} - P_{i-1,j,k}}{4\Delta x_1}\right) + 
\gamma^{32}_{i,j,k+1/2}\left(\frac{P_{i,j+1,k+1} + P_{i,j+1,k} - P_{i,j-1,k+1} - P_{i,j-1,k}}{4\Delta x_2}\right) + 
\gamma^{33}_{i,j,k+1/2}\left(\frac{P_{i,j,k+1}-P_{i,j,k}}{\Delta x_3}\right)\right)\\
-&\frac{g_{i,j,k-1/2}}{\Delta x_3}\left(\gamma^{31}_{i,j,k-1/2}\left(\frac{P_{i+1,j,k} + P_{i+1,j,k-1} - P_{i-1,j,k} - P_{i-1,j,k-1}}{4\Delta x_1}\right) + 
\gamma^{32}_{i,j,k-1/2}\left(\frac{P_{i,j+1,k} + P_{i,j+1,k-1} - P_{i,j-1,k} - P_{i,j-1,k-1}}{4\Delta x_2}\right) + 
\gamma^{33}_{i,j,k-1/2}\left(\frac{P_{i,j,k}-P_{i,j,k-1}}{\Delta x_3}\right)\right)
\end{aligned}
\label{eq:app:poisson_solve_discrete_full}
\end{equation}
\end{widetext}

This equation can be written in matrix form $\mathcal{L}x=b$ by taking $x$ as column vector $P_{i,j,k}$ and $b$ as column vector $-g_{i,j,k}{\rm div}B_{i,j,k}$, both flattened over three dimensions. Then, $\mathcal{L}$ encodes the derivative stencil in a (sparse) matrix. As written above, most of the nonzeros in the matrix are off-diagonal components of the metric, which makes the implementation more tedious and slows down the solve. However, these off-diagonal components only contribute to the matrix solve in the strong gravity region (which is irrelevant in our use case since our interpolated data is Newtonian), and the scalar divergence cleaning procedure is not unique. So, we are free to neglect the off diagonal terms in Equation \ref{eq:app:poisson_solve_discrete_full}, which leaves us with the much simpler expression
\begin{widetext}
\begin{equation}
\begin{aligned}
&-g_{i,j,k}{\rm div}B_{i,j,k}=\\
+&\frac{g_{i+1/2,j,k}}{\Delta x_1} \gamma^{11}_{i+1/2,j,k}\left(\frac{P_{i+1,j,k} - P_{i,j,k}}{\Delta x_1}\right) 
- \frac{g_{i-1/2,j,k}}{\Delta x_1}\gamma^{11}_{i-1/2,j,k}\left(\frac{P_{i,j,k} - P_{i-1,j,k}}{\Delta x_1}\right) \\
+&\frac{g_{i,j+1/2,k}}{\Delta x_2} 
\gamma^{22}_{i,j+1/2,k}\left(\frac{P_{i,j+1,k}-P_{i,j,k}}{\Delta x_2}\right) 
-\frac{g_{i,j-1/2,k}}{\Delta x_2}
\gamma^{22}_{i,j-1/2,k}\left(\frac{P_{i,j,k}-P_{i,j-1,k}}{\Delta x_2}\right)\\
+&\frac{g_{i,j,k+1/2}}{\Delta x_3}
\gamma^{33}_{i,j,k+1/2}\left(\frac{P_{i,j,k+1}-P_{i,j,k}}{\Delta x_3}\right)
-\frac{g_{i,j,k-1/2}}{\Delta x_3}
\gamma^{33}_{i,j,k-1/2}\left(\frac{P_{i,j,k}-P_{i,j,k-1}}{\Delta x_3}\right),
\end{aligned}
\end{equation}
\end{widetext}
and we are now done setting up our problem. In practice, we solve this equation on each (sub-)block using the sparse linear algebra package \verb|SuperLU| \citep{superlu99}. 

\subsection{Result of cleaning procedure}

\begin{figure*}[bht]
    \centering
    \includegraphics[width=\textwidth]{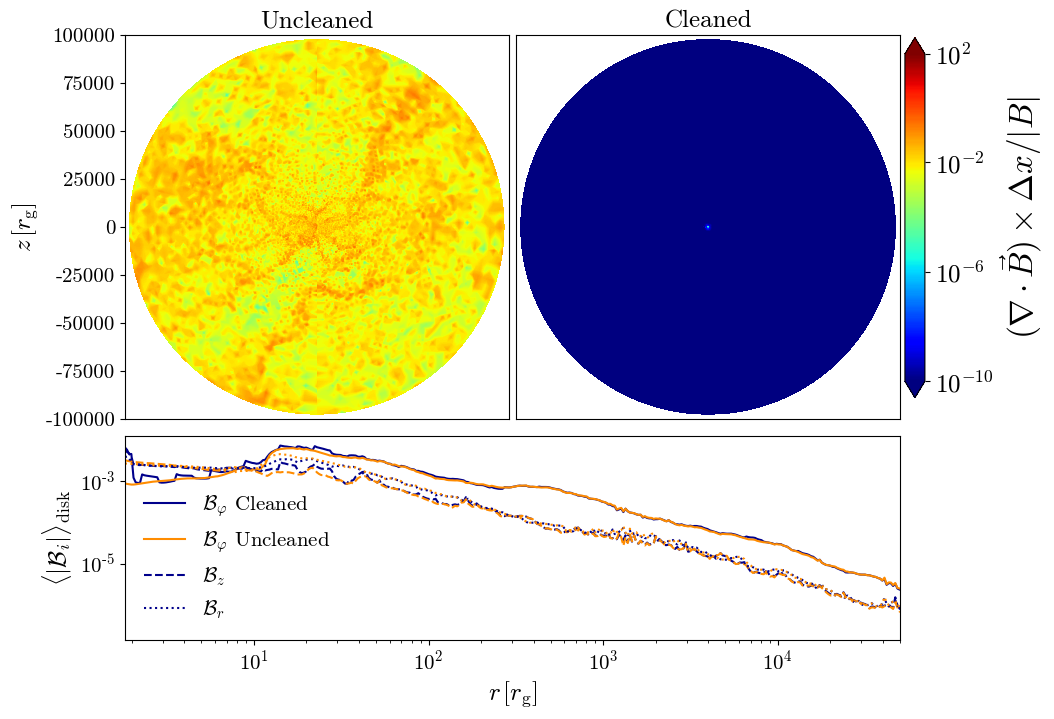}
    \caption{Results of initial interpolation and divergence cleaning. In the top, we show a snapshot of the normalized magnetic field divergence in the entire domain before and after cleaning. Before, the normalized divergence reaches values that are $\mathcal{O}(10^{-1}$), while afterwards the normalized divergence is $\mathcal{O}(10^{-10})$ everywhere except very near the BH, where is it $\mathcal{O}(10^{-7})$. In the bottom panel, we show the disk-averaged (Eq.~\ref{eq:disk_avg}) value of each component of the field (Eq.~\ref{eq:mag_vel_phys}) before and after cleaning. The change in the field is small everywhere, with the largest differences near the event horizon. We also note that since we interpolated data that extended down to $10\,r_{\rm g}$, the magnetic field within this region is totally made up.}
    \label{fig:app:divb}
\end{figure*}

In Figure \ref{fig:app:divb}, we show the result of our cleaning process. In the top panels, we show a snapshot of $\divb\times \Delta x/|B|$ in the entire domain, before and after cleaning. Before cleaning, the normalized divergence reaches values $\mathcal{O}(10^{-1})$, which is expected given the truncation errors in the interpolation scheme. After cleaning, the normalized divergences reaches $\mathcal{O}(10^{-10})$ almost everywhere. Near the BH, where the field is essentially made up\footnote{While this is not ideal, the near-BH region is small in volume, and the field added here is extremely sub-dominant to the reservoir of magnetic flux advected from large radii. This will wash away early transients.} due to a lack of \gizmo{} data within $10\,r_{\rm g}$, the normalized divergence reaches $\mathcal{O}(10^{-7})$. We consider this an extremely good result. Additionally, we show the disk average (Eq.~\ref{eq:disk_avg}) of the absolute value of the ``physical'' components of the magnetic field (Eq.~\ref{eq:mag_vel_phys}) before and after interpolation. The difference is usually very small, with the largest discrepancies happening within $10\,r_{\rm g}$. This suggests that our interpolation of the \gizmo{} data is faithful.

\section{Evolution of Initial Conditions}

\begin{figure}[bht]
    \centering
    \includegraphics[width=\textwidth]{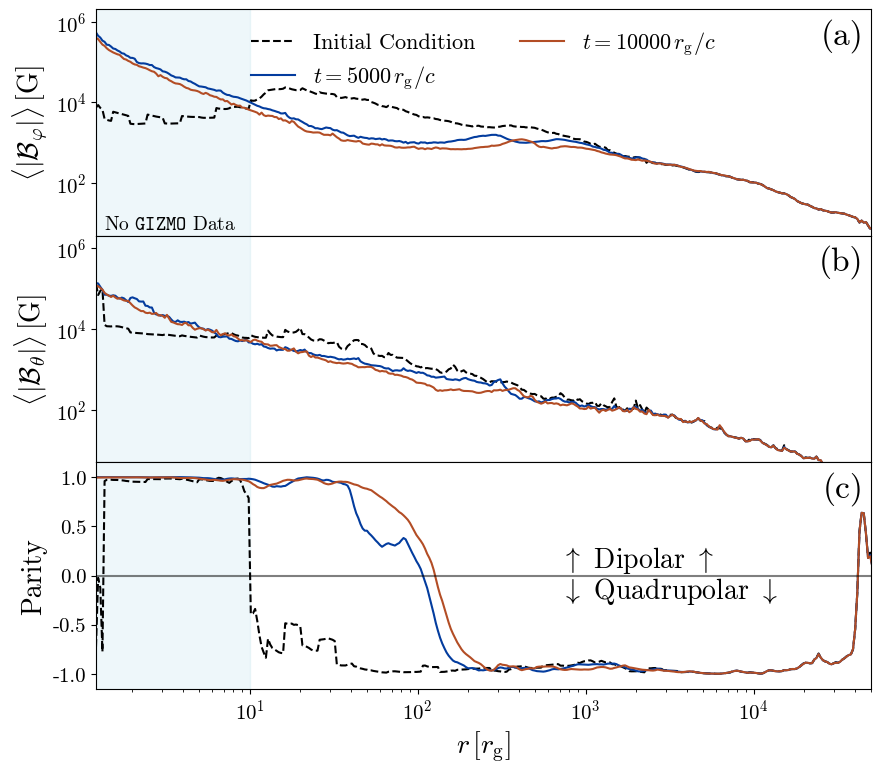}
    \caption{Initial conditions and early evolution of various magnetic properties. We show profiles at times $t=0$, $5000$ and $10000\,r_{\rm g}/c$. We have shaded the region where we did not have \gizmo{} data to interpolate, and thus made up the magnetic field artificially. \textbf{Panel a.} Disk-average (Eq.~\ref{eq:disk_avg}) of $|\mathcal{B}_\varphi$ (Eq.~\ref{eq:mag_vel_phys}). \textbf{Panel b.} Disk-average of $|\mathcal{B}_\theta$. \textbf{Panel c.}. Profiles of the parity of the magnetic field ($C(U^{\rm D},U^{\rm Q})$) see Eqs.~\ref{eq:quad_dip_energies}-\ref{eq:compare_func}).
    } 
    \label{fig:app:bquantities}
\end{figure}

In Figure \ref{fig:app:bquantities}, we show radial profiles of magnetic quantities in the disk at times $t=0$, $5000$ and $10000\,r_{\rm g}/c$ to assess how much our results diverge from the \gizmo{} results. We have shaded the region within $10\,r_{\rm g}$ to indicate that we had no \gizmo{} data within this region, and thus all properties in this region are made up by our interpolation scheme. In Fig.~\ref{fig:app:bquantities}(a), we show $\langle |\mathcal{B}_\varphi|\rangle_{\rm disk}$, which is the disk average (Eq.~\ref{eq:disk_avg}) of the absolute value of the toroidal magnetic field in ``physical'' coordinates (Eq.~\ref{eq:mag_vel_phys}). Our toroidal field decays within the inner few hundred gravitational radii, which results from the magnetic state transition to an accretion flow dominated by net vertical magnetic flux. While this is quickly established within the first $5000\,r_{\rm g}/c$, the toroidal field does not change significantly within the next $5000\,r_{\rm g}/c$. In Fig.~\ref{fig:app:bquantities}(b), we show $\langle |\mathcal{B}_\theta|\rangle_{\rm disk}$, which has changed by an order unity factor from the initial condition at the depicted times. 

\begin{figure}[bht]
    \centering
    \includegraphics[width=\textwidth]{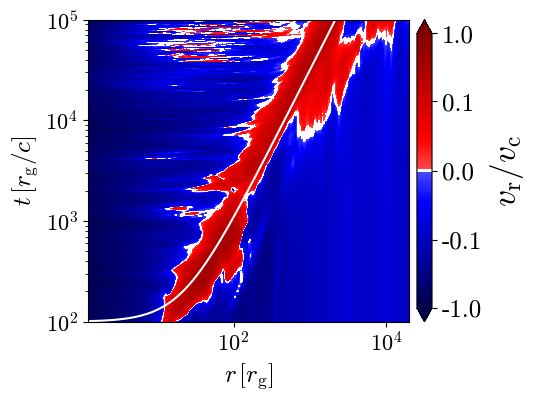}
    \caption{Spacetime diagram of radial velocity in simulation HS. We highlight the radial velocity $v_{\rm r}$ (Eq. \ref{eq:mag_vel_phys}) normalized to the local circular velocity $v_{\rm c}$ (Eq. \ref{eq:v_kep}). We have added a white curve with a $t\propto r^{3/2}$ slope to trace the local orbital timescale. We can see that there is a transient that is launched from the inner boundary of the interpolated \gizmo{} data, which begins at $r=10\,r_{\rm g}$.}
    \label{fig:app:vr_spacetime}
\end{figure}

In Fig.~\ref{fig:app:bquantities}(c), we show the parity of the magnetic field ($C(U^{\rm D},U^{\rm Q})$) see Eqs.~\ref{eq:quad_dip_energies}-\ref{eq:compare_func}) The parity is bounded between $-1$, which indicates a purely quadrupole-like magnetic field, and $1$, which indicates a purely dipole-like magnetic field. The initial condition from $\gizmo{}$ is purely quadrupolar. However, as poloidal field loops accumulated at the event horizon, the inner region becomes dipole-like, as described in Section \ref{sec:results:transition}. We attribute this difference to a difference in inner boundary condition. While the \gizmo{} simulation uses a standard sink boundary at radii well outside the horizon, which represents an unresolved spatial domain into which magnetic flux can be accretion, flux cannot be accretion across a true horizon as represented here and instead is able to accumulate in the inner accretion flow. 

In Figure \ref{fig:app:vr_spacetime}, we show a spacetime diagram of the radial velocity ($v_r$, Eq.~\ref{eq:mag_vel_phys}) normalized to the local velocity of circular orbits ($v_{\rm c}$, Eq.~\ref{eq:v_kep}). While most of the disk exhibits inflow, we can see a clear transient develop at the beginning of our simulation. It is launched from $10\,r_{\rm g}$, which is the inner radius of the interpolated \gizmo{} data, and then travels outward on the orbital timescale ($\propto r^{3/2}$, shown by the white line). As discussed in Section \ref{sec:results:transition}, this feature is both numerical and physical. It is numerical because it presumably resulted from a difference in boundary conditions between \gizmo{} and \hamr{}. However, it is also physical, because when a nascent quasar disk first accretes onto the BH, the poloidal flux must begin accumulating and would naturally launch a wave when this happens.  

While we have argued that this wave may be a result of flux accumulation, this is difficult to determine conclusively. The flow is extremely sensitive to the inner boundary condition, so even if there was no difference in flux accumulation between \gizmo{} and \hamr{}, a wave would likely be launched because of other effects such as the size of the \gizmo{} sink versus the true event horizon, or from transitioning from Newtonian to general-relativistic gravity.

\bibliographystyle{aasjournal}
\bibliography{references}
\end{document}